\documentclass[twocolumn, amssymb, nobibnotes, aps, pra,floatfix]{revtex4-1}

\usepackage{docs}
\usepackage{bm}

\usepackage{ascmac}
\usepackage{latexsym}
\usepackage{amsmath,amssymb}	
\usepackage{amsthm}
\usepackage{amsfonts}
\usepackage{bm}
\usepackage{graphicx}
\usepackage{ulem}
\usepackage{float}
\usepackage{comment}
\usepackage{color}
\usepackage{tabularx}
\usepackage{colortbl}
\usepackage{natbib}
\usepackage{subfig}

\newcommand{\ket}[1]{ |{#1}\rangle}
\newcommand{\bra}[1]{ \langle{#1}|}

\newcommand{\tr}[0]{{\rm Tr}}

\newcommand{\kket}[1]{ |{#1}\rangle\!\rangle}
\newcommand{\bbra}[1]{ \langle\!\langle{#1}|}

\newtheorem{lem}{Lemma}
\newtheorem{dfn}{Definition}
\newtheorem{thm}{Theorem}
\newtheorem{prop}{Proposition}

\allowdisplaybreaks[3]

\def\Proof{\textbf{Proof.} }

\input{qcircuit}

\begin{document}

\title{Equivalence determination of unitary operations}

\author{Atsushi Shimbo}
\email[Electronic address: ]{shimbo@eve.phys.s.u-tokyo.ac.jp}
\affiliation{Department of Physics, Graduate School of Science, The University of Tokyo}

\author{Akihito Soeda}
\email[Electronic address: ]{soeda@phys.s.u-tokyo.ac.jp}
\affiliation{Department of Physics, Graduate School of Science, The University of Tokyo}

\author{Mio Murao}
\email[Electronic address: ]{murao@phys.s.u-tokyo.ac.jp}
\affiliation{Department of Physics, Graduate School of Science, The University of Tokyo}

\date{\today}

%%%%%%%%%%%%%%%%%%%%%%%%% ABSTRACT %%%%%%%%%%%%%%%%%%%%%%%%%
\begin{abstract}
We study equivalence determination of unitary operations, a task analogous to quantum state discrimination.  The candidate states are replaced by unitary operations given as a quantum sample, \textit{i.e.}, a black-box device implementing a candidate unitary operation, and the discrimination target becomes another black-box.  The task is an instance of higher-order quantum computation with the black-boxes as input.  The optimal error probability is calculated by semidefinite programs.  Arbitrary quantum operations applied between the black-boxes in a general protocol provide advantages over protocols restricted to parallelized use of the black-boxes.  We provide a numerical proof of such an advantage.  In contrast, a parallelized scheme is analytically shown to exhibit the optimal performance of general schemes for a particular number of quantum samples of the candidates.  We find examples of finite-sample equivalence determination that achieve the same performance as when a classical description of the candidates are provided, although an exact classical description cannot be obtained from finite quantum samples.  

\end{abstract}

\maketitle

\captionsetup[figure]{
justification=RaggedRight,
}

%%%%%%%%%%%%%%%%%%%%%%%%% INTRODUCTIION %%%%%%%%%%%%%%%%%%%%%%%%%
\section{Introduction}
A typical discrimination task constitutes a ``candidate set'' and ``target object''.  The target object is equivalent to an element in the candidate set.  We are given the target object and informed of the candidate set.  The goal then is to decide which of the candidates is actually given.  Typically, the candidate set is enumerated and the aim becomes to guess the number assigned to the candidate corresponding to the given target object.  Discrimination tasks are a simplified information processing task and, conversely, various information processing tasks can be rephrased in terms of discrimination.

In quantum state discrimination, the candidate set consists of quantum states.  The target object is a quantum system prepared in a candidate state.  %A quantum measurement is performed on the target state and and the candidate is guessed based on the outcome of the measurement.  
Quantum state discrimination has been investigated for two candidate states \cite{helstrom1976quantum}, unambiguous discrimination \cite{PhysRevA.66.032315}, relations to the no-signaling principle and the no-cloning theorem~\cite{DIEKS1982271,wootters1982single, GISIN19981, PhysRevLett.107.170403}, mixed-state candidates \cite{PhysRevA.70.012308,PhysRevA.77.012328}, candidate states with geometric symmetries \cite{1193807,1302298,1176618}, bi- and multi-partite candidate states under local operations and classical communication \cite{PhysRevA.59.1070, VIRMANI200162,  PhysRevA.71.032323, PhysRevLett.85.4972, PhysRevLett.89.147901}, and the change point detection \cite{PhysRevLett.117.150502}.  See Ref.\,\cite{1751-8121-48-8-083001} for a review.

The candidate set may consist of quantum operations.  
%of discrimination are not limited to quantum states.  Quantum operations may form the candidate set, in which case 
The target object is a quantum device, provided as a black-box that implements a candidate operation.  The task is to determine which operation is performed by the target box.  %One way to discriminate quantum operations is first to choose an initial state and then to perform the target box on the state.  %The target box converts the initial state to a different state that depends on the operation implemented by the target box.  At this point, the operation discrimination reduces to a state discrimination, which is achieved by applying a quantum measurement.  
Although the state discrimination may be seen as a special case of an operation discrimination, these two types of discrimination tasks should be considered as separate problems.  For instance, a perfect discrimination is not possible for any finite number of non-orthogonal quantum states with finite copies of the target state, but it is shown that a perfect discrimination is possible for a finite candidate set of unitary operations by using the target box for a finite number of times \cite{PhysRevLett.87.177901,PhysRevLett.87.270404, PhysRevLett.98.100503}.  

Discrimination of quantum operations can be considered as an information processing task taking quantum operations as an input.  More generally, it is possible to imagine scenarios where a quantum operation is also the output of the task\,\cite{1367-2630-20-1-013004,PhysRevA.89.030303,zhou2011adding,PhysRevA.88.022318,miyazaki2017universal}.  
Conventionally, quantum operations are treated as a means to convert the states which represent the input quantum information.  
%Conventionally, the input and output of quantum information tasks have been represented by quantum states and quantum operations are treated as a means to convert the states.  
In contrast, the types of quantum information tasks that allow quantum operations to be the input and/or output belongs to what may be called \textit{functional quantum computing}\,\cite{PhysRevA.93.052321} or \textit{higher-order quantum computation} \cite{Perinotti2017}.

Quantum discrimination tasks assume that the candidate set is informed \textit{a priori}.   
%but the degree to which it is informed can vary.  
Most typically, a full \textit{classical} description of the candidates is assumed to be given.  On the other hand, it may be that the candidates are provided as a quantum object.  These quantum objects are a \textit{quantum sample} of the candidates.  Quantum state discrimination with candidate states given as a quantum sample has been investigated under various settings 
%, quantum samples are a quantum system prepared in a candidate state provided along with the number associated to the candidate\,
\cite{PhysRevA.73.062334,PhysRevLett.94.160501,HE2006103, PhysRevA.94.062320, PhysRevA.75.032316, 1367-2630-12-12-123032,PhysRevA.82.042312,PhysRevA.83.052328,PhysRevA.83.039909,PhysRevA.89.014301}.  A quantum sample in operation discrimination is another black-box implementing a candidate operation and labelled with the associated number.  %Fundamentally, these two distinct methods of informing the candidate set reveals the value of classical description in quantum information processing.  
Generally, a full classical description may be obtained from quantum samples by quantum tomography, consuming an infinite number of copies of the samples for each candidate (see Ref.\,\cite{1367-2630-15-12-125020} and references therein for a review of quantum tomography).  %In addition, to utilize a classical description of the candidates, say, to optimize whatever figure of merit we choose, implies that the necessary information processing resources must be computed and prepared for each possible candidate set, which, of course, demands separate computational and preparation cost.  A candidate-independent strategy has the advantage of avoiding these additional costs at the expense of possible loss in the optimality compared against a candidate-specific strategy.  

A figure of merit for a discrimina	tion task measures how well a discrimination protocol performs.  Commonly, there is a probability distribution defined on the candidate set with which the target object is chosen.  
%An error function is defined to quantify the error of the guess produced by the protocol.  
An optimal discrimination protocol minimizes the guessing error averaged over the candidate distribution.  We may also impose  ``unambiguousness'', namely, that we allow no mistakes with our guesses.   An unambiguous discrimination protocol is designed to declare ``inconclusive'', whenever the employed discrimination strategy fails to single out the correct candidate.  Typically in the literature, ``minimum-error'' tasks in quantum discrimination accept incorrect guesses.

Discrimination of quantum operations with a full classical description of the candidates has been investigated when the candidates are unitary operations \cite{PhysRevLett.87.177901,PhysRevLett.87.270404,PhysRevLett.98.100503}, non-unitary quantum channels \cite{PhysRevA.71.062340,PhysRevA.73.042301,PhysRevA.81.032339,PhysRevLett.101.180501,PhysRevLett.102.250501,PhysRevLett.103.210501}, and  quantum measurements \cite{PhysRevLett.96.200401,PhysRevA.90.052312}.  Both minimum-error \cite{PhysRevA.71.062340,PhysRevLett.87.177901,PhysRevLett.102.250501} and unambiguous discrimination \cite{PhysRevA.73.042301,PhysRevA.90.052312} have been studied, in addition to error-free \textit{i.e.}, perfect discrimination \cite{PhysRevLett.98.100503, PhysRevLett.103.210501,PhysRevA.81.032339,PhysRevA.90.052312,PhysRevLett.96.200401,PhysRevLett.101.180501}.  Especially for minimum-error discrimination of two unitary operations with full classical description, the optimal average success probability is derived as a closed formula for unitary operations in $\text{SU}(2)$ \cite{PhysRevLett.87.177901} and $\text{SU}(d)$ for an arbitrary dimension $d$ \cite{PhysRevLett.87.270404}.

%Unlike the case of state discrimination, 
In this paper, we analyze quantum operation discrimination with candidates presented as a quantum sample.  More specifically, our goal is \textit{equivalence determination} of quantum operations, \textit{i.e.}, to determine the quantum sample equivalent to the target box.  %We treat the quantum samples and target box as a quantum gate.  A protocol for equivalence determination is a quantum circuit using the samples and target along with other circuit elements of our choice.   
%We analyze equivalence determination of unitary operations in which an extension of the discrimination of two unitary operations with full classical descriptions of the candidate unitary operations\,\cite{PhysRevLett.87.177901,PhysRevLett.87.270404}.  
For simplicity, the candidate set consists of two single-qubit unitary operations, $U_{1} $ and $U_{2}$ in $\text{SU}(2)$, each distributed according to the Haar measure.  
%We consider that three black-boxes (a target box and two reference boxes) implementing unknown unitary operations are given as physical systems.  The target box is guaranteed to implement one of the two unitary operations.  As is the case with equivalence determination of quantum states, the two reference boxes implement the two unitary operations, respectively, instead of being given classical descriptions of the two unitary operations.  Thus the equivalence determination task for unitary operations is to determine which of the two reference boxes implements the same unitary operation of the target box.  Equivalence determination of unitary operations can be considered as an extension of comparison of unitary operations \cite{0305-4470-36-9-310, doi:10.1080/09500340903203129}, the task of which is to decide that two given black-boxes implementing unitary operations are identical or not.  
%Equivalence determination of unitary operations can be regarded as higher-order quantum computation taking three unitary operations as inputs and a binary number representing the reference box identical to the target box as an output.  
%We analyze the optimal average success probability for the case that the unitary operations are uniform-randomly sampled for a qubit system, namely the special unitary group $\text{SU}(2)$.  
The \textit{reference box} $j$ implements $U_{j}$ for $j = 1,2$, while the \textit{target box} implements either $U_{1}$ or $U_{2}$ with probability $1/2$.  An $(N_{1}, N_{2})$\textit{-equivalence determination} task allows  $N_{j}$ samples of $U_{j}$ and a single use of the target box.  Otherwise, any quantum states and operations may be employed without any cost.  The comparison of unitary operations \cite{0305-4470-36-9-310, doi:10.1080/09500340903203129} and the pattern matching \cite{doi:10.1080/09500340903203129} are $(1,0)$- and $(1,1)$-equivalence determination with restriction, respectively.  Reference \cite{PhysRevA.80.052102} investigates the comparison of quantum measurements.  

%One significant difference between equivalence determination of states and unitary operations is an extra freedom for choosing an initial state for the case of unitary operations.  The initial state can be entangled with ancillary systems.  In addition, the initial state can be an entangled state on between the systems where the target box and the reference boxes are applied.  This property does not appear in discrimination of unitary operations with the classical descriptions of candidate unitary operations.  

In general, an arbitrary quantum operation of our choice can be used in between each use of black-boxes.  Some of the black-boxes may be used concurrently.  It is known that general schemes outperform the parallelized schemes \cite{PhysRevLett.101.180501,PhysRevA.73.042301,PhysRevA.81.032339,hayashi2009discrimination}, but parallelized schemes are more efficient in terms of circuit depth.  
%Another difference is the schemes of how to use the black-boxes in the quantum circuit.  In general, the black-boxes can be used in arbitrary order and any quantum operations can be applied between the uses of the black-boxes.  We consider two types of schemes. One is parallelized schemes, in which the black-boxes is used in a parallel way without introducing the order and no quantum operation is applied between the use of the black-boxes.  The other is general schemes, in which arbitrary quantum operations can be inserted between the use of the black-boxes.  To pursue efficient quantum information processing, clarifying when the general schemes outperform the parallelized schemes is important \cite{PhysRevLett.101.180501,PhysRevA.73.042301,PhysRevA.81.032339,hayashi2009discrimination}.  
In addition to concurrency, the quantum circuit used for equivalent determination introduces an ordering on the black-boxes, which is another degree of freedom to exploit.  
%Further, in equivalence determination of unitary operations, there is a room to choose the order of the reference boxes and target box in quantum circuit.  To the best of our knowledge, such a task considering two different operations is only the problem of implementing quantum switch \cite{PhysRevA.88.022318}.  In quantum mechanics, when the order of operations is changed, the resulting operations may be different since quantum operations do not commute each other in general.  Therefore the order of the use of the black-boxes is expected to affect the performance of equivalence determination of unitary operations.  

The paper is organized as follows.  In Sec.\,\ref{sec:quantum_tester}, we review {\it quantum testers }\cite{PhysRevLett.101.060401,PhysRevA.80.022339} as generalized POVM measurements on quantum operations.  The necessary properties of irreducible representations of $\text{SU}(2)$ are given in Sec.\,\ref{sec:irreducible}.  Section~\ref{single use of reference boxes} discusses $(1, 1)$-equivalence determination and analytically derives the optimal average success probability, both in parallelized and general schemes. Section~\ref{sec:parallel restrict ent} investigates the effect of the entanglement in the initial state.  In Sec.\,\ref{sec:one known}, we derive the optimal average success probability when a classical description is given for one of the candidates.  Section~\ref{sec:ordered strategies} numerically analyzes the optimal average success probability for $(2, 1)$-equivalence determination under all possible orderings of the black-boxes.  We conclude in Sec.\,\ref{sec:conclude}.

%%%%%%%%%%%%%%%%%%%%%%%%% PRELIMINARY %%%%%%%%%%%%%%%%%%%%%%%%%
\section{Preliminary}\label{sec:preliminary}

%%%%%%%%%%%%%%%%%%%%%%%%%%%%%%%%%%%%%%%%%%%%%%%%%%%%%%%%%%%%%%%%%%%%
\subsection{Quantum testers}\label{sec:quantum_tester}
Let $\mathcal{H}$ and $\mathcal{K}$ be Hilbert spaces and $\mathcal{L}(\mathcal{H} )$ be the set of bounded linear operators on $\mathcal{H}$.  Let $\mathcal{M}$ be a completely positive and trace-preserving (CPTP) map from $\mathcal{L}(\mathcal{H} )$ to $\mathcal{L}(\mathcal{K} )$.  References \cite{PhysRevLett.101.060401, PhysRevA.80.022339} introduce  \textit{quantum testers}, which may be interpreted as a quantum measurement on CPTP maps.

Denote a positive operator-valued measure (POVM) as $\{\Pi_{i}\}_{i = 1}^{L}$ satisfying $\Pi_{i} \geq 0$ for $i = 1,2, \dots, L$ and $\sum_{i=1}^{L} \Pi_{i} = I_{\mathcal{K}\mathcal{H}_{A}}$, where $I_{\mathcal{K}\mathcal{H}_{A}}$ is the identity operator on $\mathcal{K} \otimes \mathcal{H}_{A}$.  
We define a quantum $2$-tester $\{\widetilde{\Pi}_{i}\}_{i = 1}^{L}$ by
\begin{equation}
\widetilde{\Pi}_{i} := (I_{\mathcal{K}} \otimes \sqrt{X}) \Pi_{i} (I_{\mathcal{K}} \otimes \sqrt{X}),
\end{equation}
where $X$ is a positive semidefinite operator with unit trace.  

Measuring $\{\widetilde{\Pi}_{i}\}$ on $\mathcal{M}$ corresponds to applying $\mathcal{M} \otimes \mathcal{I} $ on $\ket{\psi} = I \otimes \sqrt{X} \kket{I}$, where $\kket{I} := \sum_{i = 1}^{\dim\mathcal{H}} \ket{i}_{\mathcal{H}} \ket{i}_{\mathcal{H}_{A}}$ with the computational basis $\{\ket{i}\}_{i = 1}^{\dim \mathcal{H}}$, and then measuring $\{\Pi_{i}\}$ on the resulting state.  
The probability $q_{i}$ of obtaining the outcome $i$ is 
\begin{equation}
q_{i} = \tr [M \widetilde{\Pi}_{i} ],  
\end{equation}
where $M$ is the Choi operator of $\mathcal{M}$ defined by 
\begin{equation}
M := (\mathcal{M} \otimes \mathcal{I}) (\kket{I}\bbra{I}).  
\end{equation}

We consider $N - 1$ CPTP maps $\mathcal{M}_{i}$ from $\mathcal{L}(\mathcal{H}_{i} )$ to $\mathcal{L}(\mathcal{K}_{i} )$ for $i = 1,2, \cdots ,N-1$.  $M_{i}$ appears before $M_{i + 1}$ in the quantum circuit.  A generalized POVM measurement on the $N-1$ CPTP maps can be described by quantum $N$-tester and its definition is given as follows.  
\begin{dfn}\label{def:quantum n tester}
Quantum $N$-tester is a set of operators $\{ \widetilde{\Pi}_{i} \} $ when $ \widetilde{\Pi}_{i} \in \mathcal{L}(\bigotimes_{j = 1}^{N-1}\mathcal{K}_{j} \otimes \bigotimes_{j =1}^{N-1}\mathcal{H}_{j})$ satisfy 
\begin{align}
\widetilde{\Pi}_{i} &\geq 0, \label{eq:tester_1} \\
\sum_{i} \widetilde{\Pi}_{i} &= I_{\mathcal{K}_{N-1}} \otimes Y^{\{N - 1\}},\label{eq:tester_2}\\
\tr_{\mathcal{H}_{j}} Y^{\{j\}} &=  I_{\mathcal{K}_{j - 1}} \otimes Y^{\{j - 1\}}, \text{ for } j = 2, \cdots, N - 1, \label{eq:tester_3}\\
\tr Y^{\{1\}} &= 1,  \label{eq:tester_4}
\end{align}
for some positive semidefinite operators $Y^{\{j\}} \in \mathcal{L}(\bigotimes_{l = 1}^{j-1}\mathcal{K}_{l} \otimes \bigotimes_{l =1}^{j}\mathcal{H}_{l})$ for $j = 2,\cdots, N-1$ and $Y^{\{1\}} \in \mathcal{L}(\mathcal{H}_{1})$.  When the quantum $N$-tester $\{\widetilde{\Pi}_{i}\}$ is applied on $N -1 $ CPTP maps $\{\mathcal{M}_{j}\}_{j = 1}^{N -1}$, the probability of obtaining the outcome $i$ is given by 
\begin{equation}
q_{i} = \tr \left[ \widetilde{\Pi}_{i} \bigotimes_{j = 1}^{N -1} M_{j}\right].  
\end{equation}
\end{dfn}

\begin{figure}[h]
\begin{align}
  \begin{aligned}
    \Qcircuit @C=0.65em @R=1.3em {
      &\multiprepareC{1}{}&\ustick{\mathcal{H}_{1}}\qw& \gate{\mathcal{M}_{1}}& \ustick{\mathcal{K}_{1}} \qw& \multigate{1}{}&\ustick{\mathcal{H}_{2}}\qw& \gate{\mathcal{M}_{2}}& \ustick{\mathcal{K}_{2}} \qw  \\ 
      & \pureghost{}&\qw & \qw & \qw &\ghost{}&\qw & \qw & \qw 
    }
  \end{aligned} \cdots \begin{aligned} \Qcircuit @C=1.5em @R=1.3em {
  & \multigate{1}{}&\ustick{\mathcal{H}_{N-1}}\qw& \gate{\mathcal{M}_{N-1}}& \ustick{\mathcal{K}_{N-1}}\qw &\multimeasureD{1}{}\\
  &\ghost{}&\qw & \qw & \qw &\ghost{}
  }
  \end{aligned}\notag
\end{align}
\caption{A circuit representation for a quantum $N$-tester.}\label{fig:quantum n tester gene}
\end{figure}
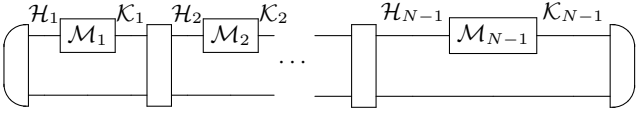
A quantum tester is a special case of {\it quantum comb} \cite{PhysRevLett.101.060401, PhysRevA.80.022339} or \textit{quantum strategy} \cite{Gutoski:2007:TGT:1250790.1250873} and can be realized by a quantum circuit (Fig.\,\ref{fig:quantum n tester gene}).  The details of quantum testers are given in Refs.\,\cite{PhysRevLett.101.060401, PhysRevA.80.022339}.   We often abbreviate $ Y^{\{j\}}$ with the largest $j$ in the range as $Y$.

%%%%%%%%%%%%%%%%%%%%%%%%%%%%%%%%%%%%%%%%%%%%%%%%%%%%%%%%%%%%%%%%%%%%

\subsection{Relaxing ordering constraint}\label{sed:reduce order complexity}
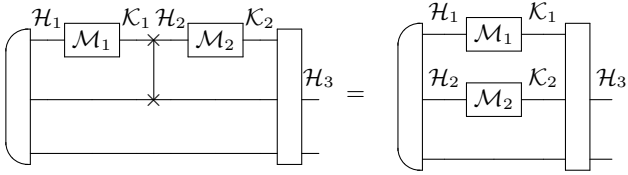
\begin{figure}[h]
\begin{align}
  \begin{aligned}
    \Qcircuit @C=0.7em @R=1.3em {
      &\multiprepareC{2}{}&\ustick{\mathcal{H}_{1}}\qw& \gate{\mathcal{M}_{1}}& \ustick{\mathcal{K}_{1}} \qw& \qswap& \ustick{\mathcal{H}_{2}} \qw &\gate{\mathcal{M}_{2}} & \ustick{\mathcal{K}_{2}} \qw&\multigate{2}{}& \\
      &\pureghost{}&\qw& \qw& \qw& \qswap \qwx &\qw &\qw&\qw &\ghost{} &\ustick{\mathcal{H}_{3}}\qw\\
      & \pureghost{}&\qw & \qw & \qw & \qw &\qw &\qw&\qw &\ghost{} &\qw
    }
  \end{aligned} \,\,\,\,\, = 
   \begin{aligned}
    \Qcircuit @C=0.9em @R=1.3em {
      &\multiprepareC{2}{}&\ustick{\mathcal{H}_{1}}\qw& \gate{\mathcal{M}_{1}}& \ustick{\mathcal{K}_{1}} \qw& \multigate{2}{}&\\
      & \pureghost{}&\ustick{\mathcal{H}_{2}}\qw& \gate{\mathcal{M}_{2}}& \ustick{\mathcal{K}_{2}} \qw& \ghost{}&\ustick{\mathcal{H}_{3}}\qw\\
      & \pureghost{}&\qw & \qw & \qw &\ghost{}&\qw 
    }
  \end{aligned}\notag
\end{align}
\caption{Relaxation of ordering constraint.}\label{fig:reducing order}
\end{figure}

%Any set of operator $\{\widetilde{\Pi}_{i}\}$ that satisfies the conditions of a general quantum tester is realizable by a quantum circuit.  
Equations~(\ref{eq:tester_2}) and (\ref{eq:tester_3}) in general imply that the input CPTP maps are applied in a particular order, for example $\mathcal{M}_{1}$ must be used before $\mathcal{M}_{2}$.
When a quantum $N$-tester defined in Def.\,\ref{def:quantum n tester} satisfies additional conditions, the ordering constraint is relaxed.  Especially for the quantum testers in this paper, the first two uses of black-boxes can be parallelized.  The necessary and sufficient condition for the parallelization is 
\begin{equation}
Y^{\{1\}} = I_{\mathcal{K}_{1}} \otimes Y'^{\{1\}}\label{eq:reducing condition}
\end{equation}
for a positive semidefinite operator $Y'^{\{1\}}$.  This condition implies that the quantum operation between $\mathcal{M}_{1}$ and $\mathcal{M}_{2}$ can be substituted by a swap operation as given in Fig.\,\ref{fig:reducing order}.  

%%%%%%%%%%%%%%%%%%%%%%%%%%%%%%%%%%%%%%%%%%%%%%%%%%%%%%%%%%%%%%%%%%%%
%\subsection{$(N_{1}, N_{2})$-equivalence determination}\label{sec:equivalence det}
%We denote the unitary operation implemented by the reference box $j$ as $U_{j}$ for $j = 1,2$ and assume that the target box implements either $U_{1}$ or $U_{2}$ with probability $1/2$.  An $(N_{1}, N_{2})$-equivalence determination task is to determine which reference box is equivalent to the target box by using $N_{j}$ samples of the $j$-th reference box and the target box only once.  

%%%%%%%%%%%%%%% IRREDUCIBLE REPRESENTATION OF SU(2) %%%%%%%%%%%%%%%%%%
\subsection{Irreducible representation of $\text{SU}(2)$} \label{sec:irreducible}
Let $\mathcal{K}_{i}$ $(i = 1,2,3)$ be any two-dimensional Hilbert space whose computational basis is $\{\ket{0}, \ket{1}\}$.  We define the following basis of the three-qubit system $\mathcal{K} := \mathcal{K}_{1} \otimes \mathcal{K}_{2} \otimes \mathcal{K}_{3} $,  
\begin{align}
\ket{v_{1}} &= \textstyle \ket{(\frac{1}{2}\frac{1}{2})0 \frac{1}{2};\frac{1}{2}\frac{-1}{2}} =  \frac{1}{\sqrt{2}} (\ket{001} - \ket{100}), \\
\ket{v_{2}} &= \textstyle \ket{(\frac{1}{2}\frac{1}{2})0 \frac{1}{2};\frac{1}{2}\frac{1}{2}} = \frac{1}{\sqrt{2}} (\ket{011} - \ket{110}), \\
\ket{v_{3}} &= \textstyle \ket{(\frac{1}{2}\frac{1}{2})1 \frac{1}{2};\frac{1}{2}\frac{-1}{2}} = \sqrt{\frac{2}{3}} \ket{010} - \sqrt{\frac{1}{6}} (\ket{001} + \ket{100}), \\
\ket{v_{4}} &= \textstyle \ket{(\frac{1}{2}\frac{1}{2})1 \frac{1}{2};\frac{1}{2}\frac{1}{2}} = -\sqrt{\frac{2}{3}} \ket{101} + \sqrt{\frac{1}{6}} (\ket{011} + \ket{110})
\end{align}
and 
\begin{align}
\ket{v_{5}} &= \textstyle \ket{(\frac{1}{2}\frac{1}{2})1 \frac{1}{2};\frac{3}{2}\frac{-3}{2}} = \ket{000}, \\
\ket{v_{6}} &= \textstyle \ket{(\frac{1}{2}\frac{1}{2})1 \frac{1}{2};\frac{3}{2}\frac{-1}{2}} = \sqrt{\frac{1}{3}}(\ket{010} + \ket{001} +\ket{100}), \\
\ket{v_{7}} &= \textstyle \ket{(\frac{1}{2}\frac{1}{2})1 \frac{1}{2};\frac{3}{2}\frac{1}{2}} = \sqrt{\frac{1}{3}}(\ket{011} + \ket{110} +\ket{101}), \\
\ket{v_{8}} &= \textstyle \ket{(\frac{1}{2}\frac{1}{2})1 \frac{1}{2};\frac{3}{2}\frac{3}{2}} = \ket{111},  
\end{align}
where $\ket{(j_{1}j_{2})j_{13} j_{2}; jm}$ represents a state with total spin angular momentum $j$ and spin angular momentum along the $z$-axis $m$, in which the spin-$j_{1}$ in $\mathcal{K}_{1}$ and the spin-$j_{3}$ in $\mathcal{K}_{3}$ are coupled to be a spin-$j_{13} $ followed by the coupling with the spin-$j_{2}$.

We introduce two-dimensional subspaces $\mathcal{U}_{\frac{1}{2}}$ and $\mathcal{V}_{\frac{1}{2}}^{[3]}$, whose basis is $\{\ket{i}^{\mathcal{U}}_{\frac{1}{2}}\}_{i = 0}^{1}$ and $\{\ket{j}^{\mathcal{V}^{[3]}}_{\frac{1}{2}}\}_{j = 0}^{1}$, respectively.  
The bases are defined so that 
\begin{align}
\ket{0}^{\mathcal{U}}_{\frac{1}{2}}\ket{0}^{\mathcal{V}^{[3]}}_{\frac{1}{2}} = \ket{v_{1}}, \\
\ket{1}^{\mathcal{U}}_{\frac{1}{2}}\ket{0}^{\mathcal{V}^{[3]}}_{\frac{1}{2}} = \ket{v_{2}}, \\
\ket{0}^{\mathcal{U}}_{\frac{1}{2}}\ket{1}^{\mathcal{V}^{[3]}}_{\frac{1}{2}} = \ket{v_{3}}, \\
\ket{1}^{\mathcal{U}}_{\frac{1}{2}}\ket{1}^{\mathcal{V}^{[3]}}_{\frac{1}{2}} = \ket{v_{4}}.   
\end{align}
Similarly, we introduce the basis of the four-dimensional subspace $\mathcal{U}_{\frac{3}{2}}$ and one-dimensional subspace $\mathcal{V}_{\frac{3}{2}}^{[3]}$, whose basis is $\{\ket{i}^{\mathcal{U}}_{\frac{3}{2}}\}_{i = 0}^{3}$ and $\{\ket{0}^{\mathcal{V}^{[3]}}_{\frac{3}{2}}\}$, respectively, so that
\begin{align}
\ket{0}^{\mathcal{U}}_{\frac{3}{2}}\ket{0}^{\mathcal{V}^{[3]}}_{\frac{3}{2}} = \ket{v_{5}}, \\
\ket{1}^{\mathcal{U}}_{\frac{3}{2}}\ket{0}^{\mathcal{V}^{[3]}}_{\frac{3}{2}} = \ket{v_{6}}, \\
\ket{2}^{\mathcal{U}}_{\frac{3}{2}}\ket{0}^{\mathcal{V}^{[3]}}_{\frac{3}{2}} = \ket{v_{7}}, \\
\ket{3}^{\mathcal{U}}_{\frac{3}{2}}\ket{0}^{\mathcal{V}^{[3]}}_{\frac{3}{2}} = \ket{v_{8}}.   
\end{align}
For economy of notation, the superscript of $\ket{i}^{\mathcal{V}^{[3]}}_{J}$ will be omitted hereafter.  The value of the subscript $J$ corresponds to the total angular momentum of the state, when interpreting each qubit as a spin-$1/2$.

The basis $\{\ket{v_{i}}\}_{i = 1}^{8}$ decomposes any unitary operator $U^{\otimes 3}$ for any $U$ in $\text{SU}(2)$ as
\begin{equation}
U^{\otimes 3} = U_{\frac{1}{2}} \otimes I_{\mathcal{V}_{\frac{1}{2}}^{[3]}} \oplus U_{\frac{3}{2}} \otimes I_{\mathcal{V}_{\frac{3}{2}}^{[3]}},  
\end{equation}
\textit{i.e.}, into the irreducible representations of $\text{SU}(2)$.  
The vectors from $\ket{v_{1}}$ to $\ket{v_{4}} $ have total angular momentum $J = 1/2$ and the vectors from $\ket{v_{5}}$ to $\ket{v_{8}} $ with $J = 3/2$.  The particular choice of the basis corresponds to a composition starting from $\mathcal{K}_{1}$ and $\mathcal{K}_{3}$ followed by $\mathcal{K}_{2}$.

Alternative decompositions are obtained depending on the order in which the three qubits are composed.  
We use the basis $\{\ket{\hat{i}}_{\frac{1}{2}}\}_{i = 0}^{1}$ instead of $\{\ket{i}_{\frac{1}{2}}\}_{i = 0}^{1}$ when interested in a composition starting from $\mathcal{K}_{1}$ and $\mathcal{K}_{2}$.   Another basis $\{\ket{\tilde{i}}_{\frac{1}{2}}\}_{i = 0}^{1} $ is for a composition starting from $\mathcal{K}_{2}$ and $\mathcal{K}_{3}$.

Calculating Wigner's 6$j$ coefficients \cite{messiah1961quantum}, we obtain the relations of these three bases
\begin{align}
\ket{\hat{0}}_{\frac{1}{2}} &= \frac{1}{2} \ket{0}_{\frac{1}{2}} + \frac{\sqrt{3}}{2} \ket{1}_{\frac{1}{2}} \label{eq: unitary1 12 and 13},\\
\ket{\hat{1}}_{\frac{1}{2}} &= \frac{\sqrt{3}}{2} \ket{0}_{\frac{1}{2}} - \frac{1}{2} \ket{1}_{\frac{1}{2}}\label{eq: unitary2 12 and 13},\\
\ket{\tilde{0}}_{\frac{1}{2}} &= - \frac{1}{2} \ket{0}_{\frac{1}{2}} + \frac{\sqrt{3}}{2} \ket{1}_{\frac{1}{2}}\label{eq: unitary1 23 and 13},\\
\ket{\tilde{1}}_{\frac{1}{2}} &= \frac{\sqrt{3}}{2} \ket{0}_{\frac{1}{2}} + \frac{1}{2} \ket{1}_{\frac{1}{2}}\label{eq: unitary2 23 and 13}.  
\end{align}

A linear operator $\rho \in \mathcal{L}(\mathcal{K})$ satisfying 
\begin{equation}
[\rho , A^{\otimes 3}] = 0 \label{eq:commute rho}
\end{equation}
for all $A \in \text{SU}(2)$ has the form
\begin{equation}
\rho = \bigoplus_{J = \frac{1}{2}}^{\frac{3}{2}} \frac{I_{J}}{d_{J}} \otimes \rho_{J}, 
\end{equation}
from Schur's lemma, where $I_{J}$ is the identity operator on $\mathcal{U}_{J}$ and $\rho_{J} = \tr_{\mathcal{U}_{J}} \rho $.

The reduced operator $\sigma = \tr_{\mathcal{K}_{3}}\rho$ 
satisfies 
\begin{equation}
[\sigma , A^{\otimes 2}] = 0,  
\end{equation}
and therefore
\begin{equation}
\sigma = \bigoplus_{J = 0}^{1} \frac{I_{J}}{d_{J}} \otimes \sigma_{J}.
\end{equation}
Here we define the basis of the two-qubit system as 
\begin{align}
\ket{w_{1}} &= \frac{1}{\sqrt{2}}(\ket{01} - \ket{10}),\\
\ket{w_{2}} &= \ket{00},\\
\ket{w_{3}} &= \frac{1}{\sqrt{2}}(\ket{01} + \ket{10}),\\
\ket{w_{4}} &= \ket{11}.  
\end{align}
The basis of the irreducible subspace $\mathcal{U}_{1}$ and the multiplicity subspace $\mathcal{V}_{1}^{[2]}$ are denoted as $\{\ket{i}^{\mathcal{U}}_{1}\}_{i = 0}^{2}$ and $\{\ket{0}^{\mathcal{V}^{[2]}}_{1}\}$.  The bases are chosen so that 
\begin{align}
\ket{0}^{\mathcal{U}}_{1}\ket{0}^{\mathcal{V}^{[2]}}_{1} = \ket{w_{2}}, \\
\ket{1}^{\mathcal{U}}_{1}\ket{0}^{\mathcal{V}^{[2]}}_{1} = \ket{w_{3}}, \\
\ket{2}^{\mathcal{U}}_{1}\ket{0}^{\mathcal{V}^{[2]}}_{1} = \ket{w_{4}}.  
\end{align}
Finally, $\mathcal{U}_{0} = \text{span}\{ \ket{0}^{\mathcal{U}}_{0} \}$ and $\mathcal{V}_{0}^{[2]} =  \text{span}\{ \ket{0}^{\mathcal{V}^{[2]}}_{0} \}$ and 
\begin{equation}
\ket{0}^{\mathcal{U}}_{0}\ket{0}^{\mathcal{V}^{[2]}}_{0} = \ket{w_{1}}.  
\end{equation}

The elements of the multiplicity subspaces of $\sigma$ are given by
\begin{equation}
\sigma_{0} = \bra{\hat{0}} \rho_{\frac{1}{2}} \ket{\hat{0}}_{\frac{1}{2}}, \quad \sigma_{1} = \bra{\hat{1}} \rho_{\frac{1}{2}} \ket{\hat{1}}_{\frac{1}{2}} + \rho_{\frac{3}{2}}.  \label{eq:partial trace}
\end{equation}
The operator $\sigma \otimes I_{\mathcal{K}_{3}} $ satisfies Eq.\,(\ref{eq:commute rho}) and 
\begin{align}
&\sigma \otimes I_{\mathcal{K}_{3}}\\ 
&\quad = \left(I_{0} \otimes I_{\mathcal{K}_{3}} \otimes \frac{\sigma_{0}}{d_{0}}\right) \oplus \left(I_{1} \otimes I_{\mathcal{K}_{3}} \otimes \frac{\sigma_{1}}{d_{1}} \right) \\
&\quad = I_{\frac{1}{2}} \otimes \frac{\sigma_{0}}{d_{0}}\ket{0}\bra{0}_{\frac{1}{2}} \oplus I_{\frac{1}{2}} \otimes \frac{\sigma_{1}}{d_{1}}\ket{1}\bra{1}_{\frac{1}{2}} \oplus I_{\frac{3}{2}} \otimes \frac{\sigma_{1}}{d_{1}}\ket{0}\bra{0}_{\frac{3}{2}} \\
&\quad = I_{\frac{1}{2}} \otimes \left(\frac{\sigma_{0}}{d_{0}}\ket{0}\bra{0}_{\frac{1}{2}} + \frac{\sigma_{1}}{d_{1}}\ket{1}\bra{1}_{\frac{1}{2}} \right) \oplus I_{\frac{3}{2}} \otimes \frac{\sigma_{1}}{d_{1}}\ket{0}\bra{0}_{\frac{3}{2}}. \label{eq:coupling}
\end{align}

%%%%%%%%%%%%%%%%%%%%%%%% (1,1)- EQUIVALENCE DETERMINATION %%%%%%%%%%%%%%%%%%%
\section{(1,1)-equivalence determination}\label{single use of reference boxes}
In this section, we consider the simplest case, $(1,1)$-equivalence determination of unitary operations.  We denote input and output Hilbert spaces of the the reference box $i$ by $\mathcal{H}_{i}$ and $ \mathcal{K}_{i}$, respectively for $i = 1,2$ and input and output spaces of the target box by $\mathcal{H}_{3}$ and $ \mathcal{K}_{3}$.  For simplicity, we define $\mathcal{H} := \bigotimes_{j = 1}^{3} \mathcal{H}_{j}$ and $\mathcal{K} := \bigotimes_{j = 1}^{3} \mathcal{K}_{j}$.  We focus on qubit systems and therefore assume $\mathcal{H}_{i} = \mathcal{K}_{i} \cong \mathbb{C}^{2}$.  For a given  quantum tester $\{\widetilde{\Pi}_{1}, \widetilde{\Pi}_{2}\}$, the success probability of obtaining the correct answer $p_{U_{1},U_{2}}$ is given by 
\begin{equation}
p_{U_{1},U_{2}} := \frac{1}{2} \tr[ \kket{W_{1}}\bbra{W_{1}} \widetilde{\Pi}_{1}+ \kket{W_{2}}\bbra{W_{2}} \widetilde{\Pi}_{2} ],  
\end{equation}
where $W_{i} := U_{1} \otimes U_{2} \otimes U_{i}$ for $ i = 1,2$ and $\kket{W_{i}} = (W_{i} \otimes I)\kket{I}$.  In other words, $(1,1)$-equivalence determination is to determine which unitary operation, $W_{1}$ or $W_{2}$, is implemented.

The success probability above depends on the specific unitary operations $U_{1}$ and $U_{2}$.  Therefore we adopt the {\it average } success probability (ASP)  over the Haar measure as a figure of merit of equivalence determination of unitary operations.  ASP $p_{ave}$ is given by 
\begin{equation}
p_{ave} = \frac{1}{2} \tr[ M_{1} \widetilde{\Pi}_{1} + M_{2} \widetilde{\Pi}_{2} ],
\end{equation}
where $M_{i}$ are
\begin{equation}
M_{i} := \int d\mu(U_{1}) \int d\mu(U_{2}) \kket{W_{i} } \bbra{W_{i} },\label{eq:n1 parallel def mi}
\end{equation}
for the Haar measure $d \mu(U)$.

\subsection{Parallelized schemes}\label{sec:parallel strategies N1}
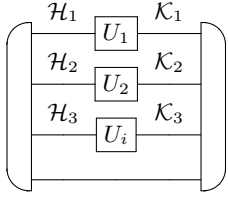
\begin{figure}[t]
\begin{equation}
  \begin{aligned}
    \Qcircuit @C=1.3em @R=0.7em {
    &\multiprepareC{3}{}&\ustick{\mathcal{H}_{1}}\qw& \gate{U_{1}}& \ustick{\mathcal{K}_{1}} \qw& \multimeasureD{3}{} \\
    &\pureghost{}&\ustick{\mathcal{H}_{2}}\qw& \gate{U_{2}}& \ustick{\mathcal{K}_{2}} \qw& \ghost{}\\
    &\pureghost{}&\ustick{\mathcal{H}_{3}}\qw& \gate{U_{i}}& \ustick{\mathcal{K}_{3}} \qw& \ghost{}\\
    &\pureghost{}&\qw & \qw & \qw &\ghost{}
    }
  \end{aligned} \notag
\end{equation}
  \caption{ The quantum circuit for $(1,1)$-equivalence determination of unitary operations under parallelized schemes.  }\label{fig:parallel_N1_general input}
\end{figure}

First we consider the parallelized schemes, in which all of the black-boxes are applied simultaneously.  A circuit representation of equivalence determination under parallelized schemes is given in Fig.\,\ref{fig:parallel_N1_general input}.  The equivalence determination is to determine which unitary operation, $W_{1}$ or $W_{2}$, is implemented.  

Within parallelized schemes, a quantum tester $\{\widetilde{\Pi}_{1}, \widetilde{\Pi}_{2}\}$ with $\widetilde{\Pi}_{i} \in \mathcal{L}(\mathcal{K} \otimes \mathcal{H})$ is a set of positive semidefinite operators satisfying
\begin{align}
& \widetilde{\Pi}_{i} \geq 0, \ i = 1,2\notag \\
&\widetilde{\Pi}_{1} + \widetilde{\Pi}_{2} = I_{\mathcal{K}} \otimes X\notag \\
&\tr X = 1,   
\end{align}
for some $X \in \mathcal{L}(\mathcal{H})$.

\begin{thm}\label{thm:n1para_opt}
The optimal average success probability of $(1,1)$-equivalence determination under parallelized schemes is $7/8$ when unitary operations are chosen from the Haar measure.  
\end{thm}

The optimal ASP for $(1,1)$-equivalence determination under parallelized schemes is given by a semidefinite program (SDP)
\begin{align}
\text{maximize}\quad & p_{ave} = \frac{1}{2} \tr \left[  M_{1} \widetilde{\Pi}_{1} +  M_{2} \widetilde{\Pi}_{2} \right], \notag \\
\text{subject to}\quad & \widetilde{\Pi}_{i} \geq 0, \ i = 1,2, \notag \\
&\widetilde{\Pi}_{1} + \widetilde{\Pi}_{2} = I_{\mathcal{K}} \otimes X, \notag \\
&\tr X = 1,  
\end{align}
where $M_{i}$ are given by Eq.\,(\ref{eq:n1 parallel def mi})

Due to the symmetry introduced by averaging over the Haar measure, the following lemma can be proven (Appx.\,\ref{prf:commutex}).  
\begin{lem}\label{lem:commutex}
The optimal average success probability of $(1,1)$-equivalence determination can be achieved with $X$ satisfying
\begin{equation}
[A^{ \otimes 3}, X ] = 0,	
\end{equation}
for any unitary operator $A \in \text{SU}(2)$.  
\end{lem}

Since the target box is chosen among $U_{1}$ and $U_{2}$ with the same probability, we may assume an additional symmetry on $X$.  
\begin{lem}\label{lem:commuteswap}
Let $S_{\mathcal{H}_{12}} $ be the swap operator between $\mathcal{H}_{1}$ and $\mathcal{H}_{2}$.  
The optimal average success probability of $(1,1)$-equivalence determination is obtained by $X$ satisfying
\begin{equation}
[S_{\mathcal{H}_{12}} \otimes I_{\mathcal{H}_{3}}, X ] = 0.  \label{eq:para commmute}
\end{equation}
\end{lem}
\noindent \Proof Suppose that a set of positive semidefinite operators $\{\tilde{\Pi}_{1}, \tilde{\Pi}_{2} \}$ gives the success probability $p$. 
By using a tensor product of the swap operators $S_{\mathcal{K}_{12}} \otimes S_{\mathcal{H}_{12}}$, where $S_{\mathcal{K}_{12}}$ acts on $\mathcal{K}_{1} \otimes \mathcal{K}_{2}$ as $S_{\mathcal{K}_{12}} (\ket{\psi} \otimes \ket{\phi}) = \ket{\phi} \otimes \ket{\psi}$ for any $\ket{\psi} \in \mathcal{K}_{1}$ and $\ket{\phi} \in \mathcal{K}_{2}$, and $S_{\mathcal{H}_{12}}$ acts similarly on $\mathcal{H}_{1} \otimes \mathcal{H}_{2}$, we define $\widetilde{\Pi}'_{i}$ as 
\begin{equation}
\widetilde{\Pi}'_{i} := \frac{1}{2} \{ \widetilde{\Pi}_{i} + (S_{\mathcal{K}_{12}} \otimes S_{\mathcal{H}_{12}} \otimes I ) \widetilde{\Pi}_{\bar{i}} (S_{\mathcal{K}_{12}} \otimes S_{\mathcal{H}_{12}} \otimes I )  \},
\end{equation}
where $\bar{1} = 2$ and $\bar{2} = 1 $.  
Then we have 
\begin{align}
\widetilde{\Pi}'_{1} + \widetilde{\Pi}'_{2} &= I_{\mathcal{K}} \otimes  X'_{\mathcal{H}},
\end{align}
where $X'_{\mathcal{H}} = (X_\mathcal{H} + (S_{\mathcal{H}_{12}} \otimes I) X_{\mathcal{H}}(S_{\mathcal{H}_{12}} \otimes I))/2$ satisfying $\tr X'_{\mathcal{H}} = 1$.  The set $\{\widetilde{\Pi}'_{1}, \widetilde{\Pi}'_{2}\}$ is also a quantum $2$-tester, which gives the same success probability $p$ since 
    \begin{align}
    \frac{1}{2} \tr \left[ M_{1} \widetilde{\Pi}'_{1}  +  M_{2} \widetilde{\Pi}'_{2} \right]&= \frac{1}{2} \tr \left[  M_{1} \widetilde{\Pi}_{1} + M_{2} \widetilde{\Pi}_{2}  \right].   
    \end{align}
The equality is derived by using 
\begin{equation}
(S_{\mathcal{K}_{12}} \otimes S_{\mathcal{H}_{12}} \otimes I ) M_{i} (S_{\mathcal{K}_{12}} \otimes S_{\mathcal{H}_{12}} \otimes I ) = M_{\bar{i}},
\end{equation}
for $i = 1,2$.  By definition of $X'_{\mathcal{H}}$, $[S_{\mathcal{H}_{12}} \otimes I_{\mathcal{H}_{3}}, X'_{\mathcal{H}} ] = 0$ holds.  Therefore we can always choose $X$ satisfying Eq.\,(\ref{eq:para commmute}). \qed

By the above argument, $(1,1)$-equivalence determination under parallelized schemes reduces to a discrimination of two (known) random unitary channels $\mathcal{M}_{1}$ and $\mathcal{M}_{2}$,
\begin{align}
&\mathcal{M}_{i}(\rho):= \notag \\
& \quad \int d\mu(U_{1})\int d\mu(U_{2}) (U_{1} \otimes U_{2} \otimes U_{i}) \rho  ( U^{\dagger}_{1} \otimes U^{\dagger}_{2} \otimes U^{\dagger}_{i} )\notag,
\end{align}
for $i = 1,2$.  The optimal ASP $p^{opt}_{ave}$ of discriminating two channels is represented in terms of the diamond norm $\| \cdot \|_{\diamond}$ \cite{PhysRevA.71.062340} as 
\begin{align}
&p^{opt}_{ave} = \frac{1}{2} + \frac{1}{4} \| \mathcal{M}_{1} - \mathcal{M}_{2} \|_{\diamond} \label{eq:diamondnorm_1} \\
&= \frac{1}{2} + \frac{1}{4}\max_{\substack{\tr X = 1,\\ X \geq 0}}\| (I_{\mathcal{K}} \otimes \sqrt{X}) (M_{1} - M_{2} ) (I_{\mathcal{K}} \otimes \sqrt{X}) \|_{1}.\label{eq:diamondnorm}
\end{align}

Proof outline of Theorem \ref{thm:n1para_opt}: Lemma~\ref{lem:commutex} and Schur's lemma imply that the non-trivial elements of $X$ are only in the multiplicity subspaces of the irreducible representation of $U^{\otimes 3}$ for $U \in \text{SU}(2)$.  Lemma~\ref{lem:commuteswap} guarantees that we can assume that $X$ restricted to $\mathcal{V}_{\frac{1}{2}}^{[3]}$ is diagonalized in the basis $\{\ket{\hat{0}}, \ket{\hat{1}}\}$.  Performing the maximization gives the optimal ASP $7/8$.

\noindent \textbf{Proof of Theorem \ref{thm:n1para_opt}}. 
From Lemma~\ref{lem:commutex}, $X $ can be chosen as
\begin{equation}
X = \frac{I_{\frac{1}{2}}}{2} \otimes p X_{\frac{1}{2}} \oplus \frac{I_{\frac{3}{2}}}{4} \otimes (1-p)\ket{0}\bra{0}_{\frac{3}{2}},\label{eq:irrd x}
\end{equation}
where $X_{\frac{1}{2}} $ is a $2 \times 2$ positive semidefinite operator on the multiplicity subspace $\mathcal{V}^{[3]}_{\frac{1}{2}}$ with unit trace and $0 \leq p \leq 1$.

In order to utilize Lemma~\ref{lem:commuteswap}, the basis $\{\ket{\hat{0}}, \ket{\hat{1}}\}$ of the multiplicity subspace $\mathcal{V}^{[3]}_{\frac{1}{2}}$ satisfies   
\begin{align}
\ket{\hat{i}} \rightarrow (-1)^{i+1}\ket{\hat{i}} 
\end{align}
for $i = 0,1$ under application of $S_{\mathcal{H}_{12}}$.  
Therefore the condition of Lemma~\ref{lem:commuteswap}, {\it i.e.}, $[S_{\mathcal{H}_{12}}, X] = 0$, implies that $X_{\frac{1}{2}}$ is diagonalized in the basis $\{\ket{\hat{0}}, \ket{\hat{1}}\}$, namely,   
\begin{equation}
X_{\frac{1}{2}} = q \ket{\hat{0}}\bra{\hat{0}} + ( 1- q ) \ket{\hat{1}}\bra{\hat{1}} =: X_{q},  
\end{equation}
where $0 \leq q \leq 1$.

We have
\begin{align}
I_{K} \otimes X &= \bigoplus_{J = \frac{1}{2}}^{\frac{3}{2}} \Bigg\{ I_{J}^{\mathcal{K}} \otimes \frac{I_{\frac{1}{2}}^{\mathcal{H}}}{d_{\frac{1}{2}}} \otimes I_{\mathcal{V}_{J}^{[3]}} \otimes pX_{\frac{1}{2}}\notag \\
& \quad \oplus I_{J}^{\mathcal{K}} \otimes \frac{I_{\frac{3}{2}}^{\mathcal{H}}}{d_{\frac{3}{2}}} \otimes I_{\mathcal{V}_{J}^{[3]}} \otimes (1-p)\ket{0}\bra{0}_{\frac{3}{2}} \Bigg\},  
\end{align}
where $I_{J}^{\mathcal{K}}$ is the identity operator on the irreducible subspace $\mathcal{U}_{J}$ of $\mathcal{K} = \bigotimes_{i= 1}^{3} \mathcal{K}_{i}$ and $I_{L}^{\mathcal{H}}$ for $\mathcal{U}_{L}$ of $\mathcal{H} = \bigotimes_{i= 1}^{3} \mathcal{H}_{i}$.  
By substituting Eq.\,(\ref{eq:irrd x}) and $ M^{(i)}_{JL}$ given in Lemma~\ref{lem:derivation of Mi N1} in Appx.\,\ref{prf:n1para_opt}, the diamond norm $\|\mathcal{M}_{1} - \mathcal{M}_{2} \|_\Diamond $ in Eq.\,(\ref{eq:diamondnorm_1}) is calculated as 
\begin{align}
\|\mathcal{M}_{1} - \mathcal{M}_{2} \|_\Diamond &=\max_{0 \leq p,q \leq 1} \left\{p \left(\Delta_{q} + \Delta'_{q}\right) + ( 1- p ) \Delta''\right\}, \label{eq:diamond_norm}  
\end{align}
where
\begin{align}
\Delta_{q} &:= \left\| \left(I_{\mathcal{V}^{[3]}_{\frac{1}{2}}} \otimes \sqrt{X_{q}}\right) \left( M^{(1)}_{\frac{1}{2}\frac{1}{2}} - M^{(2)}_{\frac{1}{2}\frac{1}{2}}\right)  \left(I_{\mathcal{V}^{[3]}_{\frac{1}{2}}} \otimes \sqrt{X_{q}}\right)\right\|_{1}, \\
\Delta'_{q} &:= \frac{2}{3} \left\|\sqrt{X_{q}} \left( M^{(1)}_{\frac{3}{2}\frac{1}{2}} - M^{(2)}_{\frac{3}{2}\frac{1}{2}} \right) \sqrt{X_{q}}\right\|_{1}, \\
\Delta'' &:= \frac{1}{3} \left\| M^{(1)}_{\frac{1}{2}\frac{3}{2}} - M^{(2)}_{\frac{1}{2}\frac{3}{2}}\right\|_{1}.  
\end{align}

To maximize the diamond norm, we can assume $p = 0 $ or $ p = 1 $.  
When $ p = 1 $, ASP is 
\[
p_{ave} = \frac{1}{2} + \frac{1}{4} \max_{0 \leq t \leq \pi/2} \frac{2}{\sqrt{3}} (\sin t )(1 + \cos t) = 7/8, 
\]
where $t$ is defined as $ q =: \sin^{2} t $ and the maximization is achieved with $ t = \pi/3 $.  
When $ p = 0 $, ASP is 
\begin{equation}
p_{ave} = \frac{1}{2} + \frac{1}{12}\Bigl\|\ket{1}\bra{1} - \ket{\tilde{1}}\bra{\tilde{1}}\Bigl\|_{1} = \frac{1}{2} + \frac{\sqrt{3}}{12} < \frac{7}{8}.  
\end{equation}
Thus the optimal ASP $ p_{ave}^{opt}$ is given by $ 7/8 $.  \qed

%%%%%%%%%%%%%%%%%%%%%%%%%%%%%%%%%%%%%%%%%%%%%%%%%%%%%%%%%%%%%%%%%%%%
\subsection{Parallelized schemes with restricted entanglement}\label{sec:parallel restrict ent}
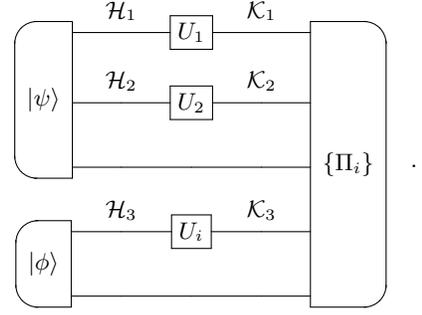
\begin{figure}[t]
\begin{equation}
  \begin{aligned}
    \Qcircuit @C=2em @R=1.5em {
    &\multiprepareC{2}{\ket{\psi}}&\ustick{\mathcal{H}_{1}}\qw& \gate{U_{1}}& \ustick{\mathcal{K}_{1}} \qw& \multimeasureD{4}{\{\Pi_{i}\}} \\
    &\pureghost{\ket{\psi}}&\ustick{\mathcal{H}_{2}}\qw& \gate{U_{2}}& \ustick{\mathcal{K}_{2}} \qw& \ghost{\{\Pi_{i}\}}\\
    &\pureghost{\ket{\psi}}&\qw & \qw & \qw &\ghost{\{\Pi_{i}\}}\\
    &\multiprepareC{1}{\ket{\phi}}&\ustick{\mathcal{H}_{3}}\qw& \gate{U_{i}}& \ustick{\mathcal{K}_{3}} \qw& \ghost{\{\Pi_{i}\}}\\
    &\pureghost{\ket{\phi}}&\qw & \qw & \qw &\ghost{\{\Pi_{i}\}}
    }
  \end{aligned}\quad. \notag
\end{equation}
  \caption{ The quantum circuit for equivalence determination of unitary operations under parallelized schemes with restricted entanglement in the initial state.  }\label{fig:parallel_N1_unentangled input}
\end{figure}

The optimal ASP under parallelized schemes is obtained using an initial state entangled between the systems on which the reference boxes and target box act on.  We prove that this entanglement is necessary.  In particular we restrict the initial state to the form of 
\begin{equation}
\ket{\psi} \otimes \ket{\phi} = \sqrt{X_{1}} \otimes \sqrt{X_{2}} \otimes I_{\mathcal{H}} \kket{I},
\end{equation}
where $X_{1}$ and $X_{2}$ are positive semidefinite operators on $\mathcal{H}_{1} \otimes \mathcal{H}_{2}$ and $\mathcal{H}_{3}$, respectively, satisfying $ \tr X_{1} = \tr X_{2} = 1$ and $\kket{I}$ is an unnormalized maximally entangled vector in $(\mathcal{H}_{1} \otimes \mathcal{H}_{2} \otimes \mathcal{H}_{3})^{\otimes 2}$ (see Fig.\,\ref{fig:parallel_N1_unentangled input}).

This imposes an extra restriction $X = X_{1} \otimes X_{2}$ to the discussion in the previous subsection.  From Lemma~\ref{lem:commutex}, $[X, A^{\otimes 3} \otimes B^{\otimes 3} ]  = [X_{1} \otimes X_{2}, A^{\otimes 3} \otimes B^{\otimes 3} ] = 0$ for arbitrary unitary operators $A,B \in \text{SU}(2)$ and 
\begin{align}
X_{1} &= q I_{0} \oplus (1 - q) \frac{I_{1}}{3}, \\
X_{2} &= \frac{I_{\mathcal{H}_{3}}}{2}.  
\end{align}
Therefore we have 
\begin{multline}
X = \frac{I_{\frac{1}{2}}}{2} \otimes \left(r \ket{\hat{0}}\bra{\hat{0}}_\frac{1}{2} + \frac{(1 - r)}{3} \ket{\hat{1}}\bra{\hat{1}}_\frac{1}{2} \right) \\ \oplus \frac{I_{\frac{3}{2}}}{4} \otimes \frac{2}{3} (1 - r)\ket{0}\bra{0}_{\frac{3}{2}}.   
\end{multline}

Thus the optimal ASP is derived from maximizing
\begin{align}
p_{ave} = \frac{1}{2} + \frac{1}{4} \left(\frac{1}{3} \sin 2 t +\frac{2 \cos^{2} t}{3 \sqrt{3}}+\frac{2
   \cos t \sqrt{2-\cos 2 t} }{3 \sqrt{3}}\right).  
\end{align}
The optimal ASP is numerically derived to be $p_{succ}^{opt} \simeq 0.746399 < 0.875 =7/8$.  Hence, the entanglement in the initial state between the systems of the target and reference boxes is crucial for achieving the optimal ASP.

%%%%%%%%%%%%%%%%%%%%%%%%%%%%%%%%%%%%%%%%%%%%%%%%%%%%%%%%%%%%%%%%%%%%
\subsection{Optimality under general schemes}\label{sec:ordered strategies N1}
In general, arbitrary quantum operations can be applied between the black-boxes, which impose an ordering on the black-boxes in the quantum circuit.  In this section, we show that the optimal ASP of $(1,1)$-equivalence determination under general schemes is $7/8$.

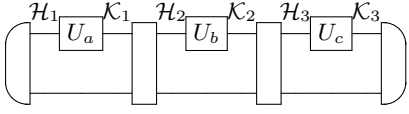
\begin{figure}[t]
\begin{align}
  \begin{aligned}
    \Qcircuit @C=0.6em @R=1.3em {
      &\multiprepareC{1}{}&\ustick{\mathcal{H}_{1}}\qw& \gate{U_{a}}& \ustick{\mathcal{K}_{1}} \qw& \multigate{1}{}&\ustick{\mathcal{H}_{2}}\qw& \gate{U_{b}}& \ustick{\mathcal{K}_{2}} \qw& \multigate{1}{}&\ustick{\mathcal{H}_{3}}\qw& \gate{U_{c}}& \ustick{\mathcal{K}_{3}} \qw& \multimeasureD{1}{}\\
      & \pureghost{}&\qw & \qw & \qw &\ghost{}&\qw & \qw & \qw &\ghost{}&\qw & \qw & \qw &\ghost{}
    }
  \end{aligned}\notag
  \end{align}
  \caption{The most general quantum circuit for $(1,1)$-equivalence determination when the optimal average success probability is concerned.}\label{fig:quatum 4-tester}
\end{figure}

Three different orderings can be considered.  
We assign the Hilbert spaces $\mathcal{H}_{1}$ and $\mathcal{K}_{1}$ to first black-box used in the circuit.  $\mathcal{H}_{2}$ and $\mathcal{K}_{2}$ are assigned to the second black-box, while $\mathcal{H}_{3}$ and $\mathcal{K}_{3}$ to the third.  Each black-box is either a reference box or the target box (see Fig.\,\ref{fig:quatum 4-tester}).  The number of independent orderings is three, because the probability of the target box being $U_{1}$ and $U_{2}$ are equal.  The independent orderings are characterized by the location of the target box.  
%Due to the symmetry of the probability of the target box being one of the candidates, the three independent orderings are characterized by the location of the target box in the quantum circuit.  

The success probability of obtaining the correct answer is given by 
\begin{align} 
p_{U_{1}, U_{2}}^{\langle j\rangle} = \frac{1}{2} \tr \left[ \kket{W_{1}^{\langle j \rangle}}\bbra{W_{1}^{\langle j \rangle}} \widetilde{\Pi}_{1}  + \kket{W_{2}^{\langle j\rangle }}\bbra{W_{2}^{\langle j \rangle}} \widetilde{\Pi}_{2}  \right],  
\end{align}
where $\kket{W_{i}^{\langle j \rangle}}$ defined by
\begin{align}
\kket{W_{i}^{\langle 1 \rangle}}_{\mathcal{KH}} &:= \kket{U_{i}} \otimes \kket{U_{1}} \otimes \kket{U_{2}}, \\
\kket{W_{i}^{\langle 2 \rangle}}_{\mathcal{KH}} &:= \kket{U_{2}} \otimes \kket{U_{i}} \otimes \kket{U_{1}}, \\
\kket{W_{i}^{\langle 3 \rangle}}_{\mathcal{KH}} &:= \kket{U_{1}} \otimes \kket{U_{2}} \otimes \kket{U_{i}},
\end{align}
correspond to the three orderings of the target box being used the first, second, and last, respectively.  
This success probability depends on the choice of $U_{1}$ and $U_{2}$. By taking the average over the Haar measure, we obtain the following proposition.  

\begin{prop}\label{prop:sed ordered}
The optimal average success probability for $(1,1)$-equivalence determination under general schemes is given as an SDP 
\begin{align}  
\text{maximize}\quad & p^{\langle j \rangle}_{ave} = \frac{1}{2} \tr \left[ \widetilde{\Pi}_{1} M_{1}^{\langle j \rangle} +  \widetilde{\Pi}_{2} M_{2}^{(j)} \right] \label{sdp:n1-co-primal_1}, \\
\text{subject to}\quad & \widetilde{\Pi}_{i} \geq 0, \ i = 1,2 \label{sdp:n1-co-primal_2},\\
& \widetilde{\Pi}_{1} + \widetilde{\Pi}_{2} = I_{\mathcal{K}_{3}} \otimes Y,  \\
& \tr_{\mathcal{H}_{3}} Y = I_{\mathcal{K}_{2}} \otimes Y^{\{1\}},  \\
& \tr_{\mathcal{H}_{2}} Y^{\{1\}} = I_{\mathcal{K}_{1}} \otimes Y^{\{0\}},  \\
& \tr Y^{\{0\}} = 1, \label{sdp:n1-co-primal_3}
\end{align}
where $Y$, $Y^{\{1\}}$ and $Y^{\{0\}}$ are positive semidefinite operators and $M^{\langle j \rangle}_{i}$ are given by 
\begin{equation}
M^{\langle j \rangle}_{i} := \int d\mu(U_{1})\int d\mu(U_{2}) \kket{W^{\langle j \rangle}_{i}}\bbra{W^{\langle j \rangle}_{i}}.  
\end{equation}
\end{prop}

The Haar random sampling $U_{1}$ and $U_{2}$ demand the following constraints on the variables of the SDP in Proposition\,\ref{prop:sed ordered}.  
\begin{lem}\label{lem:n1-co-primal-sdp}
The quantum $4$-tester $\{\widetilde{\Pi}_{i}\}$ and positive semidefinite operators $Y$, $Y^{\{1\}}$ and $Y^{\{0\}}$ can be chosen to satisfy
\begin{align}
&[\widetilde{\Pi}_{i}, (A^{\otimes 3})_{\mathcal{K}} \otimes (B^{\otimes 3})_{\mathcal{H}}] = 0, \\
&[Y, (A^{\otimes 2})_{\mathcal{K}_{1}\mathcal{K}_{2}} \otimes (B^{\otimes 3})_{\mathcal{H}}] = 0, \\
&[Y^{\{1\}}, A_{\mathcal{K}_{1}} \otimes (B^{\otimes 2})_{\mathcal{H}_{1}\mathcal{H}_{2}}] = 0, \\
&[Y^{\{0\}}, B_{\mathcal{H}_{1}}] = 0, 
\end{align}
for $i = 1,2$ and arbitrary $A,B \in \text{SU}(2)$.  
\end{lem}
\noindent The proof of the lemma is given in Appx.\,\ref{sec-prf:n1-co-primal-sdp}.

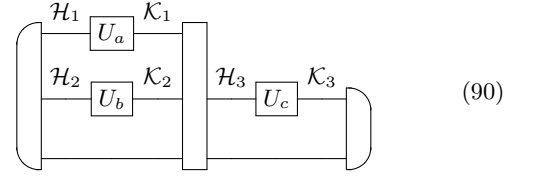
\begin{figure}[t]
\begin{align}
  \begin{aligned}
    \Qcircuit @C=1.0em @R=1.3em {
      &\multiprepareC{2}{}&\ustick{\mathcal{H}_{1}}\qw& \gate{U_{a}}& \ustick{\mathcal{K}_{1}} \qw& \multigate{2}{}&&&&\\
      & \pureghost{}&\ustick{\mathcal{H}_{2}}\qw& \gate{U_{b}}& \ustick{\mathcal{K}_{2}} \qw& \ghost{}&\ustick{\mathcal{H}_{3}}\qw& \gate{U_{c}}& \ustick{\mathcal{K}_{3}} \qw& \multimeasureD{1}{}\\
      & \pureghost{}&\qw & \qw & \qw &\ghost{}&\qw & \qw & \qw &\ghost{}
    }
  \end{aligned}
  \end{align}
  \caption{The quantum $3$-tester giving the same average success probability as that drawn in Fig.\,\ref{fig:quatum 4-tester}.  }\label{fig:quatum 3-tester}
\end{figure}

Lemma~\ref{lem:n1-co-primal-sdp} and Schur's lemma imply Eq.\,(\ref{eq:reducing condition}).  Therefore quantum $3$-testers described in Fig.\,\ref{fig:quatum 3-tester} is sufficient.  There are only two cases of non-trivial orderings, {\it i.e.}, the target box being used the first or last.

Since we formulated the optimization problem as an SDP, there exists a dual SDP.  
A solution to the dual gives an upper bound of the primal \cite{boyd2004convex}.    
A {\it lower} bound to the primal is $7/8$ since the general schemes include the parallelized schemes.  
In the following, we give a feasible solution to the dual that achieves the value $7/8$.

\begin{lem}\label{lem:n1-co-dual-sdp}
A dual SDP of the primal SDP given in Eqs.\,(\ref{sdp:n1-co-primal_1}) - \,(\ref{sdp:n1-co-primal_3}) is expressed as 
\begin{align} 
\text{minimize}\quad & \lambda,  \notag \\
\text{subject to}\quad& \frac{M^{\langle j \rangle}_{1}}{2} - \Omega \leq 0, \label{eq:sdp_dual_1}\\
& \frac{M^{\langle j \rangle}_{2}}{2} - \Omega \leq 0, \label{eq:sdp_dual_2}\\
& \tr_{\mathcal{K}_{3}} \Omega - I_{\mathcal{H}_{3}} \otimes \Omega^{\{1\}} \leq 0,\label{eq:sdp_dual_3}\\
& \tr_{\mathcal{K}_{2}} \Omega^{\{1\}} - I_{\mathcal{H}_{2}} \otimes \Omega^{\{0\}} \leq 0,\label{eq:sdp_dual_4}\\
& \tr_{\mathcal{K}_{1}} \Omega^{\{0\}} - \lambda I_{\mathcal{H}_{1}} \leq 0.  \label{eq:sdp_dual_5}
\end{align}
\end{lem}
\noindent This can be derived by introducing Lagrange multipliers \cite{boyd2004convex} (Appx.\,\ref{sec-prf:n1-co-primalrodual}).

\begin{lem}\label{lem:n1-co-dual-multi}
The dual SDP given in Lemma \ref{lem:n1-co-dual-sdp} is equivalent to the following SDP on the multiplicity subspaces.  
\begin{align} 
\text{minimize}\quad & \lambda,  \label{eq:sdp_dual_multi_first_0}\\
\text{subject to}\quad & \Omega_{JL} - \frac{M^{[i]}_{JL}}{2} \geq 0, \\
&\text{ for } J, L = 1/2, 3/2 \text{ and } i = 1,2, \label{eq:sdp_dual_multi_first}  \\
&\Omega^{\{1\}}_{00}\ket{\hat{0}}\bra{\hat{0}}_{\frac{1}{2}} + \Omega^{\{1\}}_{01}\ket{\hat{1}}\bra{\hat{1}}_{\frac{1}{2}} \notag \\
& \qquad \qquad \qquad - \Omega_{\frac{1}{2}\frac{1}{2}}^{0\rightarrow 1/2} \geq 0, \\
&\Omega^{\{1\}}_{01} - \Omega_{\frac{1}{2}\frac{3}{2}}^{0\rightarrow 1/2} \geq 0, \\
&\Omega^{\{1\}}_{10}\ket{\hat{0}}\bra{\hat{0}}_{\frac{1}{2}} + \Omega^{\{1\}}_{11}\ket{\hat{1}}\bra{\hat{1}}_{\frac{1}{2}} \notag \\
&\qquad \qquad - \Omega_{\frac{1}{2}\frac{1}{2}}^{1\rightarrow 1/2}- \Omega_{\frac{3}{2}\frac{1}{2}} \geq 0, \\
&\Omega^{\{1\}}_{11} - \Omega_{\frac{1}{2}\frac{3}{2}}^{1\rightarrow 1/2} - \Omega_{\frac{3}{2}\frac{3}{2}} \geq 0,\\
& \lambda - \Omega^{\{1\}}_{00} - \Omega^{\{1\}}_{10} \geq 0, \\
& \lambda - \Omega^{\{1\}}_{01} - \Omega^{\{1\}}_{11} \geq 0,\label{eq:sdp_dual_multi_last}  
\end{align}
where we define 
\begin{align}
\Omega &= \bigoplus_{J = \frac{1}{2}}^{\frac{3}{2}}\bigoplus_{L = \frac{1}{2}}^{\frac{3}{2}} \frac{I_{J}^{\mathcal{K}}}{d_{J}} \otimes I_{L}^{\mathcal{H}} \otimes \Omega_{JL},\\
\Omega^{\{1\}} &= \bigoplus_{J = 0}^{1}\bigoplus_{L = 0}^{1} \frac{I_{J}^{\mathcal{K}_{1}\mathcal{K}_{2}}}{d_{J}} \otimes I_{L}^{\mathcal{H}_{1}\mathcal{H}_{2}}\otimes \Omega^{\{1\}}_{JL},
\end{align}
and $\Omega_{\frac{1}{2}\frac{1}{2}}^{j\rightarrow 1/2} = (\bra{\hat{j}}_{\frac{1}{2}} \otimes I_{\mathcal{V}^{[3]}_{L}})\Omega_{\frac{1}{2}L}(\ket{\hat{j}}_{\frac{1}{2}} \otimes I_{\mathcal{V}^{[3]}_{L}})$.  
\end{lem}
\noindent The proof of this lemma is given in Appx.\,\ref{sec:n1-co-dual-multi}.

\begin{thm}\label{thm:n1-co-opt}
The optimal average success probability of $(1,1)$-equivalence determination under general schemes is 7/8 when unitary operations are chosen from the Haar measure.
\end{thm}
\noindent \textbf{Proof of Theorem \ref{thm:n1-co-opt}.}  The optimal ASP by the general schemes is at least $7/8$ since the general schemes include the parallelized schemes.  The dual SDP given in Lemmas\,\ref{lem:n1-co-dual-sdp} and \ref{lem:n1-co-dual-multi} gives an upper bound of the primal SDP, whose answer gives the optimal ASP in the general schemes.  The optimal ASPs coincide for $M^{\langle 1 \rangle}_{i} $ and $M^{\langle 2 \rangle}_{i} $.  In Appx.\,\ref{sec:prf of n1 order}, we give a feasible set of parameters for $\lambda = 7/8$ for two nontrivial orderings of the black-boxes $M^{\langle 2 \rangle}_{i} $ and $M^{\langle 3 \rangle}_{i}$.  Hence the optimal solution to the dual SDP is at most $7/8$.  This concludes the proof.  \qed

%%%%%%%%%%%%%%%%%%%%%%%%% When $U_{1}$ is known %%%%%%%%%%%%%%%%%%%%%%%
\section{When $U_{1}$ is known}\label{sec:one known}
In this section, we assume that a classical description of one of the reference boxes, $U_{1}$, is given hence we may optimize the choice of quantum operations based on the description.  
A classical description of $U_{1}$ is obtainable if there is an infinite number of quantum samples of $U_{1}$.  
Conversely, any number of quantum samples of $U_{1}$ can be generated whenever its classical description is available.  Hence a classical description and infinite quantum samples are interchangeable resources.

%%%%%%%%%%%%%%%%%%%%%%%%%%%%%%%%%%%%%%%%%%%%%%%%%%%%%%%%%%%%%%%%%%%%%%%%
\subsection{No quantum sample for $U_{2}$}\label{sec: one op known no use}
First we consider the case in which only the target box is given without any quantum sample of $U_{2}$ or its classical description.  Contrary to the difference in the resources, we show that the optimal ASP is still $7/8$ if $U_{2}$ is distributed according to the Haar measure.

We denote the input and output space of the target box as $\mathcal{H}$ and $\mathcal{K}$, respectively.  The ASP can always be attained with an initial state $\ket{\psi} \in \mathcal{H} \otimes \mathcal{H} $ of the form $\ket{\psi} = I \otimes \sqrt{X}\kket{I}$, with a positive semidefinite operator $X$ with unit trace on $\mathcal{H}$ and maximally entangled vector $\kket{I}$ in $\mathcal{H} \otimes \mathcal{H}$.

The equivalence determination in this case reduces to the state discrimination of $U_{1} \otimes I \ket{\psi}$ and $U_{2} \otimes I \ket{\psi}$.   
Without loss of generality, we may use the classical description of $U_{1}$ to apply $U_{1}^{\dagger}$ before performing the measurement and retain the same success probability.  For mathematical convenience, we assume that $U^{\dagger}_{1} \otimes I$ maps $\mathcal{K} \otimes \mathcal{H}$ to $\mathcal{K} \otimes \mathcal{H}$.  The POVM is denoted as $\{\Pi_{1}^{U_{1}}, \Pi_{2}^{U_{1}}\}$, which does not depend on $U_{2}$.

The ASP over $U_{2}$ is 
\begin{align}
p_{U_{1}} &= \frac{1}{2} \int d\mu(U_{2})\tr\big[ \ket{\psi}\bra{\psi} \Pi_{1}^{U_{1}}\notag \\
&\qquad + (U_{1}^{\dagger}U_{2} \otimes I) \ket{\psi}\bra{\psi} (U_{2}^{\dagger}U_{1} \otimes I) \Pi_{2}^{U_{1}} \big] \\
 &= \frac{1}{2} \tr [ \ket{\psi} \bra{\psi} \Pi_{1}^{U_{1}} + \widetilde{E} \Pi_{2}^{U_{1}}] \label{eq: one op known discr},
\end{align}
where 
\begin{equation}
\widetilde{E} = \frac{I_{\mathcal{K}}}{2} \otimes X_{\mathcal{H}}.
\end{equation}
Therefore, it suffices to find a POVM that optimally distinguishes $\ket{\psi}\bra{\psi}$ and $\widetilde{E}$.  To maximize ASP, we define a quantum $2$-tester $\widetilde{\Pi}_{i} = (I \otimes \sqrt{X}) \Pi_{i} (I \otimes \sqrt{X})$ and obtain
\begin{align}
p_{ave} &= \frac{1}{2} \tr \left[\kket{I} \bbra{I} \widetilde{\Pi}_{1} + \left(\frac{I}{2} \otimes I \right)  \widetilde{\Pi}_{2}  \right],
\end{align}
where $\widetilde{\Pi}_{1} + \widetilde{\Pi}_{2} = I_{\mathcal{K}} \otimes X_{\mathcal{H}}$.

For a given $\{\widetilde{\Pi}_{1}, \widetilde{\Pi}_{2}\}$ realizing ASP of $p_{ave} $, a quantum $2$-tester 
\begin{equation}
\widetilde{\Pi}'_{i} = \int d \mu(A) (A \otimes A^{*}) \widetilde{\Pi}_{i}(A \otimes A^{*})^{\dagger}
\end{equation}
also achieve the same ASP $p_{ave}$, since $(A \otimes A^{*})\kket{I} = \kket{I} $ for any $A \in \text{SU}(2)$.  
By definition, $\widetilde{\Pi}'_{i} $ satisfy $[\widetilde{\Pi}'_{i}, A \otimes A^{*}] = 0 $ for any $A \in \text{SU}(2)$.  
Thus the optimal APS can be obtained assuming this commutation relation.

The relation $\widetilde{\Pi}'_{1} + \widetilde{\Pi}'_{2} = I \otimes X$ and the commutation relation imply that 
\begin{equation}
[X, A] = 0,
\end{equation}
for any $A \in \text{SU}(2)$.  This implies that without loss of generality $X = I/2$.  
Moreover, we have
\begin{equation}
\widetilde{\Pi}'_{i} = \alpha_{i} \frac{\kket{I}\bbra{I}}{2} + \beta_{i} Q,
\end{equation}
where $Q$ is the projector onto the subspace orthogonal to $\kket{I}\bbra{I}$ defined as $Q := I - \kket{I}\bbra{I}/2$ and $\alpha_{i}, \beta_{i} \geq 0$ for $ i = 1,2$.  
From the condition $\widetilde{\Pi}'_{1} + \widetilde{\Pi}'_{2} = I\otimes I/2$ we obtain
\begin{equation}
\alpha_{1} + \alpha_{2} = \beta_{1} + \beta_{2} = \frac{1}{2}.  
\end{equation}
Hence, $p_{ave}$ satisfies
\begin{align}
p_{succ} &= \frac{1}{2} \tr \left[\kket{I} \bbra{I} \widetilde{\Pi}_{1} + \left(\frac{I}{2} \otimes I \right) \widetilde{\Pi}_{2} \right] \notag\\
&= \frac{1}{4} (\alpha_{2} + 3 \beta_{2} + 4\alpha_{1})\\
&\leq \frac{7}{8},
\end{align}
where the inequality saturates when $\alpha_{1} = \beta_{2} = 1/2 $ and $\alpha_{2} = \beta_{1} = 0$.

%%%%%%%%%%%%%%%%%%%%%%%%%%%%%%%%%%%%%%%%%%%%%%%%%%%%%%%%%%%%%%%%%%%%%%%%%%%
\subsection{Single quantum sample for $U_{2}$}\label{sec: one op known one use}
As discussed in the previous section, providing a complete classical description of $U_{1}$ implies an ability to prepare any number of its quantum samples.  
Nevertheless, the result shown in the previous subsection indicates that the classical description of reference box $1$ alone without a quantum sample of the other candidate does not improve the optimal ASP.  
In this section we show that the classical description of $U_{1}$ increases the optimal ASP, compared to $(1,1)$-equivalence determination, when a single quantum sample of the reference box $2$ is provided.

Let $U_{2}$ be distributed according to the Haar measure.  For simplicity, we employ a parallelized scheme.  Repeating a similar argument made in the previous subsection, the equivalence determination under the said conditions reduces to a discrimination of unitary operations $U_{2} \otimes U_{2} $ and $U_{2} \otimes I$.  ASP $p_{ave}$ is  
\begin{equation}
p_{ave} = \frac{1}{2} \tr[ E_{1} \widetilde{\Pi}_{1} + E_{2} \widetilde{\Pi}_{2}],
\end{equation}
where $E_{1}$ and $E_{2}$ are defined by
\begin{align}
E_{1} &= \frac{I_{\mathcal{K}_{1}}}{2} \otimes I_{\mathcal{H}_{1}} \otimes \kket{I}\bbra{I}_{\mathcal{K}_{2}\mathcal{H}_{2}},\\
E_{2} &= I_{0}^{\mathcal{K}_{1}\mathcal{K}_{2}} \otimes I_{0}^{\mathcal{H}_{1}\mathcal{H}_{2}} \oplus \frac{I_{1}^{\mathcal{K}_{1}\mathcal{K}_{2}}}{3} \otimes I_{1}^{\mathcal{H}_{1}\mathcal{H}_{2}},
\end{align}
while $\widetilde{\Pi}_{1}, \widetilde{\Pi}_{2} \geq 0$ and $\widetilde{\Pi}_{1} + \widetilde{\Pi}_{2} = I_{\mathcal{K}} \otimes X_{\mathcal{H}} $.  
Without loss of generality, we have 
\begin{equation}
X = (\sin^{2}t) I_{0} \oplus (\cos^{2}t) \frac{I_{1}}{3} =: X_{t},
\end{equation}
with $0 \leq t \leq \pi$.

The optimal ASP is calculated as
\begin{equation}
p^{opt}_{ave} = \frac{1}{2} + \frac{1}{4}\max_{0 \leq t \leq \pi/2}\| (I_{\mathcal{K}} \otimes \sqrt{X_{t}}) (E_{1} - E_{2} ) (I_{\mathcal{K}} \otimes \sqrt{X_{t}}) \|_{1}.  \label{eq:opt succ one known}
\end{equation}	
The maximization term can be calculated as 
\begin{align}
& \| (I_{\mathcal{K}} \otimes \sqrt{X}) (E_{1} - E_{2} ) (I_{\mathcal{K}} \otimes \sqrt{X}) \|_{1} \\
& \quad = \frac{5 \cos^{2} t}{36}+\frac{3}{144} \sqrt{87 -4 \cos 2t -10 \cos 4t} \notag\\ 
& \qquad + \frac{1}{36} \sqrt{357 -352 \cos 2t + 20 \cos 4t}.  
\end{align}
The above equation is derived by a symbolic calculation of Mathematica \cite{Mathematica}.  
The eigenvalues consist of $\frac{\cos^{2}t}{9}$ with $5$-fold degeneracy, 
\begin{equation}
\frac{1}{72} \left(11 -16 \cos 2t \pm \sqrt{357 - 352 \cos 2t + 20 \cos 4t}\right)
\end{equation}
with $3$-fold degeneracy, and non-degenerate
\begin{equation}
\frac{1}{72} \left(-7 + 2 \cos 2t \pm \sqrt{87 - 4\cos 2t - 10 \cos 4t }\right).  
\end{equation}
The rests are zero.  
The optimal ASP is numerically obtained as $p_{ave}^{opt} \simeq 0.902127 > 0.875 =7/8$.

%%%%%%%%%%%%%%%%%%% (2,1)-equivalence determination %%%%%%%%%%%%%%%%%%%%%%%%%%
\section{$(2,1)$-equivalence determination}\label{sec:ordered strategies}
There are $12$ distinct orderings of the reference and target boxes in the most general scheme for $(2,1)$-equivalence determination, \textit{i.e.}, fully ordered case, given by \\  
%\begin{figure}[h]
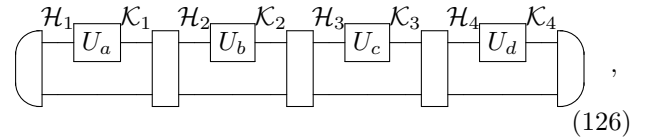
\begin{align}
  \begin{aligned}
    \Qcircuit @C=0.6em @R=0.8em {
      &\multiprepareC{1}{}&\ustick{\mathcal{H}_{1}}\qw& \gate{U_{a}}& \ustick{\mathcal{K}_{1}} \qw& \multigate{1}{}&\ustick{\mathcal{H}_{2}}\qw& \gate{U_{b}}& \ustick{\mathcal{K}_{2}} \qw& \multigate{1}{}&\ustick{\mathcal{H}_{3}}\qw& \gate{U_{c}}& \ustick{\mathcal{K}_{3}} \qw& \multigate{1}{}&\ustick{\mathcal{H}_{4}}\qw& \gate{U_{d}}& \ustick{\mathcal{K}_{4}}\qw &\multimeasureD{1}{}\\
      & \pureghost{}&\qw & \qw & \qw &\ghost{}&\qw & \qw & \qw &\ghost{}&\qw & \qw & \qw &\ghost{}&\qw & \qw & \qw &\ghost{}
    }
  \end{aligned}\quad,\label{eq:sdp-1111-4}
\end{align}
%\caption{A configuration of the black-box in the general scheme for $(2,1)$-equivalence determination.  }\label{eq:sdp-1111-4}
%\end{figure}
depending on how we assign the reference and target boxes to $U_{a}$, $U_{b}$, $U_{c}$, and $U_{d}$.  

It is expected that increasing the concurrency of black-boxes by using more of them simultaneously before applying the next quantum operation causes to lower the optimal ASP.  
We divide the orderings according to the concurrency pattern.  The number of black-boxes in the first layer, {\it i.e.}, after the initial state preparation and before the first quantum operation, is between one and four.  The black-boxes in the first two layers can always be parallelized without sacrificing the optimal ASP if the first layer contains only a single black-box, due to the symmetry of quantum testers induced by averaging over the Haar measure.  Therefore, the most general scheme of Concurrency Pattern (\ref{eq:sdp-1111-4}) is replaceable by 
%\begin{figure}[h]
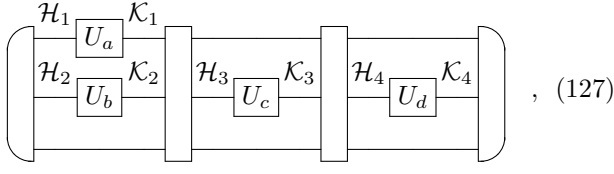
\begin{align}
  \begin{aligned}
    \Qcircuit @C=0.8em @R=0.8em {
      &\multiprepareC{2}{}&\ustick{\mathcal{H}_{1}}\qw& \gate{U_{a}}& \ustick{\mathcal{K}_{1}} \qw& \multigate{2}{}&\qw& \qw&  \qw& \multigate{2}{}&\qw& \qw& \qw& \multimeasureD{2}{}\\
      &\pureghost{}&\ustick{\mathcal{H}_{2}}\qw& \gate{U_{b}}& \ustick{\mathcal{K}_{2}} \qw& \ghost{}&\ustick{\mathcal{H}_{3}}\qw& \gate{U_{c}}& \ustick{\mathcal{K}_{3}} \qw& \ghost{}&\ustick{\mathcal{H}_{4}}\qw& \gate{U_{d}}& \ustick{\mathcal{K}_{4}} \qw&\ghost{}\\
      & \pureghost{}&\qw & \qw & \qw &\ghost{}&\qw & \qw & \qw &\ghost{}&\qw & \qw & \qw &\ghost{}
    }
  \end{aligned}\quad,\label{fig:multi211}
\end{align}
which is abbreviated as
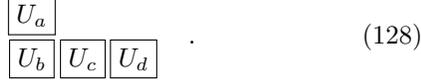
\begin{align}
\begin{aligned} \Qcircuit @C=0.2em @R=0.2em {
& \gateno{U_{a}} & &&    \\
& \gateno{U_{b}} & \gateno{U_{c}} &  \gateno{U_{d}} & 
}\end{aligned} \quad.
\end{align}

%the (\ref{fig:multi211}).  The number of non-trivial configurations of the black-boxes in the quantum circuit is four, each represented in Eqs.\,(\ref{fig:multi211}) - (\ref{fig:multi4}).  

Other concurrency patterns are 
%\caption{Configuration $1$ of $(2,1)$-equivalence determination and its abbreviation used in Table \ref{table:numerical resulf N12}.}\label{fig:multi211}
%\end{figure}
%\begin{figure}[h]
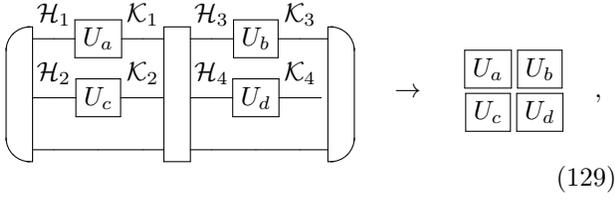
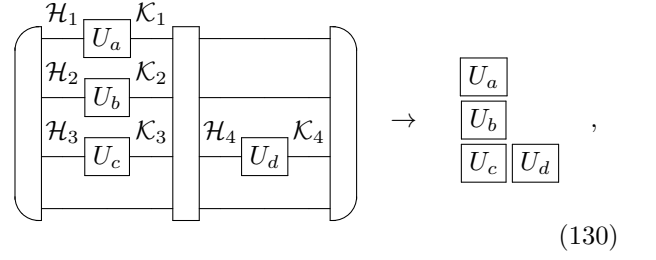
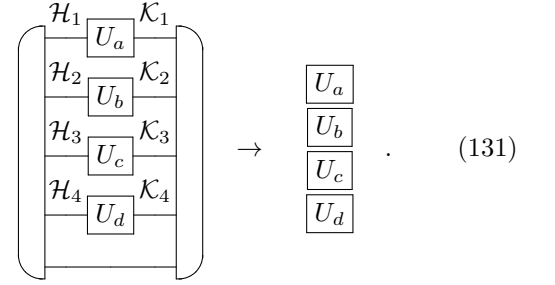
\begin{align}
  \begin{aligned}
    \Qcircuit @C=0.8em @R=0.8em {
    &\multiprepareC{2}{}&\ustick{\mathcal{H}_{1}}\qw& \gate{U_{a}}& \ustick{\mathcal{K}_{1}} \qw& \multigate{2}{}&\ustick{\mathcal{H}_{3}}\qw& \gate{U_{b}}& \ustick{\mathcal{K}_{3}} \qw& \multimeasureD{2}{} \\
      &\pureghost{}&\ustick{\mathcal{H}_{2}}\qw& \gate{U_{c}}& \ustick{\mathcal{K}_{2}} \qw& \ghost{}&\ustick{\mathcal{H}_{4}}\qw& \gate{U_{d}}& \ustick{\mathcal{K}_{4}} \qw& \ghost{1}{}\\
      & \pureghost{}&\qw & \qw & \qw &\ghost{}&\qw & \qw & \qw &\ghost{}
    }
  \end{aligned} \quad\rightarrow\quad \begin{aligned} \Qcircuit @C=0.2em @R=0.2em {
& \gateno{U_{a}} & \gateno{U_{b}} &    \\
& \gateno{U_{c}} & \gateno{U_{d}} & 
}\end{aligned}\quad,\label{fig:multi22}
\end{align}
%\caption{Configuration $2$ of $(2,1)$-equivalence determination and its abbreviation used in Table \ref{table:numerical resulf N12}.}\label{fig:multi22}
%\end{figure}  
%\begin{figure}[h]
\begin{align}
  \begin{aligned}
    \Qcircuit @C=0.8em @R=0.8em {
    &\multiprepareC{3}{}&\ustick{\mathcal{H}_{1}}\qw& \gate{U_{a}}& \ustick{\mathcal{K}_{1}} \qw& \multigate{3}{}&\qw& \qw &  \qw& \multimeasureD{3}{} \\
    &\pureghost{}&\ustick{\mathcal{H}_{2}}\qw& \gate{U_{b}}& \ustick{\mathcal{K}_{2}} \qw& \ghost{}&\qw& \qw&  \qw& \ghost{} \\
      &\pureghost{}&\ustick{\mathcal{H}_{3}}\qw& \gate{U_{c}}& \ustick{\mathcal{K}_{3}} \qw& \ghost{}&\ustick{\mathcal{H}_{4}}\qw& \gate{U_{d}}& \ustick{\mathcal{K}_{4}} \qw& \ghost{}\\
      & \pureghost{}&\qw & \qw & \qw &\ghost{}&\qw & \qw & \qw &\ghost{}
    }
  \end{aligned} \quad\rightarrow\quad \begin{aligned} \Qcircuit @C=0.2em @R=0.2em {
& \gateno{U_{a}} & & \\
& \gateno{U_{b}} & & \\
& \gateno{U_{c}} & \gateno{U_{d}} & 
}\end{aligned}\quad,\label{fig:multi31}
\end{align}
and
%\caption{Configuration $3$ of $(2,1)$-equivalence determination and its abbreviation used in Table \ref{table:numerical resulf N12}.}
%\end{figure}
%\begin{figure}[h]
\begin{align}
  \begin{aligned}
    \Qcircuit @C=0.8em @R=0.8em {
    &\multiprepareC{4}{}&\ustick{\mathcal{H}_{1}}\qw& \gate{U_{a}}& \ustick{\mathcal{K}_{1}} \qw& \multimeasureD{4}{} \\
    &\pureghost{}&\ustick{\mathcal{H}_{2}}\qw& \gate{U_{b}}& \ustick{\mathcal{K}_{2}} \qw& \ghost{} \\
    &\pureghost{}&\ustick{\mathcal{H}_{3}}\qw& \gate{U_{c}}& \ustick{\mathcal{K}_{3}} \qw& \ghost{}\\
    &\pureghost{}&\ustick{\mathcal{H}_{4}}\qw& \gate{U_{d}}& \ustick{\mathcal{K}_{4}} \qw& \ghost{}\\
    &\pureghost{}&\qw & \qw & \qw &\ghost{}
    }
  \end{aligned}\quad\rightarrow\quad \begin{aligned} \Qcircuit @C=0.2em @R=0.2em {
& \gateno{U_{a}} & \\
& \gateno{U_{b}} & \\
& \gateno{U_{c}} & \\
& \gateno{U_{d}} & 
}\end{aligned}\quad .  \label{fig:multi4}
\end{align}
%\caption{Configuration $4$ of $(2,1)$-equivalence determination and its abbreviation used in Table ref{table:numerical resulf N12}.}\label{fig:multi4}
%\end{figure} 

The optimal ASP is obtained for all concurrency patterns and assignments by numerically solving the relevant SDP.  The results are summarized in Table~\ref{table:numerical resulf N12}.  The derivation of the Choi operators and SDPs corresponding to each ordering are given in Supplemental Material\,\cite{supplement}.  The SDPs are rewritten in terms of the multiplicity subspaces.

\onecolumngrid 
\vspace{7mm}
\captionsetup[table]{
justification=RaggedRight,
}
\begin{center}
\begin{table}[h]
%\subfloat{
\begin{tabular}{c|c}
Class $1$: $p_{ave}^{opt} \simeq 0.910516 $ & Class $2$: $p_{ave}^{opt} \simeq 0.902127$ \\ \hline
$\begin{aligned}
{} \\
& \Qcircuit @C=0.2em @R=0.2em {
& \gateno{U_{i}}    &                &               &    \\
& \gateno{U_{1}}&  \gateno{U_{1}}&  \gateno{U_{2}} &    
} \quad \Qcircuit @C=0.2em @R=0.2em {
& \gateno{U_{1}}    &                &               &    \\
& \gateno{U_{2}}&  \gateno{U_{i}}&  \gateno{U_{1}} &    
} \quad \Qcircuit @C=0.2em @R=0.2em {
& \gateno{U_{1}}    &                &               &    \\
& \gateno{U_{2}}&  \gateno{U_{1}}&  \gateno{U_{i}} &    
} \\
{} \\
&\Qcircuit @C=0.2em @R=0.2em {
& \gateno{U_{i}}    &                &               &    \\
& \gateno{U_{1}}&  \gateno{U_{2}}&  \gateno{U_{1}} &    
} \quad \Qcircuit @C=0.2em @R=0.2em {
& \gateno{U_{i}}    & \gateno{U_{2}} &     \\
& \gateno{U_{1}}& \gateno{U_{1}} &    
} \quad \Qcircuit @C=0.2em @R=0.2em {
& \gateno{U_{2}}& \gateno{U_{i}}            &     \\
& \gateno{U_{1}}& \gateno{U_{1}} &    
} \\{} \end{aligned}$   &
$\begin{aligned} \\
&\Qcircuit @C=0.2em @R=0.2em {
& \gateno{U_{1}}    &                &               &    \\
& \gateno{U_{1}}&  \gateno{U_{i}}&  \gateno{U_{2}} &    
} \quad \Qcircuit @C=0.2em @R=0.2em {
& \gateno{U_{1}}    &                &               &    \\
& \gateno{U_{1}}&  \gateno{U_{2}}&  \gateno{U_{i}} &    
} \quad \Qcircuit @C=0.2em @R=0.2em {
& \gateno{U_{i}}    &                &               &    \\
& \gateno{U_{2}}&  \gateno{U_{1}}&  \gateno{U_{1}} &    
} \quad \Qcircuit @C=0.2em @R=0.2em {
& \gateno{U_{1}} &  \gateno{U_{i}    }&     \\
& \gateno{U_{1}} &  \gateno{U_{2}}&    
} \\ {} \\&\Qcircuit @C=0.2em @R=0.2em {
& \gateno{U_{i}}    &         \\
& \gateno{U_{1}}&         \\
& \gateno{U_{1}}& \gateno{U_{2}} 
} \qquad \Qcircuit @C=0.2em @R=0.2em {
& \gateno{U_{2}}    &         \\
& \gateno{U_{1}}&         \\
& \gateno{U_{1}}& \gateno{U_{i}} 
} \quad \Qcircuit @C=0.2em @R=0.2em {
& \gateno{U_{i}}    &         \\
& \gateno{U_{1}}&         \\
& \gateno{U_{2}}& \gateno{U_{1}} 
} \quad \Qcircuit @C=0.2em @R=0.2em {
& \gateno{U_{i}    }&         \\
& \gateno{U_{2}}&         \\
& \gateno{U_{1}}&         \\
& \gateno{U_{1}}&     
} \quad \Qcircuit @C=0.2em @R=0.2em {
& \gateno{U_{i}}    & \gateno{U_{1}} &     \\
& \gateno{U_{2}}& \gateno{U_{1}} &    
}  \end{aligned}$\\{}\\ \hline
  \end{tabular}
%  } %subfloat

\caption{Numerical results of the optimal average success probability $p_{ave}^{opt}$ for $(2,1)$-equivalence determination.  
Each four-block group corresponds to a particular use of black-boxes indicated by the subscripts.  The corresponding quantum circuit for each black-box ordering is given in Concurrency Patterns (\ref{fig:multi211}) - (\ref{fig:multi4}).  The orderings are divided into two classes according to $p_{ave}^{opt}$. }\label{table:numerical resulf N12}  

%\subfloat{
\end{table}
\end{center}
\begin{center}
%\twocolumngrid
%\onecolumngrid

\begin{table}[H]
\begin{tabular}{|c||c|c|c|c|c|c|c|c|c|}
\hline Analyzed in:  & Sec.\,\ref{sec:parallel restrict ent} & Sec.\,\ref{sec:parallel strategies N1} & Sec.\,\ref{sec:ordered strategies N1} & Sec.\,\ref{sec: one op known no use} & Refs.\,\cite{0305-4470-36-9-310,doi:10.1080/09500340903203129} & Sec.\,\ref{sec: one op known one use} & \multicolumn{2}{c|}{Sec.\,\ref{sec:ordered strategies}}& Ref.\,\cite{PhysRevLett.87.177901}, Appx.\,\ref{sec:discrimination known unitary}\\  \hline \hline 
$N_{1}$ & \multicolumn{3}{c|}{1} & known  & 1 & known  & \multicolumn{2}{c|}{2}  & known      \\ \hline
$N_{2}$ & \multicolumn{3}{c|}{1} & \multicolumn{2}{c|}{0}  & \multicolumn{3}{c|}{1}& known  \\ \hline
Initial entanglement & R & \multicolumn{8}{c|}{G}   \\ \hline
Ordering & P & P & G & G = P & P & P & C2 & C1 & G = P\\ \hline \rule[-5pt]{0em}{15pt}
$p_{ave}^{opt}$ & $\simeq 0.746399$ &\multicolumn{4}{c|}{$7/8 = 0.875$} & \multicolumn{2}{c|}{$\simeq 0.902127$} & $\simeq 0.910516$ & $\frac{1}{2} + \frac{4}{3\pi} \simeq 0.924413 $ \\ \hline
\end{tabular}
% } %subfloat
\caption{A comparison of the optimal average success probabilities of $(N_{1}, N_{2})$-equivalence determination.  $N_{i}$ are the number of quantum samples for $U_{i}$. ``known'' indicates that a classical descriptions of $U_{1}$ is given.  ``R'' in the row ``initial entanglement'' implies that the initial entanglement is restricted and ``G'' otherwise.  In the row ``ordering'', ``P'' is for parallelized, ``G'' for general, and ``C1'' and ``C2'' for Class $1$ and  Class $2$, respectively.  }\label{table: comparison of parallel strategies}\label{table:sameasbox}
\end{table}
\end{center}
\twocolumngrid

% ここにAcinさんの結果を入れておく

%%%%%%%%%%%%%%%%%%%%%%%%% CONCLUSION %%%%%%%%%%%%%%%%%%%%%%%%%%%%%%%%%%%

\section{Conclusion}\label{sec:conclude}
In this paper, we introduced $(N_{1}, N_{2})$-equivalence determination of unitary operations, which is a discrimination task with two candidate unitary operations, $U_{1}$ and $U_{2}$.  Classical descriptions of $U_{i}$ are not available, but $N_{i}$ quantum samples are given.  The optimal average success probability (ASP) obtained under each setting is summarized in Table~\ref{table:sameasbox}.  
%Three black-boxes (one target box and two reference boxes) each implementing a unitary operation are given as a physical system.  The unitary operation implemented by the target box is guaranteed to be one of the unitary operations implemented by the two reference boxes but the classical description of the unitary operations are unknown.  
%If $i$-th reference box is allowed to be used $N_{i}$ times and the target box once, then is is called $(N_{1}, N_{2})$-equivalence determination of unitary operations.  

We derived the optimal ASP for $(1,1)$-equivalence determination in both parallelized and general schemes.  The problem was formulated as a semidefinite program (SDP).  The SDP was used for the parallelized schemes to reduce the number of degrees of freedom in the choice of the initial state.  The optimal ASP under the parallelized schemes is $7/8$.  We also showed that $7/8$ cannot be achieved when the entanglement of the initial state is restricted.  For the general schemes, a dual SDP was derived, for which we found a feasible set of parameters establishing that the optimal ASP under general schemes is at most $7/8$.  Therefore, the parallelized schemes achieve the optimal ASP of the general schemes.

We investigated when a classical description of one of the candidates $U_{1}$ is given.  With no quantum sample of $U_{2}$, the optimal ASP is analytically derived to be $7/8$ in this case.  The numerics shows that the probability increases to $\simeq 0.902127$ with a single quantum sample of $U_{2}$.  

%We considered two types of schemes called parallelized schemes and general schemes and investigated when one outperforms the other.  In the parallelized schemes, the black-boxes are used simultaneously during the computation and no quantum operation is applied between the uses of the black-boxes.  In contrast, the general schemes permit arbitrary quantum operations to be inserted between the uses of the black-boxes.  To perform equivalence determination of unitary operations, we used quantum testers that generalize quantum measurements to higher-order quantum computation.  A tester takes quantum operations as inputs of computation.  We formulated optimization problems to maximize the success probability for obtaining the correct guess in terms of semidefinite programmings.  

%There are $44$ in $(2,1)$-equivalence determination.  
In $(2,1)$-equivalence determination, the symmetry induced by averaging over the Haar measure reduces non-trivial orderings of the black-boxes to $15$.  From numerics, they divide into two classes according to the optimal ASP, \textit{i.e.}, Class $1$ with $p_{ave}^{opt} \simeq 0.910516$ and Class $2$ with $p_{ave}^{opt} \simeq 0.902127$.  
%showed that the optimal performance of the task varies depending on the order of the operation.  %To the best of our knowledge, such a task considered before is only the quantum switch \cite{PhysRevA.88.022318}, where the order dependence is trivial as its exact implementation is impossible with the single use of each operation in the quantum circuit model.

%General conditions for improvements under the general schemes, however, are not known.  
%The general method to find the order to obtain the higher success probability is still an open problem.  

The optimal ASP of $7/8$ in $(1,1)$-equivalence determination has been obtained in the context of the comparison of unitary operations~\cite{0305-4470-36-9-310,doi:10.1080/09500340903203129}, which is a restricted $(1,0)$-equivalence determination.  Therefore, one of the quantum samples does not contribute in $(1,1)$-equivalence determination.  Contrasting the results obtained in Sec.\,\ref{sec: one op known no use} and Refs.\,\cite{0305-4470-36-9-310,doi:10.1080/09500340903203129}, the optimal ASP for $(N_{1}, 0)$-equivalence determination under the parallelized schemes can be achieved with $ N_{1} = 1 $.  The optimal ASP does not increase with the additional $N_{1} - 1$ quantum samples.  Similarly, the results obtained in Secs.\,\ref{sec: one op known one use} and \ref{sec:ordered strategies} indicate that $(N_{1}, 1)$-equivalence determination under the parallelized schemes can be achieved with $ N_{1} = 2 $.  

The adaptive operations allowed in the general schemes provide advantages over the parallelized schemes in optimization~\cite{PhysRevLett.101.180501,PhysRevA.73.042301,PhysRevA.81.032339,hayashi2009discrimination}.  
Indeed, the general schemes in $(2,1)$-equivalence determination outperform the parallelized.  
In contrast, the general schemes in $(1,1)$-equivalence determination do not give improvements over the parallelized.  Moreover, an exact classical description of an unknown unitary operation implemented by a black-box cannot be determined by finite uses of the black-box.  Nevertheless, finite quantum samples were sufficient to achieve the same performance as with a classical description given.  Equivalence determination has revealed unexpected properties of resourcefulness of input quantum operations and their orderings in higher-order quantum computation.

\section*{Acknowledgment}
This work is supported by the Project for Developing Innovation Systems of MEXT, Japan, and JSPS KAKENHI (Grants No. 26330006, No\,15H01677, No\,16H01050, and No\,17H01694). %We also acknowledge the ELC project [Grant-in-Aid for Scientific Research on Innovative Areas MEXT KAKENHI (Grant No\,24106009)]. 

%%%%%%%%%%%%%%%%%%%%%%%%%%% APPENDIX %%%%%%%%%%%%%%%%%%%%%%%%%%%%%%%%%%%%%%%
\appendix

%%%%%%%%%%%%%%%%%%%%%%%%%%%%%%%%%%%%%%%%%%%%%%%%%%%%%%%%%%%%%%%%%%%%
\section{Discrimination of two unitary operations with full classical descriptions}\label{sec:discrimination known unitary}
We summarize the relevant results in Ref.\,\cite{PhysRevLett.87.270404} on minimum-error discrimination of two unitary operations with their full classical description given.  Consider unitary operations $U_{1}$ and $U_{2}$ in $\text{SU}(d)$ acting on $\mathcal{H}$ and a black-box implementing $U_{1}$ and $U_{2}$ with probability $\eta_{1}$ and $\eta_{2}$, respectively. We denote an initial state as $\ket{\psi}_{\mathcal{H}\mathcal{H}_{A}}$ where $\mathcal{H} \cong \mathcal{H}_{A}$.  Then the two candidate states $\ket{\psi_{1}} = U_{1} \otimes I \ket{\psi} $ and $\ket{\psi_{2}} = U_{2} \otimes I \ket{\psi} $ are obtained after applying the unitary operation implemented by the black-box.

The optimal success probability for minimum-error discrimination is derived as 
\begin{equation}
p^{opt}_{U_{1},U_{2}} = \begin{cases}
    1 & (\theta_{d} - \theta_{1} \geq \pi) \\
    \frac{1}{2}( 1 + \sqrt{1 - 4 \eta_{1} \eta_{2} \cos^{2}{\frac{\theta_{d} - \theta_{1}}{2}}}) & (otherwise),  
  \end{cases}
\end{equation}%\label{eq:opt unitary discriminate}
where $\{\theta_{i}\}_{i = 1}^{d}$ are defined by the spectral decomposition $U_{1}^{\dagger} U_{2} = \sum_{j = 1}^{d} e^{i \theta_{j}} \ket{\zeta_{i}}\bra{\zeta_{i}} $ satisfying $-\pi \leq \theta_{i} < \pi $ and $\theta_{1} \leq \theta_{2} \leq \cdots \leq \theta_{d}$.

For the case of $\text{SU}(2)$, we can denote $ U = (\cos t)  I + i (\sin t) (\sum_{j = 1}^{3} v_{j} \sigma_{j})$, where $(v_{1}, v_{2}, v_{3})$ is a normalized real vector and $\{\sigma\}_{j=1}^{3}$ are the Pauli operators.  The optimal ASP $p^{opt}_{ave}$ over the Haar measure is given by
\begin{equation}
p^{opt}_{ave} = \frac{1}{2} + \frac{1}{\pi}\int_{0}^{\pi} dt |\sin t| \sin^{2}t = \frac{1}{2} + \frac{4}{3\pi},	 
\end{equation}
for $\eta_{1} = \eta_{2} = 1/2$.

%%%%%%%%%%%%%%%%%%%%%%%%%%% PROOF OF LEMMA 1 %%%%%%%%%%%%%%%%%%%%%%%%%%%%%%%%%%%%%%%
\section{Proof of Lemma~\ref{lem:commutex}}\label{prf:commutex}
Suppose that a quantum $2$-tester $\{\widetilde{\Pi}_{1}, \widetilde{\Pi}_{1} \}$ gives ASP $p_{ave}$, satisfying $ \Pi_{i} \geq 0$ for $ i = 1,2$ and $\widetilde{\Pi}_{1} + \widetilde{\Pi}_{2} = I_{\mathcal{K}} \otimes X_{\mathcal{H}} $ with $\tr X = 1 $.  Let us define an averaged operator of $\widetilde{\Pi}_{i}$  as
\begin{multline}
    \widetilde{\Pi}'_{i} := \int d\mu(A) \int d\mu(B)\\ 
    ((A^{\otimes 3})_{\mathcal{K}}\otimes (B^{\otimes 3})_{\mathcal{H}}) \widetilde{\Pi}_{i}  ((A^{\dagger \otimes 3})_{\mathcal{K}} \otimes (B^{\dagger \otimes 3})_{\mathcal{H}}).  
\end{multline}
We have
    \begin{align}
    \widetilde{\Pi}'_{1} + \widetilde{\Pi}'_{2} &= I_{\mathcal{K}} \otimes X'_{\mathcal{H}}, 
    \end{align}
where $X'_{\mathcal{H}} $ is defined as 
\begin{equation}
X'_{\mathcal{H}} := \int d\mu(B) B^{\otimes 3} X_{\mathcal{H}} B^{\dagger \otimes 3}. \label{eq:def-of-x}
\end{equation}
For any unitary operator $T$, $[X'_{\mathcal{H}}, T^{\otimes 3}] = 0 $, since
\begin{align}
T^{\otimes 3}  X' T^{\dagger \otimes 3} &= \int d\mu(B) (TB)^{\otimes 3} X_{\mathcal{H}} (TB)^{\dagger \otimes 3} \\
&= \int d\mu(TB) (TB)^{\otimes 3} X_{\mathcal{H}} (TB)^{\dagger \otimes 3} \\
&= \int d\mu(B') (B')^{\otimes 3} X_{\mathcal{H}} (B')^{\dagger \otimes 3} \\
&= X',
\end{align}
where we used the property of the Haar measure $d\mu(AB) = d\mu(B) $ for arbitrary unitary operators $A$ and $B$ in $\text{SU}(2)$.

Finally, $\{\widetilde{\Pi}'_{i}\}$ gives the same ASP as $\{\widetilde{\Pi}_{i}\}$, because  
    \begin{align}
\tr \left[ M_{1} \widetilde{\Pi}'_{1} + M_{2} \widetilde{\Pi}'_{2} \right] &=  \tr \left[ M_{1} \widetilde{\Pi}_{1} + M_{2} \widetilde{\Pi}_{2} \right] 
    \end{align}
from
\begin{align}
((A^{\dagger \otimes 3})_{\mathcal{K}} \otimes (B^{\dagger \otimes 3})_{\mathcal{H}})M_{i}((A^{\otimes 3})_{\mathcal{K}} \otimes (B^{\otimes 3})_{\mathcal{H}} ) = M_{i}\label{eq:n1_m1_int},
\end{align}
for $i = 1,2$.  
Hence we may assume without loss of generality  that $[T^{\otimes 3}, X] = 0$ for an arbitrary unitary operator $T$.  
\qed

%%%%%%%%%%%%%%%%%%%%%%%%%%% LEMMA 6 %%%%%%%%%%%%%%%%%%%%%%%%%%%%%%%%%%%%%%%
\section{Lemma \ref{lem:derivation of Mi N1}} \label{prf:n1para_opt}
\begin{lem}\label{lem:derivation of Mi N1}
$M_{1}$ is represented as 
    \begin{align}
    M_{1} &= \frac{I_{\frac{1}{2}}^{\mathcal{K}}}{2} \otimes I_{\frac{1}{2}}^{\mathcal{H}} \otimes \left(\ket{00}\bra{00}_{\frac{1}{2}\frac{1}{2}} + \frac{1}{3}\ket{11}\bra{11}_{\frac{1}{2}\frac{1}{2}}\right) \notag\\
    &\quad \oplus \frac{I_{\frac{1}{2}}^{\mathcal{K}}}{2}  \otimes I_{\frac{3}{2}}^{\mathcal{H}} \otimes  \frac{1}{3}\ket{10}\bra{10}_{\frac{1}{2}\frac{3}{2}} \oplus \frac{ I_{\frac{3}{2}}^{\mathcal{K}}}{4} \otimes I_{\frac{1}{2}}^{\mathcal{H}} \otimes  \frac{2}{3}\ket{01}\bra{01}_{\frac{3}{2}\frac{1}{2}} \notag\\
    &\quad \oplus \frac{I_{\frac{3}{2}}^{\mathcal{K}}}{4}  \otimes I_{\frac{3}{2}}^{\mathcal{H}} \otimes \frac{2}{3}\ket{00}\bra{00}_{\frac{3}{2}\frac{3}{2}}    \label{eq:n1-test-last-m1}
    \end{align}
$M_{2}$ is obtained by transforming $\{\ket{0}, \ket{1}\} \rightarrow \{\ket{\tilde{0}}, \ket{\tilde{1}}\}$ in $M_{1}$.  \end{lem}

\noindent \textbf{Proof.} From Eq.\,(\ref{eq:n1_m1_int}), we have
\begin{equation}
[M_{i} , (A^{\otimes 3})_{\mathcal{K}} \otimes (B^{\otimes 3})_{\mathcal{H}}] = 0, \label{eq:n1-m-commute}
\end{equation}
for any unitary operators $A, B$ in $\text{SU}(2) $ and $ i = 1,2$.  

$\mathcal{K} \otimes \mathcal{H}$ is decomposed as 
\begin{equation}
\mathcal{K} \otimes \mathcal{H} = \bigoplus_{J = \frac{1}{2}}^{\frac{3}{2}}\bigoplus_{L = \frac{1}{2}}^{\frac{3}{2}} \mathcal{U}_{J} \otimes \mathcal{U}_{L} \otimes \mathcal{V}^{[3]}_{J} \otimes \mathcal{V}^{[3]}_{L}.
\end{equation}
Here we changed the order of the spaces for convenience.  
In terms of irreducible representation, the tensor products of unitary operators are given as 
\begin{equation}
(A^{\otimes 3})_{\mathcal{K}} \otimes (B^{\otimes 3})_{\mathcal{H}} = \bigoplus_{J = \frac{1}{2}}^{\frac{3}{2}}\bigoplus_{L = \frac{1}{2}}^{\frac{3}{2}} A_{J} \otimes B_{L} \otimes I_{\mathcal{V}^{[3]}_{J}\mathcal{V}^{[3]}_{L}},  
\end{equation}
where $A_{J}$ and $B_{L}$ are the irreducible representations acting on $\mathcal{U}_{J} $ and $ \mathcal{U}_{L}$, respectively, and $I_{\mathcal{V}^{[3]}_{J}\mathcal{V}^{[3]}_{L}}$ are the identity operator on $\mathcal{V}^{[3]}_{J} \otimes \mathcal{V}^{[3]}_{L}$.

From Schur's lemma and Eq.\,(\ref{eq:n1-m-commute}), $M_{i}$ are represented as 
\begin{equation}
M_{i} = \bigoplus_{J = \frac{1}{2}}^{\frac{3}{2}}\bigoplus_{L = \frac{1}{2}}^{\frac{3}{2}} \frac{I_{J}^{\mathcal{K}}}{d_{J}} \otimes I_{L}^{\mathcal{H}} \otimes M^{(i)}_{JL} \label{ea:form of M}, 
\end{equation}
where $M^{(i)}_{JL}$ are linear operators on $\mathcal{V}^{[3]}_{J} \otimes \mathcal{V}^{[3]}_{L}$ and $d_{J} := 2 J + 1$.

The next step is to derive $M^{(i)}_{JL}$ for $i = 1,2$.  
Define $\eta^{[N]}$ as 
\begin{equation}
\eta^{[N]} = \int d\mu(U) \kket{U^{\otimes N}}\bbra{U^{\otimes N}}. \label{eq:def_eta}
\end{equation}
$M_{1}$ and $M_{2}$ are represented as 
\begin{align}
M_{1} &= \eta^{[2]}_{\mathcal{K}_{1}\mathcal{K}_{3}\mathcal{H}_{1}\mathcal{H}_{3}} \otimes \eta^{[1]}_{\mathcal{K}_{2}\mathcal{H}_{2}},\\
M_{2} &= \eta^{[1]}_{\mathcal{K}_{1}\mathcal{H}_{1}} \otimes \eta^{[2]}_{\mathcal{K}_{2}\mathcal{K}_{3}\mathcal{H}_{2}\mathcal{H}_{3}}.  
\end{align}
By inserting Eq.\,(\ref{eq:def_eta}), we obtain
\begin{align} 
\eta^{[2]}_{\mathcal{K}_{1}\mathcal{K}_{3}\mathcal{H}_{1}\mathcal{H}_{3}} &= I_{0}^{\mathcal{K}_{1}\mathcal{K}_{3}} \otimes I_{0}^{\mathcal{H}_{1}\mathcal{H}_{3}} \oplus \frac{1}{3} I_{1}^{\mathcal{K}_{1}\mathcal{K}_{3}}  \otimes I_{1}^{\mathcal{H}_{1}\mathcal{H}_{3}}  \\
\eta^{[1]}_{\mathcal{K}_{2}\mathcal{H}_{2}} &= \frac{I_{\frac{1}{2}}^{\mathcal{K}_{2}}}{2} \otimes I_{\frac{1}{2}}^{\mathcal{H}_{2}}.  
\end{align}
Therefore, $M_{1}$ is decomposed as  
    \begin{align}
    M_{1} &= \left(I_{0}^{\mathcal{K}_{1}\mathcal{K}_{3}} \otimes I_{0}^{\mathcal{H}_{1}\mathcal{H}_{3}} \oplus \frac{1}{3} I_{1}^{\mathcal{K}_{1}\mathcal{K}_{3}} \otimes I_{1}^{\mathcal{H}_{1}\mathcal{H}_{3}} \right) \otimes \frac{1}{2} I_{\frac{1}{2}}^{\mathcal{K}_{2}} \otimes I_{\frac{1}{2}}^{\mathcal{H}_{2}} \notag\\
    &= \frac{I_{\frac{1}{2}}^{\mathcal{K}}}{2} \otimes I_{\frac{1}{2}}^{\mathcal{H}} \otimes \left(\ket{00}\bra{00}_{\frac{1}{2}\frac{1}{2}} + \frac{1}{3}\ket{11}\bra{11}_{\frac{1}{2}\frac{1}{2}}\right) \notag\\
    &\quad \oplus \frac{I_{\frac{1}{2}}^{\mathcal{K}}}{2}  \otimes I_{\frac{3}{2}}^{\mathcal{H}} \otimes  \frac{1}{3}\ket{10}\bra{10}_{\frac{1}{2}\frac{3}{2}} \oplus \frac{ I_{\frac{3}{2}}^{\mathcal{K}}}{4} \otimes I_{\frac{1}{2}}^{\mathcal{H}} \otimes  \frac{2}{3}\ket{01}\bra{01}_{\frac{3}{2}\frac{1}{2}} \notag\\
    &\quad \oplus \frac{I_{\frac{3}{2}}^{\mathcal{K}}}{4}  \otimes I_{\frac{3}{2}}^{\mathcal{H}} \otimes \frac{2}{3}\ket{00}\bra{00}_{\frac{3}{2}\frac{3}{2}}.  \label{eq:n1-test-last-m1}
    \end{align}

The swap operation on $\mathcal{K}_{1} \otimes \mathcal{H}_{1}$ and $\mathcal{K}_{2} \otimes \mathcal{H}_{2}$ transforms $M_{1}$ to $M_{2}$.  The transformation corresponds to $\{\ket{0}_{\frac{1}{2}}, \ket{1}_{\frac{1}{2}}\} \rightarrow \{\ket{\tilde{0}}_{\frac{1}{2}}, \ket{\tilde{1}}_{\frac{1}{2}}\}$ in the multiplicity subspaces in $M_{1}$.  \qed

%%%%%%%%%%%%%%%%%%%%%%%%%%% PROOF OF LEMMA 3 %%%%%%%%%%%%%%%%%%%%%%%%%%%%%%%%%%%%%%%
\section{Proof of Lemma~\ref{lem:n1-co-primal-sdp}} \label{sec-prf:n1-co-primal-sdp}
Suppose that a quantum $4$-tester $\{\widetilde{\Pi}_{i}\}$, operators $Y$, $Y^{\{1\}} $ and $Y^{\{0\}}$ give ASP $p_{ave}$, satisfying Eqs.\,(\ref{sdp:n1-co-primal_2}) - (\ref{sdp:n1-co-primal_3}).  
Let us define averaged operators by
\begin{widetext}
    \begin{align}
    \widetilde{\Pi}'_{i} &:= \int d\mu(U) \int d\mu(V) ((U^{\otimes 3})_{\mathcal{K}}\otimes (V^{\otimes 3})_{\mathcal{H}}) \widetilde{\Pi}_{i}  ((U^{\dagger \otimes 3})_{\mathcal{K}} \otimes (V^{\dagger \otimes 3})_{\mathcal{H}}), \\
    Y' &:= \int d\mu(U) \int d\mu(V) ((U^{\otimes 2})_{\mathcal{K}_{1}\mathcal{K}_{2}}\otimes (V^{\otimes 3})_{\mathcal{H}}) Y  ((U^{\dagger \otimes 3})_{\mathcal{K}_{1}\mathcal{K}_{2}} \otimes (V^{\dagger \otimes 3})_{\mathcal{H}}), \\
    Y'^{\{1\}} &:= \int d\mu(U) \int d\mu(V) (U_{\mathcal{K}_{1}} \otimes (V^{\otimes 2})_{\mathcal{H}_{1}\mathcal{H}_{2}}) Y^{\{1\}}  (U^{\dagger}_{\mathcal{K}_{1}} \otimes (V^{\dagger \otimes 2})_{\mathcal{H}_{1}\mathcal{H}_{2}}), \\
    Y'^{\{0\}} &:= \int d\mu(V) V Y^{\{0\}} V^{\dagger}.   
    \end{align}
\end{widetext}
The new operators $\{\widetilde{\Pi}'_{i}\}$, $Y'$, $Y'^{\{1\}}$ and $Y'^{\{0\}}$ also satisfy Eqs.\,(\ref{sdp:n1-co-primal_2}) - (\ref{sdp:n1-co-primal_3}).  Therefore $\{\widetilde{\Pi}'_{i}\}$ is also a quantum tester.  
Similarly to the proof of Lemma~\ref{lem:commutex}, the new quantum tester $\{\widetilde{\Pi}'_{i}\}$ can achieve the same ASP $p_{ave}$.  From definition, 
\begin{align}
&[\Pi'_{i}, (A^{\otimes 3})_{\mathcal{K}} \otimes (B^{\otimes 3})_{\mathcal{H}}] = 0, \\
&[Y', (A^{\otimes 2})_{\mathcal{K}_{1}\mathcal{K}_{2}} \otimes (B^{\otimes 3})_{\mathcal{H}}] = 0, \\
&[Y'^{\{1\}}, A_{\mathcal{K}_{1}} \otimes (B^{\otimes 2})_{\mathcal{H}_{1}\mathcal{H}_{2}}] = 0, \\
&[Y'^{\{0\}}, B_{\mathcal{H}_{1}}] = 0.  
\end{align} \qed

%%%%%%%%%%%%%%%%%%%%%%%%%%%%%% PROOF OF LEMMA 4 %%%%%%%%%%%%%%%%%%%%%%%
\section{Proof of Lemma~\ref{lem:n1-co-dual-sdp}} \label{sec-prf:n1-co-primalrodual}
We derive the dual SDP using Lagrange multipliers.   
Lagrangian $L$ is defined as 
    \begin{align}
    L &:= \frac{1}{2}\tr \left[ \widetilde{\Pi}_{1} M^{\langle j \rangle}_{1} + \widetilde{\Pi}_{2} M^{\langle j \rangle}_{2} \right] \notag \\
    &\quad - \tr \left[\Omega(\widetilde{\Pi}_{1} + \widetilde{\Pi}_{2} - I_{\mathcal{K}} \otimes Y )\right] \notag \\
    &\quad - \tr \left[ \Omega^{\{1\}}(\tr_{\mathcal{H}} Y - I_{\mathcal{K}_{3}} \otimes Y^{\{1\}} )\right]\notag \\
    &\quad - \tr \left[ \Omega^{\{0\}}(\tr_{\mathcal{H}} Y^{\{1\}} - I_{1} \otimes Y^{\{0\}} ) \right] \notag \\
    &\quad - \lambda (\tr Y^{\{0\}} - 1), 
    \end{align}
where $\Omega$, $\Omega^{\{1\}}$, $\Omega^{\{0\}}$, and $ \lambda $ are Lagrange multipliers.  
If the conditions in Eqs.\,(\ref{sdp:n1-co-primal_1}) - (\ref{sdp:n1-co-primal_3}) are satisfied, from the second to the fifth term in $L$ are $0$ regardless of the choice of the values of the Lagrange multipliers.  
Rewriting $L$, we have
    \begin{align}
    L &= \tr \left[\widetilde{\Pi}_{1}\left(\frac{M^{\langle j \rangle}_{1}}{2} - \Omega\right)\right] + \tr \left[\widetilde{\Pi}_{2}\left(\frac{M^{\langle j \rangle}_{2}}{2} - \Omega\right)\right] \notag \\
    &\quad + \tr \left[ Y       (\tr_{\mathcal{K}_{3}} \Omega - I_{\mathcal{H}_{3}} \otimes \Omega^{\{1\}} ) \right] \notag \\
    &\quad + \tr \left[ Y^{\{1\}} (\tr_{\mathcal{K}_{2}} \Omega^{\{1\}} - I_{\mathcal{H}_{2}} \otimes \Omega^{\{0\}} )\right] \notag \\
    &\quad + \tr \left[ Y^{\{0\}} (\tr_{\mathcal{K}_{1}} \Omega^{\{0\}} - \lambda I_{\mathcal{H}_{1}})\right] \notag \\
    &\quad + \lambda.
    \end{align}
Note that the trace of the product of two positive semidefinite operators is non-negative.  
Therefore, if the following inequalities
\begin{align}
& \frac{M^{\langle j \rangle}_{1}}{2} - \Omega \leq 0, \label{eq:n1-cond-1}\\
& \frac{M^{\langle j \rangle}_{2}}{2} - \Omega \leq 0, \label{eq:n1-cond-2} \\
& \tr_{\mathcal{K}_{3}} \Omega - I_{\mathcal{H}_{3}} \otimes \Omega^{\{1\}} \leq 0, \label{eq:n1-cond-3} \\
& \tr_{\mathcal{K}_{2}} \Omega^{\{1\}} - I_{\mathcal{H}_{2}} \otimes \Omega^{\{0\}} \leq 0, \label{eq:n1-cond-4}\\
& \tr_{\mathcal{K}_{1}} \Omega^{\{0\}} - \lambda I_{\mathcal{H}_{1}}  \leq 0, \label{eq:n1-cond-5}
\end{align}
are satisfied, we obtain. 
\begin{equation}
L \leq \lambda.  \label{eq:L-leq-a}
\end{equation} 
Therefore, minimizing $ \lambda $ under Conditions (\ref{eq:n1-cond-1}) - (\ref{eq:n1-cond-5}) is the desired dual SDP.  \qed

%%%%%%%%%%%%%%%%%%%%%%%%%%% PROOF OF LEMMA 5 %%%%%%%%%%%%%%%%%%%%%%%%%%%%%%%%%%%%%%%
\section{Proof of Lemma \ref{lem:n1-co-dual-multi}}\label{sec:n1-co-dual-multi}

First we assume that positive semidefinite operators $\Omega$, $\Omega^{\{1\}}$, $\Omega^{\{0\}}$, and $ \lambda $ fulfill Eqs.\,(\ref{eq:sdp_dual_1}) - (\ref{eq:sdp_dual_5}).  
Then new positive semidefinite operators defined by 
\begin{widetext}
    \begin{align}
    \Omega' &:= \int d\mu(U)  \int d\mu(V) ((U^{\otimes 3})_{\mathcal{K}} \otimes (V^{\otimes 3})_{\mathcal{H}}) \Omega ((U^{\dagger \otimes 3} )_{\mathcal{K}} \otimes (V^{\dagger \otimes 3})_{\mathcal{H}}), \\
    \Omega'^{\{1\}} &:= \int d\mu(U)  \int d\mu(V) ((U^{\otimes 2})_{\mathcal{K}_{1}\mathcal{K}_{2}} \otimes (V^{\otimes 2})_{\mathcal{H}_{1}\mathcal{H}_{2}}) \Omega^{\{1\}} ((U^{\dagger \otimes 2} )_{\mathcal{K}_{1}\mathcal{K}_{2}} \otimes (V^{\dagger \otimes 2})_{\mathcal{H}_{1}\mathcal{H}_{2}}), \\
    \Omega'^{\{0\}} &:= \int d\mu(U)  \int d\mu(V) (U_{\mathcal{K}_{1}} \otimes V_{\mathcal{H}_{1}}) \Omega^{\{0\}} (U^{\dagger} _{\mathcal{K}} \otimes V^{\dagger}_{\mathcal{H}_{1}}), 
    \end{align}
\end{widetext}
also satisfy Eqs.\,(\ref{eq:sdp_dual_1}) - (\ref{eq:sdp_dual_5}).  
By definition, $\Omega'$, $\Omega'^{\{1\}}$, $\Omega'^{\{0\}}$, and $\lambda$ form a feasible set of parameters and satisfy
\begin{align}
[ \Omega', A^{\otimes 3} \otimes B^{\otimes 3} ] &= 0, \\ \label{eq:n1-dual-omega-commute}
[ \Omega'^{\{1\}}, A^{\otimes 2} \otimes B^{\otimes 2} ] &= 0, \\
[ \Omega'^{\{0\}}, A \otimes B ] &= 0, 
\end{align}
for arbitrary unitary operators $A$ and $B$ in $\text{SU}(2)$.

We can assume that $\Omega'$, $\Omega'^{\{1\}}$,  and $\Omega'^{\{0\}}$ are represented as 
\begin{align}
\Omega &= \bigoplus_{J = \frac{1}{2}}^{\frac{3}{2}}\bigoplus_{L = \frac{1}{2}}^{\frac{3}{2}} \frac{I_{J}^{\mathcal{K}}}{d_{J}} \otimes I_{L}^{\mathcal{H}} \otimes \Omega_{JL},\\
\Omega^{\{1\}} &= \bigoplus_{J = 0}^{1}\bigoplus_{L = 0}^{1} \frac{I_{J}^{\mathcal{K}_{1}\mathcal{K}_{2}}}{d_{J}} \otimes I_{L}^{\mathcal{H}_{1}\mathcal{H}_{2}}\otimes \Omega^{\{1\}}_{JL},
\end{align}
where $I_{J}$ is the identity operator on the irreducible subspace $\mathcal{U}_{J}$, $\Omega_{JL}$ an operator on $\mathcal{V}^{[3]}_{J} \otimes \mathcal{V}^{[3]}_{L}$ for $J,L = 1/2,3/2$, and $\Omega^{\{1\}}_{JL}$ for $J,L = 0,1$ and $\Omega^{\{0\}}_{\frac{1}{2}\frac{1}{2}}$ are some positive numbers.  Note that $\Omega_{\frac{3}{2}\frac{3}{2}}$ is a scalar.

We rewrite Eqs.\,(\ref{eq:sdp_dual_1}) and (\ref{eq:sdp_dual_2}) in terms of the operators on the multiplicity subspaces.  
The operators $M^{\langle j \rangle}_{i}$ are represented as 
\begin{equation}
M^{\langle j \rangle}_{i} = \bigoplus_{J = \frac{1}{2}}^{\frac{3}{2}}\bigoplus_{L = \frac{1}{2}}^{\frac{3}{2}} \frac{I_{J}^{\mathcal{K}}}{d_{J}} \otimes I_{L}^{\mathcal{H}} \otimes M^{\langle j \rangle(i)}_{JL},
\end{equation}
for $i = 1,2$.  
Thus, Eqs.\,(\ref{eq:sdp_dual_1}) and (\ref{eq:sdp_dual_2}) are rewritten as 
\begin{equation}
\Omega_{JL} - \frac{M^{\langle j \rangle(i)}_{JL}}{2} \geq 0,
\end{equation}
for $J, L = 1/2, 3/2$ and $ i = 1,2$.

Next we rewrite Eq.\,(\ref{eq:sdp_dual_3}).  
Using Eq.\,(\ref{eq:partial trace}) for $\Omega'$, we have
\begin{widetext}
    \begin{align}
    \tr_{\mathcal{K}_{3}}\Omega' &= \bigoplus_{L = \frac{1}{2}}^{\frac{3}{2}} \Bigl[\frac{I_{0}^{\mathcal{K}_{1}\mathcal{K}_{2}}}{d_{0}} \otimes I_{L}^{\mathcal{H}} \otimes (\bra{\hat{0}}_{\frac{1}{2}} \otimes I_{\mathcal{V}^{[3]}_{L}})\Omega_{\frac{1}{2}L}(\ket{\hat{0}}_{\frac{1}{2}} \otimes I_{\mathcal{V}^{[3]}_{L}}) \oplus \frac{I_{1}^{\mathcal{K}_{1}\mathcal{K}_{2}}}{d_{1}} \otimes I_{L}^{\mathcal{H}} \otimes ((\bra{\hat{1}}_{\frac{1}{2}} \otimes I_{\mathcal{V}^{[3]}_{L}})\Omega_{\frac{1}{2}L}(\ket{\hat{1}}_{\frac{1}{2}} \otimes I_{\mathcal{V}^{[3]}_{L}}) + \Omega_{\frac{3}{2}L}) \Bigr].  
    \end{align}
\end{widetext}
Using Eq.\,(\ref{eq:coupling}), we have
\begin{align}
&I_{\mathcal{H}_{3}} \otimes \Omega'^{\{1\}} = \bigoplus_{J = 0}^{1} \frac{I_{J}^{\mathcal{K}_{1}\mathcal{K}_{2}}}{d_{J}} \otimes \big[(I_{\mathcal{H}_{3}} \otimes I_{0}^{\mathcal{H}_{1}\mathcal{H}_{2}}) \otimes \Omega^{\{1\}}_{J0} \notag \\ 
&\quad \oplus (I_{\mathcal{H}_{3}} \otimes I_{1}^{\mathcal{H}_{1}\mathcal{H}_{2}}) \otimes \Omega^{\{1\}}_{J1}\big]\\
&\quad = \bigoplus_{J = 0}^{1} \frac{I_{J}^{\mathcal{K}_{1}\mathcal{K}_{2}}}{d_{J}} \otimes \big[I_{\frac{1}{2}}^{\mathcal{H}} \otimes (\Omega^{\{1\}}_{J0}\ket{\hat{0}}\bra{\hat{0}}_{\frac{1}{2}} +\Omega^{\{1\}}_{J1} \ket{\hat{1}}\bra{\hat{1}}_{\frac{1}{2}}) \notag \\ 
&\quad \oplus I_{\frac{3}{2}}^{\mathcal{H}} \otimes \Omega^{\{1\}}_{J1}\ket{0}\bra{0}_{\frac{3}{2}} \big].  
\end{align}
Equation~(\ref{eq:sdp_dual_3}) is rewritten as 
\begin{align}
&\Omega^{\{1\}}_{00}\ket{\hat{0}}\bra{\hat{0}}_{\frac{1}{2}} + \Omega^{\{1\}}_{01}\ket{\hat{1}}\bra{\hat{1}}_{\frac{1}{2}} - \Omega_{\frac{1}{2}\frac{1}{2}}^{0\rightarrow 1/2} \geq 0, \\
&\Omega^{\{1\}}_{01} - \Omega_{\frac{1}{2}\frac{3}{2}}^{0\rightarrow 1/2} \geq 0, \\
&\Omega^{\{1\}}_{10}\ket{\hat{0}}\bra{\hat{0}}_{\frac{1}{2}} + \Omega^{\{1\}}_{11}\ket{\hat{1}}\bra{\hat{1}}_{\frac{1}{2}} - \Omega_{\frac{1}{2}\frac{1}{2}}^{1\rightarrow 1/2}- \Omega_{\frac{3}{2}\frac{1}{2}} \geq 0, \\
&\Omega^{\{1\}}_{11} - \Omega_{\frac{1}{2}\frac{3}{2}}^{1\rightarrow 1/2} - \Omega_{\frac{3}{2}\frac{3}{2}} \geq 0, 
\end{align}
where we define $\Omega_{\frac{1}{2}L}^{j\rightarrow 1/2} := (\bra{\hat{j}}_{\frac{1}{2}} \otimes I_{\mathcal{V}^{[3]}_{L}})\Omega_{\frac{1}{2}L}(\ket{\hat{j}}_{\frac{1}{2}} \otimes I_{\mathcal{V}^{[3]}_{L}})$.  
In addition, we obtain
\begin{multline}
\tr_{\mathcal{K}_{2}} \Omega'^{\{1\}} = \frac{I_{\frac{1}{2}}^{\mathcal{K}_{1}}}{d_{\frac{1}{2}}} \otimes \big[ (\Omega_{00} + \Omega_{10}) I_{0}^{\mathcal{H}_{1}\mathcal{H}_{2}} \\
\oplus (\Omega_{01} + \Omega_{11}) I_{1}^{\mathcal{H}_{1}\mathcal{H}_{2}}\big],
\end{multline}
\begin{align}
I_{\mathcal{H}_{2}} \otimes \Omega'^{\{0\}} &= \frac{I_{\frac{1}{2}}^{\mathcal{K}_{1}}}{d_{\frac{1}{2}}} \otimes ( \Omega^{\{0\}}_{\frac{1}{2}\frac{1}{2}} I_{0}^{\mathcal{H}_{1}\mathcal{H}_{2}} + \Omega^{\{0\}}_{\frac{1}{2}\frac{1}{2}} I_{1}^{\mathcal{H}_{1}\mathcal{H}_{2}} )
\end{align}
and 
\begin{equation}
\tr_{\mathcal{K}_{1}} \Omega'^{\{0\}} = \Omega^{\{0\}}_{\frac{1}{2}\frac{1}{2}} I_{\frac{1}{2}}^{\mathcal{H}_{1}}. 
\end{equation}
Equations~(\ref{eq:sdp_dual_4}) and (\ref{eq:sdp_dual_5}) become
\begin{align}
& \Omega^{\{0\}}_{\frac{1}{2}\frac{1}{2}} - \Omega^{\{1\}}_{00} - \Omega^{\{1\}}_{10} \geq 0, \\
& \Omega^{\{0\}}_{\frac{1}{2}\frac{1}{2}} - \Omega^{\{1\}}_{01} - \Omega^{\{1\}}_{11} \geq 0, \\
& \lambda - \Omega^{\{0\}}_{\frac{1}{2}\frac{1}{2}} \geq 0. 
\end{align}
We can assume $\Omega^{\{0\}}_{\frac{1}{2}\frac{1}{2}} = \lambda $ without loss of generality.  
All in all, the dual SDP expressed in the multiplicity subspaces is 
\begin{align} 
\text{minimize}\quad & \lambda, \\
\text{subject to }\quad & \Omega_{JL} - \frac{M^{(i)}_{JL}}{2} \geq 0, \\
&\text{ for } J, L = 1/2, 3/2 \text{ and } i = 1,2,   \\
&\Omega^{\{1\}}_{00}\ket{\hat{0}}\bra{\hat{0}}_{\frac{1}{2}} + \Omega^{\{1\}}_{01}\ket{\hat{1}}\bra{\hat{1}}_{\frac{1}{2}} \notag \\
& \qquad \qquad \qquad - \Omega_{\frac{1}{2}\frac{1}{2}}^{0\rightarrow 1/2} \geq 0, \\
&\Omega^{\{1\}}_{01} - \Omega_{\frac{1}{2}\frac{3}{2}}^{0\rightarrow 1/2} \geq 0, \\
&\Omega^{\{1\}}_{10}\ket{\hat{0}}\bra{\hat{0}}_{\frac{1}{2}} + \Omega^{\{1\}}_{11}\ket{\hat{1}}\bra{\hat{1}}_{\frac{1}{2}} \notag \\
&\qquad \qquad - \Omega_{\frac{1}{2}\frac{1}{2}}^{1\rightarrow 1/2}- \Omega_{\frac{3}{2}\frac{1}{2}} \geq 0, \\
&\Omega^{\{1\}}_{11} - \Omega_{\frac{1}{2}\frac{3}{2}}^{1\rightarrow 1/2} - \Omega_{\frac{3}{2}\frac{3}{2}} \geq 0,\\
& \lambda - \Omega^{\{1\}}_{00} - \Omega^{\{1\}}_{10} \geq 0, \\
& \lambda - \Omega^{\{1\}}_{01} - \Omega^{\{1\}}_{11} \geq 0. 
\end{align}\qed

%%%%%%%%%%%%%% FEASIBLE SET OF THE DUAL SDP IN THEOREM 2 %%%%%%%%%%%%%%
\section{Feasible sets of of the dual SDP in Theorem \ref{thm:n1-co-opt}}\label{sec:prf of n1 order}
The Choi operators $M^{\langle 3 \rangle}_{i} $ are the same as the Choi operators for the parallelized scheme in Eq.\,(\ref{ea:form of M}).  
In the basis $\{\ket{\hat{0}}_{\frac{1}{2}}, \ket{\hat{1}}_{\frac{1}{2}}\}$,
\begin{align}
M_{\frac{1}{2}\frac{1}{2}}^{\langle 3 \rangle (1)} &= \frac{1}{4}\left(
\begin{array}{cccc}
 1 & 0 & 0 & 1 \\
 0 & 1 & 1 & \frac{2}{\sqrt{3}} \\
 0 & 1 & 1 & \frac{2}{\sqrt{3}} \\
 1 & \frac{2}{\sqrt{3}} & \frac{2}{\sqrt{3}} & \frac{7}{3} \\
\end{array}
\right) , \\ 
M_{\frac{1}{2}\frac{3}{2}}^{\langle 3 \rangle (1)} &= \frac{1}{4}\left(
\begin{array}{cc}
 1& -\frac{1}{\sqrt{3}} \\
 -\frac{1}{\sqrt{3}} & \frac{1}{3} \\
\end{array}
\right) ,\\ 
M_{\frac{3}{2}\frac{1}{2}}^{\langle 3 \rangle (1)} &= \frac{1}{2}\left(
\begin{array}{cc}
 1 & -\frac{1}{\sqrt{3}} \\
 -\frac{1}{\sqrt{3}} & \frac{1}{3} \\
\end{array}
\right) ,\\
M_{\frac{3}{2}\frac{3}{2}}^{\langle 3 \rangle (1)} &= \frac{2}{3},  
\end{align}
from Eq.\,(\ref{eq:n1-test-last-m1}).   Note that $\dim \mathcal{V}^{[3]}_{\frac{1}{2}} = 2$ and $\dim \mathcal{V}^{[3]}_{\frac{3}{2}} = 1$.  The swap operation on $\mathcal{K}_{1} \otimes \mathcal{H}_{1}$ and $\mathcal{K}_{2} \otimes \mathcal{H}_{2}$ transforms $M^{\langle 3 \rangle}_{1}$ to $M^{\langle 3 \rangle}_{2}$.  The transformation corresponds to $\{\ket{0}_{\frac{1}{2}}, \ket{1}_{\frac{1}{2}}\} \rightarrow \{\ket{\tilde{0}}_{\frac{1}{2}}, \ket{\tilde{1}}_{\frac{1}{2}}\}$ in the multiplicity subspaces in $M^{\langle 3 \rangle}_{1}$. 
In the basis $\{\ket{\hat{0}}_{\frac{1}{2}}, \ket{\hat{1}}_{\frac{1}{2}}\}$,
\begin{align}
M_{\frac{1}{2}\frac{1}{2}}^{\langle 3 \rangle (2)} &= \frac{1}{4}\left(
\begin{array}{cccc}
 1 & 0 & 0 & 1 \\
 0 & 1 & 1 & -\frac{2}{\sqrt{3}} \\
 0 & 1 & 1 & -\frac{2}{\sqrt{3}} \\
 1 & -\frac{2}{\sqrt{3}} & -\frac{2}{\sqrt{3}} & \frac{7}{3} \\
\end{array}
\right) ,~ \\ 
M_{\frac{1}{2}\frac{3}{2}}^{\langle 3 \rangle (2)} &= \frac{1}{4}\left(
\begin{array}{cc}
 1& \frac{1}{\sqrt{3}} \\
 \frac{1}{\sqrt{3}} & \frac{1}{3} \\
\end{array}
\right) ,~ \\
M_{\frac{3}{2}\frac{1}{2}}^{\langle 3 \rangle (2)} &= \frac{1}{2}\left(
\begin{array}{cc}
 1 & \frac{1}{\sqrt{3}} \\
 \frac{1}{\sqrt{3}} & \frac{1}{3} \\
\end{array}
\right), \\
M_{\frac{3}{2}\frac{3}{2}}^{\langle 3 \rangle (2)} &= \frac{2}{3}.  
\end{align}
A feasible set of parameters of the dual SDP for $M_{i}^{\langle 3 \rangle}$ is 
\begin{align}
\lambda &= \frac{7}{8}, \\
\Omega_{\frac{1}{2}\frac{1}{2}} &= \frac{1}{4} \left(
\begin{array}{cccc}
 \frac{1}{2} & 0 & 0 & \frac{1}{2} \\
 0 & 1 & 1 & 0 \\
 0 & 1 & 1 & 0 \\
 \frac{1}{2} & 0 & 0 & \frac{11}{24}, \\
\end{array}
\right), \\
\Omega_{\frac{1}{2}\frac{3}{2}} &= \left(
\begin{array}{cc}
 \frac{1}{4} & 0 \\
 0& \frac{1}{6} \\
\end{array}
\right),~ \\ 
\Omega_{\frac{3}{2}\frac{1}{2}} &= \left(
\begin{array}{cc}
 \frac{1}{2} & 0 \\
 0 & \frac{1}{6} \\
\end{array}
\right) ,~ 
\Omega_{\frac{3}{2}\frac{3}{2}} = \frac{11}{3}, \\
\Omega^{\{1\}}_{00} &= \frac{1}{8},~ \Omega^{\{1\}}_{01} = \frac{1}{4},~ \Omega^{\{1\}}_{10} = \frac{3}{4},~ \Omega^{\{1\}}_{11} = \frac{5}{8}.  
\end{align}

The Choi operator $M_{1}^{\langle 2 \rangle}$ is 
\begin{align}
M_{\frac{1}{2}\frac{1}{2}}^{\langle 2 \rangle (1)} &= \left(
\begin{array}{cccc}
 1 & 0 & 0 & 0 \\
 0 & 0 & 0 & 0 \\
 0 & 0 & 0 & 0 \\
 0 & 0 & 0 & \frac{1}{3} \\
\end{array}
\right),\\
M_{\frac{1}{2}\frac{3}{2}}^{\langle 2 \rangle (1)} &= \left(
\begin{array}{cc}
 0 & 0 \\
 0 & \frac{1}{3} \\
\end{array}
\right), \\
M_{\frac{3}{2}\frac{1}{2}}^{\langle 2 \rangle (1)} &=  \left(
\begin{array}{cc}
 0 & 0 \\
 0 & \frac{2}{3} \\
\end{array}
\right),\\ 
M_{\frac{3}{2}\frac{3}{2}}^{\langle 2 \rangle (1)} &= \frac{2}{3},   
\end{align}
and $M_{2}^{\langle 2 \rangle} = M_{2}^{\langle 3 \rangle}$.   
A feasible set of parameters of the dual SDP for $M_{i}^{\langle 2 \rangle}$ is 
\begin{align}
\lambda &= \frac{7}{8}, \\
\Omega_{\frac{1}{2}\frac{1}{2}} &= \frac{1}{4} \left(
\begin{array}{cccc}
 2 & 0 & 0 & 0 \\
 0 & \frac{1}{2} & \frac{1}{2}  & -\frac{1}{\sqrt{3}} \\
 0 & \frac{1}{2} & \frac{1}{2}  & -\frac{1}{\sqrt{3}} \\
 0 & -\frac{1}{\sqrt{3}} & -\frac{1}{\sqrt{3}} & \frac{4}{3} \\
\end{array}
\right) ,\\
\Omega_{\frac{1}{2}\frac{3}{2}} &= \frac{1}{8}\left(
\begin{array}{cc}
 1 & \frac{1}{\sqrt{3}} \\
 \frac{1}{\sqrt{3}} & 2 \\
\end{array}
\right), \\
\Omega_{\frac{3}{2}\frac{1}{2}} &= \frac{1}{4}\left(
\begin{array}{cc}
 1 & \frac{1}{\sqrt{3}} \\
 \frac{1}{\sqrt{3}} & \frac{5}{3} \\
\end{array}
\right) ,~
\Omega_{\frac{3}{2}\frac{3}{2}} = \frac{1}{2}, \\
\Omega^{\{1\}}_{00} &= \frac{1}{2},~  \Omega^{\{1\}}_{01} = \frac{1}{8},~ \Omega^{\{1\}}_{10} = \frac{3}{8},\  \Omega^{\{1\}}_{11} = \frac{3}{4}.
\end{align}
\qed

%%%%%%%%%%%%%%%%%%%%%%%%%%% REFERENCE %%%%%%%%%%%%%%%%%%%%%%%%%%%%%%%%%%%%%%%
%\bibliographystyle{apsrev4-1}
%\bibliography{Shimbo_Phd_thesis}
%

\end{document}

\title{Supplemental Material for ``Equivalence determination of unitary operations''}%

\author{Atsushi Shimbo}%
\email[Electronic address: ]{shimbo@eve.phys.s.u-tokyo.ac.jp}
\affiliation{Department of Physics, Graduate School of Science, The University of Tokyo}

\author{Akihito Soeda}%
\email[Electronic address: ]{soeda@phys.s.u-tokyo.ac.jp}
\affiliation{Department of Physics, Graduate School of Science, The University of Tokyo}

\author{Mio Murao}%
\email[Electronic address: ]{murao@phys.s.u-tokyo.ac.jp}
\affiliation{Department of Physics, Graduate School of Science, The University of Tokyo}
%\affiliation{Institute for Nano Quantum Information Electronics, University of Tokyo}

\date{\today}%

\maketitle

%%%%%%%%%%%%%%% CHOI OPERATO IN THE MULTIPLICITY SUBSPACES %%%%%%%%%%%%%%%%%%%
\section{Choi Operators in the multiplicity subspaces}\label{sec:ordered N12 choi multi}
First we consider the Choi operators $M_{i}$ corresponding to the ordering of the black-boxes where the first two are  reference box 1, followed by the test box, and end with reference box 2.  The input and output systems of the $i$-th black-box are denoted as $\mathcal{H}_{i}$ and $\mathcal{K}_{i}$, respectively.  The Choi operators for the remaining $11$  orderings of the black-boxes are calculated by introducing unitary operators on the multiplicity subspaces that represent the action of the swap operations on $\mathcal{H}_{i} \otimes \mathcal{H}_{j}$ and $\mathcal{K}_{i} \otimes \mathcal{K}_{j}$.  We denote $\mathcal{K} := \bigotimes _{i = 1}^{4} \mathcal{K}_{i}$ and $\mathcal{H} := \bigotimes _{i = 1}^{4} \mathcal{H}_{i}$. 

For $M_{i}$, we have
\begin{equation}
M_{i} := \int d\mu(U_{1}) \int d\mu(U_{2}) \kket{U_{1}^{\otimes 2} \otimes U_{i} \otimes U_{2} } \bbra{U_{1}^{\otimes 2} \otimes U_{i} \otimes U_{2}},  \label{eq:n12 parallel def mi}
\end{equation}
which have the form of
\begin{align}
M_{1} &= \eta^{[3]}_{\mathcal{K}_{1}\mathcal{K}_{2}\mathcal{K}_{3}\mathcal{H}_{1}\mathcal{H}_{2}\mathcal{H}_{3}} \otimes \eta^{[1]}_{\mathcal{K}_{4}\mathcal{H}_{4}}, \\ 
M_{2} &= \eta^{[2]}_{\mathcal{K}_{1}\mathcal{K}_{2}\mathcal{H}_{1}\mathcal{H}_{2}} \otimes \eta^{[2]}_{\mathcal{K}_{3}\mathcal{K}_{4}\mathcal{H}_{3}\mathcal{H}_{4}},
\end{align}
where
\begin{align} 
\eta^{[3]}_{\mathcal{K}_{1}\mathcal{K}_{2}\mathcal{K}_{3}\mathcal{H}_{1}\mathcal{H}_{2}\mathcal{H}_{3}} 
&= \frac{I_{\frac{1}{2}}^{\mathcal{K}_{1}\mathcal{K}_{2}\mathcal{K}_{3}}}{2} \otimes I_{\frac{1}{2}}^{\mathcal{H}_{1}\mathcal{H}_{2}\mathcal{H}_{3}} \otimes \left\langle \ket{\hat{0}\hat{0}}_{\frac{1}{2}\frac{1}{2}} + \ket{\hat{1}\hat{1}}_{\frac{1}{2}\frac{1}{2}} \right\rangle \oplus \frac{I_{\frac{3}{2}}^{\mathcal{K}_{1}\mathcal{K}_{2}\mathcal{K}_{3}}}{4} \otimes I_{\frac{3}{2}}^{\mathcal{H}_{1}\mathcal{H}_{2}\mathcal{H}_{3}} \otimes \left\langle \ket{\hat{0}\hat{0}}_{\frac{3}{2}\frac{3}{2}} \right\rangle,  \\
\eta^{[1]}_{\mathcal{K}_{4}\mathcal{H}_{4}} &= \textstyle\frac{I_{\frac{1}{2}}^{\mathcal{K}_{4}}}{2} \otimes I_{\frac{1}{2}}^{\mathcal{H}_{4}}, 
\end{align}
and $\left\langle \ket{\psi}\right\rangle := \ket{\psi}\bra{\psi}$.   We define bases of the multiplicity subspaces by 
\begin{align}
\ket{0}^{\mathcal{U}}_{0}\ket{0}^{\mathcal{V}^{[4]}}_{0} &:= \textstyle \ket{(0\frac{1}{2})\frac{1}{2}\frac{1}{2};00}  = \frac{1}{\sqrt{2}} (\ket{v_{1}}\ket{1} - \ket{v_{2}}\ket{0}), \\
\ket{0}^{\mathcal{U}}_{0}\ket{1}^{\mathcal{V}^{[4]}}_{0} &:= \textstyle \ket{(1\frac{1}{2})\frac{1}{2}\frac{1}{2};00} = \frac{1}{\sqrt{2}} (\ket{v_{3}}\ket{1} - \ket{v_{4}}\ket{0}), \\
\ket{0}^{\mathcal{U}}_{1}\ket{0}^{\mathcal{V}^{[4]}}_{1} &:= \textstyle \ket{(0\frac{1}{2})\frac{1}{2}\frac{1}{2};1 ~{-1}} = \ket{v_{1}} \ket{0}, \\
\ket{1}^{\mathcal{U}}_{1}\ket{0}^{\mathcal{V}^{[4]}}_{1} &:= \textstyle \ket{(0\frac{1}{2})\frac{1}{2}\frac{1}{2};1 0} = \frac{1}{\sqrt{2}}(\ket{v_{1}}\ket{1} + \ket{v_{2}}\ket{0}), \\
\ket{2}^{\mathcal{U}}_{1}\ket{0}^{\mathcal{V}^{[4]}}_{1} &:= \textstyle \ket{(0\frac{1}{2})\frac{1}{2}\frac{1}{2};1 1} = \ket{v_{2}}\ket{1}, \\
\ket{0}^{\mathcal{U}}_{1}\ket{1}^{\mathcal{V}^{[4]}}_{1} &:= \textstyle \ket{(1\frac{1}{2})\frac{1}{2}\frac{1}{2};1 ~{-1}} = \ket{v_{3}} \ket{0}, \\
\ket{1}^{\mathcal{U}}_{1}\ket{1}^{\mathcal{V}^{[4]}}_{1} &:= \textstyle \ket{(1\frac{1}{2})\frac{1}{2}\frac{1}{2};1 0} = \frac{1}{\sqrt{2}}(\ket{v_{3}}\ket{1} + \ket{v_{4}}\ket{0}), \\
\ket{2}^{\mathcal{U}}_{1}\ket{1}^{\mathcal{V}^{[4]}}_{1} &:= \textstyle \ket{(1\frac{1}{2})\frac{1}{2}\frac{1}{2};1 1} = \ket{v_{4}}\ket{1}, \\
\ket{0}^{\mathcal{U}}_{1}\ket{2}^{\mathcal{V}^{[4]}}_{1} &:= \textstyle \ket{(1\frac{1}{2})\frac{1}{2}\frac{1}{2};1 ~{-1}} = \frac{\sqrt{3}}{2}\ket{v_{5}}\ket{1} - \frac{1}{2}\ket{v_{4}\ket{0}}, \\
\ket{1}^{\mathcal{U}}_{1}\ket{2}^{\mathcal{V}^{[4]}}_{1} &:= \textstyle \ket{(1\frac{1}{2})\frac{3}{2}\frac{1}{2};1 0} = \frac{1}{\sqrt{2}} (\ket{v_{6}}\ket{1} - \frac{1}{\sqrt{2}}\ket{v_{7}}\ket{0}, \\
\ket{2}^{\mathcal{U}}_{1}\ket{2}^{\mathcal{V}^{[4]}}_{1} &:= \textstyle \ket{(1\frac{1}{2})\frac{3}{2}\frac{1}{2};1 1} = \frac{1}{2} \ket{v_{7}}\ket{1} - \frac{\sqrt{3}}{2}\ket{v_{8}}\ket{0}, \\
\ket{0}^{\mathcal{U}}_{2}\ket{0}^{\mathcal{V}^{[4]}}_{2} &:= \textstyle \ket{(1\frac{1}{2})\frac{3}{2}\frac{1}{2};2 ~{-2}} = \ket{v_{5}}\ket{0}, \\
\ket{1}^{\mathcal{U}}_{2}\ket{0}^{\mathcal{V}^{[4]}}_{2} &:= \textstyle \ket{(1\frac{1}{2})\frac{3}{2}\frac{1}{2};2 ~{-1}} = \frac{1}{2}\ket{v_{5}}\ket{1} + \frac{\sqrt{3}}{2} \ket{v_{6}\ket{0}}, \\
\ket{2}^{\mathcal{U}}_{2}\ket{0}^{\mathcal{V}^{[4]}}_{2} &:= \textstyle \ket{(1\frac{1}{2})\frac{3}{2}\frac{1}{2};2 0} = \frac{1}{\sqrt{2}}(\ket{v_{6}}\ket{1} + \ket{v_{7}}\ket{0}), \\
\ket{3}^{\mathcal{U}}_{2}\ket{0}^{\mathcal{V}^{[4]}}_{2} &:= \textstyle \ket{(1\frac{1}{2})\frac{3}{2}\frac{1}{2};2 1} = \frac{\sqrt{3}}{2}\ket{v_{7}}\ket{1} + \frac{1}{2}\ket{v_{8}\ket{0}}, \\
\ket{4}^{\mathcal{U}}_{2}\ket{0}^{\mathcal{V}^{[4]}}_{2} &:= \textstyle \ket{(1\frac{1}{2})\frac{3}{2}\frac{1}{2};2 2} = \ket{v_{8}}\ket{1},    
\end{align} 
where $\ket{(j_{12}j_{3})j_{123} j_{4}; jm}$ represent a state with the spin angular momentum of $m$ along the $z$-axis, obtained by first coupling the spin in $\mathcal{K}_{1}$ and $\mathcal{K}_{2}$ to form a spin-$j_{12}$, then coupled with the spin-$j_{3}$ in $\mathcal{K}_{3}$ to form a spin-$j_{123}$, and finally coupled with the spin-$j_{4}$ in $\mathcal{K}_{4}$ to form a spin-$j$.  
In these bases, 
\begin{align}
M_{1} &= 
\textstyle\frac{1}{4}I_{0}^{\mathcal{K}} \otimes I_{0}^{\mathcal{H}} \otimes \left\langle \ket{00}_{00} + \ket{11}_{00} \right\rangle \oplus 
\textstyle\frac{1}{4}I_{0}^{\mathcal{K}} \otimes I_{1}^{\mathcal{H}} \otimes \left\langle \ket{00}_{01} + \ket{11}_{01} \right\rangle \notag \\ & \quad \oplus
\textstyle\frac{1}{4}I_{1}^{\mathcal{K}} \otimes I_{0}^{\mathcal{H}} \otimes \left\langle \ket{00}_{10} + \ket{11}_{10} \right\rangle \oplus 
\textstyle\frac{1}{4}I_{1}^{\mathcal{K}} \otimes I_{1}^{\mathcal{H}} \otimes \left\langle \ket{00}_{11} + \ket{11}_{11} \right\rangle \notag \\ & \quad \oplus
\textstyle\frac{1}{8}I_{1}^{\mathcal{K}} \otimes I_{1}^{\mathcal{H}} \otimes \left\langle \ket{22}_{11} \right\rangle \oplus 
\textstyle\frac{1}{8}I_{1}^{\mathcal{K}} \otimes I_{2}^{\mathcal{H}} \otimes \left\langle \ket{20}_{12} \right\rangle \notag \\ &\quad \oplus 
\textstyle\frac{1}{8}I_{2}^{\mathcal{K}} \otimes I_{1}^{\mathcal{H}} \otimes \left\langle \ket{02}_{21} \right\rangle \oplus 
\textstyle\frac{1}{8}I_{2}^{\mathcal{K}} \otimes I_{2}^{\mathcal{H}} \otimes \left\langle \ket{00}_{22} \right\rangle \\
&= 
\textstyle I_{0}^{\mathcal{K}} \otimes I_{0}^{\mathcal{H}} \otimes \frac{1}{4} \left\langle \ket{00}_{00} + \ket{11}_{00} \right\rangle \oplus 
\textstyle I_{0}^{\mathcal{K}} \otimes I_{1}^{\mathcal{H}} \otimes \frac{1}{4} \left\langle \ket{00}_{01} + \ket{11}_{01} \right\rangle \notag \\ & \quad \oplus 
\textstyle \frac{I_{1}^{\mathcal{K}}}{3} \otimes I_{0}^{\mathcal{H}} \otimes \frac{3}{4} \left\langle \ket{00}_{10} + \ket{11}_{10} \right\rangle \oplus 
\textstyle \frac{I_{1}^{\mathcal{K}}}{3} \otimes I_{1}^{\mathcal{H}} \otimes \left( \frac{3}{4}\left\langle \ket{00}_{11} + \ket{11}_{11}\right\rangle + \frac{3}{8} \left\langle \ket{22}_{11} \right\rangle \right) \oplus
\textstyle \frac{I_{1}^{\mathcal{K}}}{3} \otimes I_{2}^{\mathcal{H}} \otimes \frac{3}{8}\left\langle \ket{20}_{12} \right\rangle  \notag \\ & \quad \oplus 
\textstyle \frac{I_{2}^{\mathcal{K}}}{5} \otimes I_{1}^{\mathcal{H}} \otimes \frac{5}{8}\left\langle \ket{02}_{21} \right\rangle \oplus 
\textstyle \frac{I_{2}^{\mathcal{K}}}{5} \otimes I_{2}^{\mathcal{H}} \otimes \frac{5}{8}\left\langle \ket{00}_{22} \right\rangle \\
&= \bigoplus_{J = 0}^{2}\bigoplus_{L = 0}^{2} \frac{I_{J}^{\mathcal{K}}}{d_{J}} \otimes I_{L}^{\mathcal{H}} \otimes M^{(1)}_{JL}.  
\end{align}

Therefore,
\begin{align}
M^{(1)}_{00} &= \frac{1}{4} \left\langle \ket{00}_{00} + \ket{11}_{00} \right\rangle ,\\
M^{(1)}_{01} &= \frac{1}{4} \left\langle \ket{00}_{01} + \ket{11}_{01} \right\rangle ,\\
M^{(1)}_{02} &= 0 ,\\
M^{(1)}_{10} &= \frac{3}{4} \left\langle \ket{00}_{10} + \ket{11}_{10} \right\rangle ,\\
M^{(1)}_{11} &= \frac{3}{4} \left\langle \ket{00}_{11} + \ket{11}_{11} \right\rangle  + \frac{3}{8} \left\langle \ket{22}_{11} \right\rangle ,\\
M^{(1)}_{12} &= \frac{3}{8} \left\langle \ket{20}_{12} \right\rangle ,\\
M^{(1)}_{20} &= 0 ,\\
M^{(1)}_{21} &= \frac{5}{8} \left\langle \ket{02}_{21} \right\rangle ,\\
M^{(1)}_{22} &= \frac{5}{8} \left\langle \ket{00}_{22} \right\rangle . 
\end{align}

For $M_{2}$, we have
    \begin{align}
    \eta^{[2]}_{\mathcal{K}_{1}\mathcal{K}_{2}\mathcal{H}_{1}\mathcal{H}_{2}} &= I_{0}^{\mathcal{K}_{1}\mathcal{K}_{2}} \otimes I_{0}^{\mathcal{H}_{1}\mathcal{H}_{2}} \oplus \frac{I_{1}^{\mathcal{K}_{1}\mathcal{K}_{2}}}{3} \otimes I_{1}^{\mathcal{H}_{1}\mathcal{H}_{2}} ,\\
    \eta^{[2]}_{\mathcal{K}_{3}\mathcal{K}_{4}\mathcal{H}_{3}\mathcal{H}_{4}} &=  I_{0}^{\mathcal{K}_{3}\mathcal{K}_{4}} \otimes I_{0}^{\mathcal{H}_{3}\mathcal{H}_{4}} \oplus \frac{I_{1}^{\mathcal{K}_{3}\mathcal{K}_{4}}}{3} \otimes I_{1}^{\mathcal{H}_{3}\mathcal{H}_{4}}.   
    \end{align}
We also define other bases of the multiplicity subspaces as 
\begin{align}
\ket{0}^{\mathcal{U}}_{0}\ket{\hat{0}}^{\mathcal{V}^{[4]}}_{0} &:= \textstyle \ket{0(\frac{1}{2}\frac{1}{2})0;00} = \ket{w_{1}}\ket{w_{1}},\\
\ket{0}^{\mathcal{U}}_{0}\ket{\hat{1}}^{\mathcal{V}^{[4]}}_{0} &:= \textstyle \ket{1(\frac{1}{2}\frac{1}{2})1;00} = \frac{1}{\sqrt{3}}(\ket{w_{2}}\ket{v_{4}} - \ket{w_{3}}\ket{w_{3}} + \ket{w_{4}}\ket{w_{2}} ,\\
\ket{0}^{\mathcal{U}}_{1}\ket{\hat{0}}^{\mathcal{V}^{[4]}}_{1} &:= \textstyle \ket{0(\frac{1}{2}\frac{1}{2})1;1~{-1}} = \ket{w_{1}}\ket{w_{2}}, \\
\ket{1}^{\mathcal{U}}_{1}\ket{\hat{0}}^{\mathcal{V}^{[4]}}_{1} &:= \textstyle \ket{0(\frac{1}{2}\frac{1}{2})1;10} =  \ket{w_{1}}\ket{w_{3}}, \\
\ket{2}^{\mathcal{U}}_{1}\ket{\hat{0}}^{\mathcal{V}^{[4]}}_{1} &:= \textstyle \ket{0(\frac{1}{2}\frac{1}{2})1;11} =  \ket{w_{1}}\ket{w_{4}}, \\
\ket{0}^{\mathcal{U}}_{1}\ket{\hat{1}}^{\mathcal{V}^{[4]}}_{1} &:= \textstyle \ket{1(\frac{1}{2}\frac{1}{2})0;1~{-1}} = \ket{w_{2}}\ket{w_{1}}, \\
\ket{1}^{\mathcal{U}}_{1}\ket{\hat{1}}^{\mathcal{V}^{[4]}}_{1} &:= \textstyle \ket{1(\frac{1}{2}\frac{1}{2})0;10} = \ket{w_{3}}\ket{w_{1}}, \\
\ket{2}^{\mathcal{U}}_{1}\ket{\hat{1}}^{\mathcal{V}^{[4]}}_{1} &:= \textstyle \ket{1(\frac{1}{2}\frac{1}{2})0;11} = \ket{w_{4}}\ket{w_{1}}, \\
\ket{0}^{\mathcal{U}}_{1}\ket{\hat{2}}^{\mathcal{V}^{[4]}}_{1} &:= \textstyle \ket{1(\frac{1}{2}\frac{1}{2})1;1~{-1}} = \frac{1}{\sqrt{2}} (\ket{w_{2}} \ket{w_{3}} - \ket{w_{3}} \ket{w_{2}}), \\
\ket{1}^{\mathcal{U}}_{1}\ket{\hat{2}}^{\mathcal{V}^{[4]}}_{1} &:= \textstyle \ket{1(\frac{1}{2}\frac{1}{2})1;10} = \frac{1}{\sqrt{2}} (\ket{w_{2}} \ket{w_{4}} - \ket{w_{4}} \ket{w_{2}}), \\
\ket{2}^{\mathcal{U}}_{1}\ket{\hat{2}}^{\mathcal{V}^{[4]}}_{1} &:= \textstyle \ket{1(\frac{1}{2}\frac{1}{2})1;11} = \frac{1}{\sqrt{2}} (\ket{w_{3}} \ket{w_{4}} - \ket{w_{4}} \ket{w_{3}}), \\
\ket{0}^{\mathcal{U}}_{2}\ket{\hat{0}}^{\mathcal{V}^{[4]}}_{2} &:= \textstyle \ket{1(\frac{1}{2}\frac{1}{2})1;1~{-2}} = \ket{w_{2}}\ket{w_{2}} \\
\ket{1}^{\mathcal{U}}_{2}\ket{\hat{0}}^{\mathcal{V}^{[4]}}_{2} &:= \textstyle \ket{1(\frac{1}{2}\frac{1}{2})1;1~{-1}} = \frac{1}{\sqrt{2}} (\ket{w_{3}}\ket{w_{2}} + \ket{w_{2}}\ket{w_{3}}), \\
\ket{2}^{\mathcal{U}}_{2}\ket{\hat{0}}^{\mathcal{V}^{[4]}}_{2} &:= \textstyle \ket{1(\frac{1}{2}\frac{1}{2})1;10} = \frac{1}{\sqrt{6}} (\ket{w_{2}}\ket{w_{4}} + \ket{w_{4}}\ket{w_{2}} + 2 \ket{w_{3}}\ket{w_{3}}), \\
\ket{3}^{\mathcal{U}}_{2}\ket{\hat{0}}^{\mathcal{V}^{[4]}}_{2} &:= \textstyle \ket{1(\frac{1}{2}\frac{1}{2})1;11} = \frac{1}{\sqrt{2}} (\ket{w_{3}}\ket{w_{4}} + \ket{w_{4}}\ket{w_{3}}), \\
\ket{4}^{\mathcal{U}}_{2}\ket{\hat{0}}^{\mathcal{V}^{[4]}}_{2} &:= \textstyle \ket{1(\frac{1}{2}\frac{1}{2})1;12} = \ket{w_{4}}\ket{w_{4}}.  
\end{align}
The definition of the bases of the multiplicity subspaces corresponds to a composition of spin-$1/2$ particles starting from composing pair $\{\mathcal{K}_{1}$, $\mathcal{K}_{2}\}$ and $\{\mathcal{K}_{3}$, $\mathcal{K}_{4}\}$, individually, and followed by composition of the composed pairs.  
In these bases,  
\begin{align}
M_{2} &= 
\textstyle I_{0}^{\mathcal{K}} \otimes I_{0}^{\mathcal{H}} \otimes \left\langle \ket{\hat{0}\hat{0}}_{00}\right\rangle + \frac{1}{9} I_{0}^{\mathcal{K}} \otimes I_{0}^{\mathcal{H}} \otimes \left\langle \ket{\hat{1}\hat{1}}_{00}\right\rangle \notag \\ & \quad \oplus 
\textstyle \frac{1}{3} I_{1}^{\mathcal{K}} \otimes I_{1}^{\mathcal{H}} \otimes \left( \left\langle \ket{\hat{0}\hat{0}}_{11} \right\rangle + \left\langle \ket{\hat{1}\hat{1}}_{11} \right\rangle \right) + \frac{1}{9} I_{1}^{\mathcal{K}} \otimes I_{1}^{\mathcal{H}} \otimes \left\langle \ket{\hat{2}\hat{2}}_{11} \right\rangle  \oplus \frac{1}{9} I_{2}^{\mathcal{K}} \otimes I_{2}^{\mathcal{H}} \otimes \left\langle \ket{\hat{0}\hat{0}}_{22} \right\rangle \notag \\ & \quad \oplus
\textstyle\frac{1}{9} I_{0}^{\mathcal{K}} \otimes I_{1}^{\mathcal{H}} \otimes \left\langle \ket{\hat{1}\hat{2}}_{01} \right\rangle \oplus 
\textstyle\frac{1}{9} I_{0}^{\mathcal{K}} \otimes I_{2}^{\mathcal{H}} \otimes \left\langle \ket{\hat{1}\hat{0}}_{02} \right\rangle \notag \\ & \quad \oplus 
\textstyle\frac{1}{9} I_{1}^{\mathcal{K}} \otimes I_{0}^{\mathcal{H}} \otimes \left\langle \ket{\hat{2}\hat{1}}_{10} \right\rangle \oplus 
\textstyle\frac{1}{9} I_{2}^{\mathcal{K}} \otimes I_{0}^{\mathcal{H}} \otimes \left\langle \ket{\hat{0}\hat{1}}_{20} \right\rangle \notag \\ & \quad \oplus 
\textstyle\frac{1}{9} I_{1}^{\mathcal{K}} \otimes I_{2}^{\mathcal{H}} \otimes \left\langle \ket{\hat{2}\hat{0}}_{12} \right\rangle \oplus 
\textstyle\frac{1}{9} I_{2}^{\mathcal{K}} \otimes I_{1}^{\mathcal{H}} \otimes \left\langle \ket{\hat{0}\hat{2}}_{21} \right\rangle \\ &=
\textstyle I_{0}^{\mathcal{K}} \otimes I_{0}^{\mathcal{H}} \otimes \left( \left\langle \ket{\hat{0}\hat{0}}_{00}\right\rangle + \frac{1}{9} \left\langle \ket{\hat{1}\hat{1}}_{00}\right\rangle \right) \oplus 
\textstyle I_{0}^{\mathcal{K}} \otimes I_{1}^{\mathcal{H}} \otimes \frac{1}{9} \left\langle \ket{\hat{1}\hat{2}}_{01} \right\rangle \oplus 
\textstyle I_{0}^{\mathcal{K}} \otimes I_{2}^{\mathcal{H}} \otimes \frac{1}{9} \left\langle \ket{\hat{1}\hat{0}}_{01} \right\rangle \notag \\ & \quad \oplus 
\textstyle \frac{I_{1}^{\mathcal{K}}}{3} \otimes I_{0}^{\mathcal{H}} \otimes \frac{1}{3} \left\langle \ket{\hat{2}\hat{1}}_{10} \right\rangle \oplus 
\textstyle \frac{I_{1}^{\mathcal{K}}}{3} \otimes I_{1}^{\mathcal{H}} \otimes \left( \left\langle \ket{\hat{0}\hat{0}}_{11} \right\rangle + \left\langle \ket{\hat{0}\hat{0}}_{11} \right\rangle + \frac{1}{3} \left\langle \ket{\hat{2}\hat{2}}_{11} \right\rangle \right) \oplus 
\textstyle\frac{I_{1}^{\mathcal{K}}}{3} \otimes I_{2}^{\mathcal{H}} \otimes \frac{1}{3} \left\langle \ket{\hat{2}\hat{0}}_{12} \right\rangle \notag \\ & \quad \oplus 
\textstyle\frac{I_{2}^{\mathcal{K}}}{5} \otimes I_{0}^{\mathcal{H}} \otimes \frac{5}{9} \left\langle \ket{\hat{0}\hat{1}}_{20} \right\rangle \oplus 
\textstyle\frac{I_{2}^{\mathcal{K}}}{5} \otimes I_{1}^{\mathcal{H}} \otimes \frac{5}{9} \left\langle \ket{\hat{0}\hat{2}}_{21} \right\rangle \oplus
\textstyle\frac{I_{2}^{\mathcal{K}}}{5} \otimes I_{2}^{\mathcal{H}} \otimes \frac{5}{9} \left\langle \ket{\hat{0}\hat{0}}_{22} \right\rangle.  
\end{align}
 
Therefore, we have
\begin{align}
M^{(2)}_{00} &= \left\langle \ket{\hat{0}\hat{0}}_{00} \right\rangle + \frac{1}{9} \left\langle \ket{\hat{1}\hat{1}}_{00} \right\rangle ,\\
M^{(2)}_{01} &= \frac{1}{9} \left\langle \ket{\hat{1}\hat{2}}_{01} \right\rangle ,\\
M^{(2)}_{02} &= \frac{1}{9} \left\langle \ket{\hat{1}\hat{0}}_{02} \right\rangle ,\\
M^{(2)}_{10} &= \frac{1}{3}\left\langle \ket{\hat{2}\hat{1}}_{10} \right\rangle ,\\
M^{(2)}_{11} &= \left\langle \ket{\hat{1}\hat{1}}_{11} \right\rangle + \left\langle \ket{\hat{0}\hat{0}}_{11} \right\rangle + \frac{1}{3} \left\langle \ket{\hat{2}\hat{2}}_{11} \right\rangle ,\\
M^{(2)}_{12} &= \frac{1}{3}\left\langle \ket{\hat{2}\hat{0}}_{12} \right\rangle ,\\
M^{(2)}_{20} &= \frac{5}{9}\left\langle \ket{\hat{0}\hat{1}}_{20} \right\rangle ,\\
M^{(2)}_{21} &= \frac{5}{9}\left\langle \ket{\hat{0}\hat{2}}_{21} \right\rangle ,\\
M^{(2)}_{22} &= \frac{5}{9}\left\langle \ket{\hat{0}\hat{0}}_{22} \right\rangle . 
\end{align}

We define a $2 \times 2$ matrix $ V(j_{1}, j_{3}, j_{13}) $ by 
    \begin{align}
    V(j_{1}, j_{3}, j_{13})_{11} &:= \frac{(-1)^{2(j_{1} + j_{3} + j_{13})}}{\sqrt{(2j_{13} + 2)(2j_{1} + 1)}}\sqrt{\left( j_{3} + \frac{1}{2} \right)^{2} - \left(j_{13} - j_{1} + \frac{1}{2}\right)^2}, \\
    V(j_{1}, j_{3}, j_{13})_{12} &:= \frac{(-1)^{2(j_{1} + j_{3} + j_{13})}}{\sqrt{(2j_{13} + 2)(2j_{1} + 1)}} \sqrt{\left(j_{13} + j_{1} + \frac{3}{2}\right)^{2} - \left( j_{3}  + \frac{1}{2}\right)^{2}}, \\ 
    V(j_{1}, j_{3}, j_{13})_{21} &:= \frac{(-1)^{2(j_{1} + j_{3} + j_{13} )}}{\sqrt{(2j_{13} + 2)(2j_{1} + 1)}} \sqrt{\left(j_{13} + j_{1} + \frac{3}{2} \right)^{2} - \left(j_{3}+ \frac{1}{2} \right)^2}, \\
    V(j_{1}, j_{3}, j_{13})_{22} &:= - \frac{(-1)^{2(j_{1} + j_{3} + j_{13} )}}{\sqrt{(2j_{13} + 2)(2j_{1} + 1)}} \sqrt{\left(j_{3} + \frac{1}{2} \right)^{2} - \left( j_{13} - j_{1} + \frac{1}{2} \right)^{2}}.  
    \end{align}
Calculating Wigner's $6j$ coefficients, the relation between the bases is
\begin{align}
\ket{\hat{0}}_{0} &= V(0,1/2,0)_{12} \ket{0}_{0} ,\\
\ket{\hat{1}}_{0} &= V(0,1/2,0)_{21} \ket{1}_{0} ,\\
\ket{\hat{0}}_{1} &= V(0,1/2,1)_{22} \ket{0}_{1} ,\\
\ket{\hat{1}}_{1} &= V(1,1/2,1)_{11} \ket{1}_{1} + V(1,1/2,1)_{12} \ket{2}_{1} ,\\
\ket{\hat{2}}_{1} &= V(1,1/2,1)_{21} \ket{1}_{1} + V(1,1/2,1)_{22} \ket{2}_{1} ,\\
\ket{\hat{0}}_{2} &= V(1,1/2,2)_{22} \ket{0}_{2}.  
\end{align}

Swap operations on $\mathcal{H}_{i} \otimes \mathcal{H}_{j}$ and $\mathcal{K}_{i} \otimes \mathcal{K}_{j}$ are applied to change the ordering of the black-boxes.  In the multiplicity subspaces, these swap operations correspond to unitary operations within each multiplicity subspace. Let $U^{i:j}_{J}$ be such a unitary operation on the multiplicity subspace $\mathcal{V}_{J}^{[4]}$ when the $i$-th and $j$-th system are exchanged.  
By calculating Wigner's $6j$ coefficients, we can derive

\begin{align}
U_{0}^{1:2} &= \left(\begin{array}{cc}-1 & 0 \\0 & 1\end{array}\right), \quad 
U_{1}^{1:2} = \left(\begin{array}{ccc}-1 & 0 & 0 \\0 & 1 & 0 \\0 & 0 & 1\end{array}\right), \quad 
U_{2}^{1:2} = 1,  \\
U_{0}^{2:3} &= V(1/2, 1/2, 0)^{T}, \quad 
U_{1}^{2:3} = \left(\begin{array}{ccc}  &   & 0 \\ & \raisebox{1.2ex}[0ex][0ex]{$V(1/2, 1/2, 0)^{T}$} & 0 \\0 & 0 & V(1,1/2,1/2)_{22} \end{array}\right), U_{2}^{2:3} = V(1/2,1/2,1)_{12}, \\
U_{0}^{3:4} &= \left(\begin{array}{cc}V(0,1/2,-1/2)_{22} & 0 \\0 & V(1,1/2,-1/2) \end{array}\right), \\
U_{1}^{3:4} &= \left(\begin{array}{ccc} V(0,1/2,1/2) & 0  & 0 \\ 0 &   \\0 & \raisebox{1ex}[0ex][0ex]{$V(1, 1/2, 1/2)^{T}$} \end{array}\right),\quad 
U_{2}^{3:4} = V(1,1/2,3/2)_{12}.  
\end{align}

We add a superscript $\langle j \rangle$ as $M_{i}^{\langle j \rangle}$  to indicate the $12$ orderings of the reference and target boxes. Different $M_{i}^{\langle j \rangle}$ can be obtained by multiplying $U_{J}^{i':j'} \otimes U_{L}^{i':j'}$ and $U_{J}^{i':j' \dagger} \otimes U_{L}^{i':j' \dagger}$ on $M_{JL}^{\langle j \rangle (i)}$.  

%%%%%%%%%%%%%% SDO FOR (2,1)-EQUIVALENCE DETERMINATION IN THE MULTIPLICITY SUBSPACE %%%%%%%%%%%%%%
\section{SDP for $(2,1)$-equivalence determination in the multiplicity subspaces}
%In this section, we provide the SDPs for Concurrency Patterns (127) to (\ref{fig:multi4}) .  
In this section, we provide the SDPs for Concurrency Patterns (127) to (131) .

%%%%%%%%%%%%%%%%%%%%%%%%%%%%%%%%%%%%%%%%%%%%%%%%%%%%%%%%

%\subsection{Concurrency Pattern (\ref{fig:multi211})}
\subsection{Concurrency Pattern (127)}
The SDP for Concurrency Pattern (127) is
\begin{align}
\text{maximize}\quad & p_{succ} = \frac{1}{2} \tr \left[  M_{1}^{\langle j\rangle} \widetilde{\Pi}_{1} +  M_{2}^{\langle j\rangle} \widetilde{\Pi}_{2} \right], \notag \\
\text{subject to}\quad & \widetilde{\Pi}_{i} \geq 0, \ i = 1,2, \label{eq:sdp-211-0} \\
& \widetilde{\Pi}_{1} + \widetilde{\Pi}_{2} = I_{\mathcal{K}_{4}} \otimes Y^{\{3\}}, \label{eq:sdp-211-1} \\
& \tr_{\mathcal{H}_{4}} Y^{\{3\}} = I_{\mathcal{K}_{3}} \otimes Y^{\{2\}}, \label{eq:sdp-211-2}  \\
& \tr_{\mathcal{H}_{3}} Y^{\{2\}} = I_{\mathcal{K}_{1}\mathcal{K}_{2}} \otimes Y^{\{1\}}, \label{eq:sdp-211-3} \\
& \tr Y^{\{1\}} = 1.   \label{eq:sdp-211-4}
\end{align}
We rewrite Eqs.\,(\ref{eq:sdp-211-1}) - (\ref{eq:sdp-211-4}) in terms of the multiplicity subspaces.  To this end, we may assume without loss of generality that 
\begin{align}
\widetilde{\Pi}_{i} &= \bigoplus_{J,L = 0}^{2} I_{J}^{\mathcal{K}} \otimes \frac{I_{L}^{\mathcal{H}}}{d_{L}} \otimes \widetilde{\Pi}^{(i)}_{JL}, \\
Y^{\{3\}} &= \bigoplus_{J = \frac{1}{2}}^{\frac{3}{2}} \bigoplus_{L = 0}^{2} I_{J}^{\mathcal{K}_{1}\mathcal{K}_{2}\mathcal{K}_{3}} \otimes \frac{I_{L}^{\mathcal{H}}}{d_{L}} \otimes Y^{\{3\}}_{JL} \\
Y^{\{2\}} &= \bigoplus_{J = 0}^{1} \bigoplus_{L = \frac{1}{2}}^{\frac{3}{2}} I_{J}^{\mathcal{K}_{1}\mathcal{K}_{2}} \otimes \frac{I_{L}^{\mathcal{H}_{1}\mathcal{H}_{2}\mathcal{H}_{3}}}{d_{L}} \otimes Y^{\{2\}}_{JL}, \\
Y^{\{1\}} &= \bigoplus_{L = 0}^{1} \frac{I_{L}^{\mathcal{H}_{1}\mathcal{H}_{2}}}{d_{L}} \otimes Y^{\{1\}}_{L}, 
\end{align}
for $i = 1,2$.  The positivity condition (\ref{eq:sdp-211-0}) is equivalent to 
\begin{equation}
\widetilde{\Pi}^{(i)}_{JL} \geq 0,
\end{equation}
for $ i = 1,2 $ and $ J,L = 0,1,2$.  
We have 
\begin{align}
\widetilde{\Pi}_{1} + \widetilde{\Pi}_{2} 
&= \bigoplus_{J,L = 0}^{2} I_{J}^{\mathcal{K}} \otimes \frac{I_{L}^{\mathcal{H}}}{d_{L}} \otimes (\widetilde{\Pi}^{(1)}_{JL} + \widetilde{\Pi}^{(2)}_{JL}), \\
I_{\mathcal{K}_{4}} \otimes Y^{\{3\}} 
&= \bigoplus_{L = 0}^{2} I_{0}^{\mathcal{K}} \otimes \frac{I_{L}^{\mathcal{H}}}{d_{L}} \otimes Y^{\{3\}}_{\frac{1}{2}L} \oplus \bigoplus_{L = 0}^{2} I_{1}^{\mathcal{K}} \otimes \frac{I_{L}^{\mathcal{H}}}{d_{L}} \otimes (Y^{\{3\}}_{\frac{1}{2}L} \oplus Y^{\{3\}}_{\frac{3}{2}L}) \oplus \bigoplus_{L = 0}^{2} I_{2}^{\mathcal{K}} \otimes \frac{I_{L}^{\mathcal{H}}}{d_{L}} \otimes Y^{\{3\}}_{\frac{3}{2}L}.  
\end{align}
Thus Eq.\,(\ref{eq:sdp-211-1}) is equivalent to 
\begin{align}
\left(P_{0,\frac{1}{2}}^{[4]} \otimes I_{\mathcal{V}^{[4]}_{L}} \right) \left(\widetilde{\Pi}^{\{1\}}_{0L} + \widetilde{\Pi}^{\{2\}}_{0L}\right) \left(P_{0,\frac{1}{2}}^{[4]} \otimes I_{\mathcal{V}^{[4]}_{L}} \right)^{\dagger} &= Y^{\{3\}}_{\frac{1}{2}L},  \\
\left(P_{1,\frac{1}{2}}^{[4]} \otimes I_{\mathcal{V}^{[4]}_{L}} \right) \left(\widetilde{\Pi}^{\{1\}}_{1L} + \widetilde{\Pi}^{\{2\}}_{1L} \right) \left(P_{1,\frac{1}{2}}^{[4]} \otimes I_{\mathcal{V}^{[4]}_{L}} \right)^{\dagger} &= Y^{\{3\}}_{\frac{1}{2}L}, \\
\left(P_{1,\frac{3}{2}}^{[4]} \otimes I_{\mathcal{V}^{[4]}_{L}} \right) \left(\widetilde{\Pi}^{\{1\}}_{1L} + \widetilde{\Pi}^{\{2\}}_{1L} \right) \left(P_{1,\frac{3}{2}}^{[4]} \otimes I_{\mathcal{V}^{[4]}_{L}} \right)^{\dagger} &= Y^{\{3\}}_{\frac{3}{2}L}, \\
\left(P_{2,\frac{3}{2}}^{[4]} \otimes I_{\mathcal{V}^{[4]}_{L}} \right) \left(\widetilde{\Pi}^{\{1\}}_{2L} + \widetilde{\Pi}^{\{2\}}_{2L}\right) \left(P_{2,\frac{3}{2}}^{[4]} \otimes I_{\mathcal{V}^{[4]}_{L}} \right)^{\dagger} &= Y^{\{3\}}_{\frac{3}{2}L},  
\end{align}
for $L = 0,1,2$, where
\begin{align}
P_{0,\frac{1}{2}}^{[4]} &:= 
\ket{0}^{\mathcal{V}^{[3]}}_{\frac{1}{2}} \bra{0}^{\mathcal{V}^{[4]}}_{0} + \ket{1}^{\mathcal{V}^{[3]}}_{\frac{1}{2}} \bra{1}^{\mathcal{V}^{[4]}}_{0}, \\
P_{1,\frac{1}{2}}^{[4]} &:= 
\ket{0}^{\mathcal{V}^{[3]}}_{\frac{1}{2}} \bra{0}^{\mathcal{V}^{[4]}}_{1} + \ket{1}^{\mathcal{V}^{[3]}}_{\frac{1}{2}} \bra{1}^{\mathcal{V}^{[4]}}_{1}, \\
P_{1,\frac{3}{2}}^{[4]} &:= 
\ket{0}^{\mathcal{V}^{[3]}}_{\frac{3}{2}} \bra{2}^{\mathcal{V}^{[4]}}_{1}, \\
P_{2,\frac{3}{2}}^{[4]} &:= 
\ket{2}^{\mathcal{V}^{[3]}}_{\frac{3}{2}} \bra{0}^{\mathcal{V}^{[4]}}_{2}. \\
\end{align}

We have
\begin{align}
\tr_{\mathcal{H}_{4}} Y^{\{3\}} = \bigoplus_{J = \frac{1}{2}}^{\frac{3}{2}} I_{J}^{\mathcal{K}_{1}\mathcal{K}_{2}\mathcal{K}_{3}} 
&\otimes \Bigg[ 
\frac{I_{\frac{1}{2}}^{\mathcal{H}_{1}\mathcal{H}_{2}\mathcal{H}_{3}}}{d_{\frac{1}{2}}} \otimes \left\{
(I_{\mathcal{V}^{[3]}_{J}} \otimes P^{[4]}_{0,\frac{1}{2}}) Y_{J0}^{\{3\}} (I_{\mathcal{V}^{[3]}_{J}} \otimes P^{[4]}_{0,\frac{1}{2}})^{\dagger} + 
(I_{\mathcal{V}^{[3]}_{J}} \otimes P^{[4]}_{1,\frac{1}{2}}) Y_{J1}^{\{3\}} (I_{\mathcal{V}^{[3]}_{J}} \otimes P^{[4]}_{1,\frac{1}{2}})^{\dagger} \right\}\notag \\
&\oplus \frac{I_{\frac{3}{2}}^{\mathcal{H}_{1}\mathcal{H}_{2}\mathcal{H}_{3}}}{d_{\frac{3}{2}}} \otimes \left\{
(I_{\mathcal{V}^{[3]}_{J}} \otimes P^{[4]}_{1,\frac{3}{2}}) Y_{J1}^{\{3\}} (I_{\mathcal{V}^{[3]}_{J}} \otimes P^{[4]}_{1,\frac{3}{2}})^{\dagger} + 
(I_{\mathcal{V}^{[3]}_{J}} \otimes P^{[4]}_{2,\frac{3}{2}}) Y_{J2}^{\{3\}} (I_{\mathcal{V}^{[3]}_{J}} \otimes P^{[4] }_{2,\frac{3}{2}})^{\dagger} \right\}.
\Bigg]\label{eq:tr-h_4}
\end{align}
and
\begin{align}
I_{\mathcal{K}_{3}} \otimes Y^{\{2\}} &= \bigoplus_{L = \frac{1}{2}}^{\frac{3}{2}} \left[I_{\frac{1}{2}}^{\mathcal{K}_{1}\mathcal{K}_{3}\mathcal{K}_{3}} \otimes \frac{I_{L}^{\mathcal{H}_{1}\mathcal{H}_{2}\mathcal{H}_{3}}}{d_{L}} \otimes (Y_{0L}^{\{2\}} \oplus Y_{1L}^{\{2\}}) \oplus \bigoplus_{L = \frac{1}{2}}^{\frac{3}{2}} I_{\frac{3}{2}}^{\mathcal{K}_{1}\mathcal{K}_{3}\mathcal{K}_{3}} \otimes \frac{I_{L}^{\mathcal{H}_{1}\mathcal{H}_{2}\mathcal{H}_{3}}}{d_{L}} \otimes Y_{1L}^{\{2\}} \right].  
\end{align}
Therefore Eq.\,(\ref{eq:sdp-211-2}) is equivalent to 
\begin{align}
(P^{[3]}_{\frac{1}{2},0} \otimes P^{[4]}_{0,\frac{1}{2}}) Y_{\frac{1}{2}0}^{\{3\}} 
(P^{[3]}_{\frac{1}{2},0} \otimes P^{[4]}_{0,\frac{1}{2}})^{\dagger} + 
(P^{[3]}_{\frac{1}{2},0} \otimes P^{[4]}_{1,\frac{1}{2}}) Y_{\frac{1}{2}1}^{\{3\}}  
(P^{[3]}_{\frac{1}{2},0} \otimes P^{[4]}_{1,\frac{1}{2}})^{\dagger} &= Y^{\{2\}}_{0\frac{1}{2}} ,\\
(P^{[3]}_{\frac{1}{2},1} \otimes P^{[4]}_{0,\frac{1}{2}}) Y_{\frac{1}{2}0}^{\{3\}} 
(P^{[3]}_{\frac{1}{2},1} \otimes P^{[4]}_{0,\frac{1}{2}})^{\dagger} + 
(P^{[3]}_{\frac{1}{2},1} \otimes P^{[4]}_{1,\frac{1}{2}}) Y_{\frac{1}{2}1}^{\{3\}}  
(P^{[3]}_{\frac{1}{2},1} \otimes P^{[4]}_{1,\frac{1}{2}})^{\dagger} &= Y^{\{2\}}_{1\frac{1}{2}} ,\\
(P^{[3]}_{\frac{1}{2},0} \otimes P^{[4]}_{1,\frac{3}{2}}) Y_{\frac{1}{2}1}^{\{3\}} 
(P^{[3]}_{\frac{1}{2},0} \otimes P^{[4]}_{1,\frac{3}{2}})^{\dagger} + 
(P^{[3]}_{\frac{1}{2},0} \otimes P^{[4]}_{2,\frac{3}{2}}) Y_{\frac{1}{2}2}^{\{3\}}  
(P^{[3]}_{\frac{1}{2},0} \otimes P^{[4]}_{2,\frac{3}{2}})^{\dagger} &= Y^{\{2\}}_{0\frac{3}{2}}
,\\
(P^{[3]}_{\frac{1}{2},1} \otimes P^{[4]}_{1,\frac{3}{2}}) Y_{\frac{1}{2}1}^{\{3\}} 
(P^{[3]}_{\frac{1}{2},1} \otimes P^{[4]}_{1,\frac{3}{2}})^{\dagger} + 
(P^{[3]}_{\frac{1}{2},1} \otimes P^{[4]}_{2,\frac{3}{2}}) Y_{\frac{1}{2}2}^{\{3\}} 
(P^{[3]}_{\frac{1}{2},1} \otimes P^{[4]}_{2,\frac{3}{2}})^{\dagger} &= Y^{\{2\}}_{1\frac{3}{2}}
,\\
(P^{[3]}_{\frac{3}{2},1} \otimes P^{[4]}_{0,\frac{1}{2}}) Y_{\frac{3}{2}0}^{\{3\}}  
(P^{[3]}_{\frac{3}{2},1} \otimes P^{[4]}_{0,\frac{1}{2}})^{\dagger} + 
(P^{[3]}_{\frac{3}{2},1} \otimes P^{[4]}_{0,\frac{1}{2}}) Y_{\frac{3}{2}1}^{\{3\}}  
(P^{[3]}_{\frac{3}{2},1} \otimes P^{[4]}_{0,\frac{1}{2}})^{\dagger} &= Y^{\{2\}}_{1\frac{1}{2}} ,\\
(P^{[3]}_{\frac{3}{2},1} \otimes P^{[4]}_{1,\frac{3}{2}}) Y_{\frac{3}{2}1}^{\{3\}}  
(P^{[3]}_{\frac{3}{2},1} \otimes P^{[4]}_{1,\frac{3}{2}})^{\dagger} + 
(P^{[3]}_{\frac{3}{2},1} \otimes P^{[4]}_{2,\frac{3}{2}}) Y_{\frac{3}{2}2}^{\{3\}}  
(P^{[3]}_{\frac{3}{2},1} \otimes P^{[4]}_{2,\frac{3}{2}})^{\dagger} &= Y^{\{2\}}_{1\frac{3}{2}},  
\end{align}
where 
\begin{align}
P_{\frac{1}{2},0}^{[3]} &:= \ket{0}^{\mathcal{V}^{[2]}}_{0} \bra{\hat{0}}^{\mathcal{V}^{[3]}}_{\frac{1}{2}}, \\
P_{\frac{1}{2},1}^{[3]} &:= \ket{0}^{\mathcal{V}^{[2]}}_{1} \bra{\hat{1}}^{\mathcal{V}^{[3]}}_{\frac{1}{2}}, \\
P_{\frac{3}{2},1}^{[3]} &:= \ket{0}^{\mathcal{V}^{[2]}}_{1} \bra{\hat{0}}^{\mathcal{V}^{[3]}}_{\frac{3}{2}}. 
\end{align} 
We have
\begin{align}
\tr_{\mathcal{H}_{3}} Y^{\{2\}} &= \bigoplus_{J = 0}^{1} I_{J}^{\mathcal{K}_{1}\mathcal{K}_{2}} \otimes \left[ I_{0}^{\mathcal{H}_{1}\mathcal{H}_{2}} \otimes P_{\frac{1}{2},0}^{[3]} Y_{J\frac{1}{2}}^{\{2\}} P_{\frac{1}{2},0}^{[3] \dagger} \oplus \frac{I_{1}^{\mathcal{H}_{1}\mathcal{H}_{2}}}{d_{1}} \otimes \left(P_{\frac{1}{2},1}^{[3]} Y_{J\frac{1}{2}}^{\{2\}} P_{\frac{1}{2},1}^{[3] \dagger} + P_{\frac{3}{2},1}^{[3]} Y_{J\frac{3}{2}}^{\{2\}} P_{\frac{3}{2},1}^{[3] \dagger} \right) \right], \\
I_{\mathcal{K}_{1}\mathcal{K}_{2}} \otimes Y^{\{1\}} &= \bigoplus_{J = 0}^{1} I_{J}^{\mathcal{K}_{1}\mathcal{K}_{2}} \otimes \left[ I_{0}^{\mathcal{H}_{1}\mathcal{H}_{2}} \otimes Y^{\{1\}}_{0} \oplus \frac{I_{1}^{\mathcal{H}_{1}\mathcal{H}_{2}}}{d_{1}} \otimes Y^{\{1\}}_{1} \right].  
\end{align}
Therefore Eq.\,(\ref{eq:sdp-211-3}) is equivalent to 
\begin{align}
P_{\frac{1}{2},0}^{[3]} Y_{J\frac{1}{2}}^{\{2\}} P_{\frac{1}{2},0}^{[3] \dagger} &= Y^{\langle 1 \rangle}_{0}, \\
P_{\frac{1}{2},1}^{[3]} Y_{J\frac{1}{2}}^{\{2\}} P_{\frac{1}{2},1}^{[3] \dagger} + P_{\frac{3}{2},1}^{[3] } Y_{J\frac{3}{2}}^{(2)} P_{\frac{3}{2},1}^{[3] \dagger} &= Y^{\langle 1 \rangle}_{1}, 
\end{align}
for $J = 0,1 $.  For Eq.\,(\ref{eq:sdp-211-4}), we have
\begin{equation}
Y^{\{1\}}_{0} + Y^{\{1\}}_{1} = 1.  
\end{equation}

%%%%%%%%%%%%%%%%%%%%%%%%%%%%%%%%%%%%%%%%%%%%%%%%%%%%%%%%%%%%%%%%%%%%%%%%%%%%%%%%%%%%
\subsection{Concurrency Pattern (129)}
The SDP for Concurrency Pattern (129) is 
\begin{align}
\text{maximize}\quad & p_{succ} = \frac{1}{2} \tr \left[  M_{1}^{\langle j\rangle} \widetilde{\Pi}_{1} +  M_{2}^{\langle j\rangle} \widetilde{\Pi}_{2} \right] \notag \\
\text{subject to}\quad & \widetilde{\Pi}_{i} \geq 0, \ i = 1,2 \label{eq:sdp-22-0} \\
&\widetilde{\Pi}_{1} + \widetilde{\Pi}_{2} = I_{\mathcal{K}_{3}\mathcal{K}_{4}} \otimes Y^{\{2\}} \label{eq:sdp-22-1}\\
&\tr_{\mathcal{H}_{3}\mathcal{H}_{4}} Y^{\{2\}} = I_{\mathcal{K}_{2}\mathcal{K}_{1}} \otimes Y^{\{1\}} \label{eq:sdp-22-2}  \\
&\tr Y^{\{1\}} = 1,   \label{eq:sdp-22-3}
\end{align}
We rewrite Eqs.\,(\ref{eq:sdp-22-1}) - (\ref{eq:sdp-22-3}) in terms of the multiplicity subspaces.  We may assume without loss of generality that 
\begin{align}
\widetilde{\Pi}_{i} &= \bigoplus_{J,L = 0}^{2} I_{J}^{\mathcal{K}} \otimes \frac{I_{L}^{\mathcal{H}}}{d_{L}} \otimes \widetilde{\Pi}^{(i)}_{JL}, \\
Y^{\{2\}} &= \bigoplus_{J = 0}^{1} \bigoplus_{L = 0}^{2} I_{J}^{\mathcal{K}_{1}\mathcal{K}_{2}} \otimes \frac{I_{L}^{\mathcal{H}}}{d_{L}} \otimes Y^{\{2\}}_{JL}, \\
Y^{\{1\}} &= \bigoplus_{L = 0}^{1} \frac{I_{L}^{\mathcal{H}_{1}\mathcal{H}_{2}}}{d_{L}} \otimes Y^{\{1\}}_{JL}, 
\end{align}
for $i = 1,2$.  The positivity condition (\ref{eq:sdp-22-0}) is equivalent to 
\begin{equation}
\widetilde{\Pi}^{(i)}_{JL} \geq 0,
\end{equation}
for $ i = 1,2 $ and $ J,L = 0,1,2$.  
We have.  
\begin{align}
\widetilde{\Pi}_{1} + \widetilde{\Pi}_{2} =& \bigoplus_{J,L = 0}^{2} I_{J}^{\mathcal{K}} \otimes \frac{I_{L}^{\mathcal{H}}}{d_{L}} \otimes (\widetilde{\Pi}^{(1)}_{JL} + \widetilde{\Pi}^{(2)}_{JL}), \\
I_{\mathcal{K}_{3}\mathcal{K}_{4}} \otimes Y^{\{2\}} = 
&\bigoplus_{j = 0}^{2} I_{0}^{\mathcal{K}} \otimes \frac{I_{L}^{\mathcal{H}}}{d_{L}} \otimes ( Y_{0L}^{\{2\}} \oplus Y_{1L}^{\{2\}} ) \\
&\oplus\bigoplus_{j = 0}^{2} I_{1}^{\mathcal{K}} \otimes \frac{I_{L}^{\mathcal{H}}}{d_{L}} \otimes ( Y_{0L}^{\{2\}} \oplus Y_{1L}^{\{2\}} \oplus Y_{1L}^{\{2\}} ) \\
&\oplus\bigoplus_{j = 0}^{2} I_{2}^{\mathcal{K}} \otimes \frac{I_{L}^{\mathcal{H}}}{d_{L}} \otimes ( Y_{1L}^{\{2\}} ).  
\end{align}
Thus Eq.\,(\ref{eq:sdp-22-1}) is represented as is equivalent to 
\begin{align}
(P^{[3]}_{\frac{1}{2},1}P^{[4]}_{0,\frac{1}{2}} \otimes I_{\mathcal{V}^{[4]}_{L}}) (\widetilde{\Pi}^{(1)}_{0L} + \widetilde{\Pi}^{(2)}_{0L}) ( P^{[4] \dagger}_{0,\frac{1}{2}}P^{[3] \dagger}_{\frac{1}{2},1} \otimes I_{\mathcal{V}^{[4]}_{L}}) &=  Y^{\{2\}}_{0L} \\
(P^{[3]}_{\frac{1}{2},0}P^{[4]}_{0,\frac{1}{2}} \otimes I_{\mathcal{V}^{[4]}_{L}}) (\widetilde{\Pi}^{(1)}_{0L} + \widetilde{\Pi}^{(2)}_{0L}) ( P^{[4] \dagger}_{0,\frac{1}{2}}P^{[3] \dagger}_{\frac{1}{2},0} \otimes I_{\mathcal{V}^{[4]}_{L}}) &= Y^{\{2\}}_{1L} \\
(P^{[3]}_{\frac{1}{2},0}P^{[4]}_{1,\frac{1}{2}} \otimes I_{\mathcal{V}^{[4]}_{L}}) (\widetilde{\Pi}^{(1)}_{1L} + \widetilde{\Pi}^{(2)}_{1L}) ( P^{[4] \dagger}_{1,\frac{1}{2}}P^{[3] \dagger}_{\frac{1}{2},0} \otimes I_{\mathcal{V}^{[4]}_{L}}) &=  Y^{\{2\}}_{0L} \\
(P^{[3]}_{\frac{1}{2},1}P^{[4]}_{1,\frac{1}{2}} \otimes I_{\mathcal{V}^{[4]}_{L}}) (\widetilde{\Pi}^{(1)}_{1L} + \widetilde{\Pi}^{(2)}_{1L}) ( P^{[4] \dagger}_{1,\frac{1}{2}}P^{[3] \dagger}_{\frac{1}{2},1} \otimes I_{\mathcal{V}^{[4]}_{L}}) &=  Y^{\{2\}}_{1L} \\
(P^{[3]}_{\frac{3}{2},1}P^{[4]}_{1,\frac{3}{2}} \otimes I_{\mathcal{V}^{[4]}_{L}}) (\widetilde{\Pi}^{(1)}_{1L} + \widetilde{\Pi}^{(2)}_{1L}) ( P^{[4] \dagger}_{1,\frac{3}{2}}P^{[3] \dagger}_{\frac{3}{2},1} \otimes I_{\mathcal{V}^{[4]}_{L}}) &=  Y^{\{2\}}_{1L} \\
(P^{[3]}_{\frac{3}{2},1} P^{[4]}_{2,\frac{3}{2}} \otimes I_{\mathcal{V}^{[4]}_{L}}) (\widetilde{\Pi}^{(1)}_{2L} + \widetilde{\Pi}^{(2)}_{2L}) (P^{[4] \dagger}_{2, \frac{3}{2}} P^{[3] \dagger}_{\frac{3}{2},1} \otimes I_{\mathcal{V}^{[4]}_{L}}) &= Y^{\{2\}}_{1L},  
\end{align}
 for $L = 0,1,2$.  
We have
\begin{align}
\tr_{\mathcal{H}_{3}\mathcal{H}_{4}} Y^{\{2\}} & = \bigoplus_{J = 0}^{1} I_{J}^{\mathcal{K}_{1}\mathcal{K}_{2}} \otimes I_{0}^{\mathcal{H}_{1}\mathcal{H}_{2}} \otimes \left[ P^{[3]}_{\frac{1}{2},0}P^{[4]}_{0,\frac{1}{2}} Y^{\{2\}}_{J0} P^{[4] \dagger}_{0,\frac{1}{2}} P^{[3] \dagger}_{\frac{1}{2},0} + P^{[3]}_{\frac{1}{2},0}P^{[4]}_{1,\frac{1}{2}} Y^{\{2\}}_{J1} P^{[4] \dagger}_{1,\frac{1}{2}} P^{[3] \dagger}_{\frac{1}{2},0} \right] \\
&\quad\oplus \bigoplus_{J = 0}^{1} I_{J}^{\mathcal{K}_{1}\mathcal{K}_{2}} \otimes \frac{I_{1}^{\mathcal{H}_{1}\mathcal{H}_{2}}}{d_{1}} \otimes \big[
P^{[3]}_{\frac{1}{2},1}P^{[4]}_{0,\frac{1}{2}} Y^{\{2\}}_{J0} P^{[4] \dagger}_{0,\frac{1}{2}} P^{[3] \dagger}_{\frac{1}{2},1} + 
P^{[3]}_{\frac{1}{2},1}P^{[4]}_{1,\frac{1}{2}} Y^{\{2\}}_{J1} P^{[4] \dagger}_{1,\frac{1}{2}} P^{[3] \dagger}_{\frac{1}{2},1} \notag \\
&\qquad\qquad\qquad\qquad \qquad \qquad + P^{[3]}_{\frac{3}{2},1}P^{[4]}_{1,\frac{3}{2}} Y^{\{2\}}_{J1} P^{[4] \dagger}_{1,\frac{3}{2}} P^{[3] \dagger}_{\frac{3}{2},1} +
P^{[3]}_{\frac{3}{2},1}P^{[4]}_{2,\frac{3}{2}} Y^{\{2\}}_{J2} P^{[4] \dagger}_{2,\frac{3}{2}} P^{[3] \dagger}_{\frac{3}{2},1} \big],
\end{align}
and 
\begin{equation}
I_{\mathcal{K}_{1}\mathcal{K}_{2}} \otimes Y^{\{1\}} = \bigoplus_{J = 0}^{1} I_{J}^{\mathcal{K}_{1}\mathcal{K}_{2}} \otimes \left[ I_{0}^{\mathcal{H}_{1}\mathcal{H}_{2}} \otimes Y^{\{1\}}_{0} \oplus \frac{I_{1}^{\mathcal{H}_{1}\mathcal{H}_{2}}}{d_{1}} \otimes Y^{\{1\}}_{1} \right].
\end{equation}

Therefore Eq.\,(\ref{eq:sdp-22-2}) is equivalent to 
\begin{align} 
P^{[3]}_{\frac{1}{2},0}P^{[4]}_{0,\frac{1}{2}} Y^{\{2\}}_{J0} P^{[4] \dagger}_{0,\frac{1}{2}} P^{[3] \dagger}_{\frac{1}{2},0} + P^{[3]}_{\frac{1}{2},0}P^{[4]}_{1,\frac{1}{2}} Y^{\{2\}}_{J1} P^{[4] \dagger}_{1,\frac{1}{2}} P^{[3] \dagger}_{\frac{1}{2},0}  &= Y^{\{1\}}_{0} \\
P^{[3]}_{\frac{1}{2},1}P^{[4]}_{0,\frac{1}{2}} Y^{\{2\}}_{J0} P^{[4] \dagger}_{0,\frac{1}{2}} P^{[3] \dagger}_{\frac{1}{2},1} + 
P^{[3]}_{\frac{1}{2},1}P^{[4]}_{1,\frac{1}{2}} Y^{\{2\}}_{J1} P^{[4] \dagger}_{1,\frac{1}{2}} P^{[3] \dagger}_{\frac{1}{2},1} + 
P^{[3]}_{\frac{3}{2},1}P^{[4]}_{1,\frac{3}{2}} Y^{\{2\}}_{J1} P^{[4] \dagger}_{1,\frac{3}{2}} P^{[3] \dagger}_{\frac{3}{2},1} +
P^{[3]}_{\frac{3}{2},1}P^{[4]}_{2,\frac{3}{2}} Y^{\{2\}}_{J2} P^{[4] \dagger}_{2,\frac{3}{2}} P^{[3] \dagger}_{\frac{3}{2},1} &= Y^{\{1\}}_{1} 
\end{align}
for $J = 0,1$.  Eq.\,(\ref{eq:sdp-22-2}) is equivalent to
\begin{equation}
Y^{\{1\}}_{0} + Y^{\{1\}}_{1} = 1.  
\end{equation}

%%%%%%%%%%%%%%%%%%%%%%%%%%%%%%%%%%%%%%%%%%%%%%%%%%%%%%%%%%%%%%%%%%%%%%
\subsection{Concurrency Pattern (130)}
The SDP for Concurrency Pattern (130) is 
\begin{align}
\text{maximize}\quad & p_{succ} = \frac{1}{2} \tr \left[  M_{1}^{\langle j\rangle} \widetilde{\Pi}_{1} +  M_{2}^{\langle j\rangle} \widetilde{\Pi}_{2} \right] \notag \\
\text{subject to}\quad & \widetilde{\Pi}_{i} \geq 0, \ i = 1,2 \label{eq:sdp-31-0}\\
&\widetilde{\Pi}_{1} + \widetilde{\Pi}_{2} = I_{\mathcal{K}_{4}} \otimes Y^{\{2\}} \label{eq:sdp-31-1}\\
&\tr_{\mathcal{H}_{4}} Y^{\{2\}} = I_{\mathcal{K}_{1}\mathcal{K}_{2}\mathcal{K}_{2}} \otimes Y^{\{1\}} \label{eq:sdp-31-2}  \\
&\tr Y^{\{1\}} = 1,  \label{eq:sdp-31-3}
\end{align}
We rewrite Eqs.\,(\ref{eq:sdp-31-1}) - (\ref{eq:sdp-31-3}) in terms of the multiplicity subspaces.  We can assume without loss of generality that
\begin{align}
\widetilde{\Pi}_{i} &= \bigoplus_{J,L = 0}^{2} I_{J}^{\mathcal{K}} \otimes \frac{I_{L}^{\mathcal{H}}}{d_{L}} \otimes \widetilde{\Pi}^{(i)}_{JL} \\
Y^{\{2\}} &= \bigoplus_{J = \frac{1}{2}}^{\frac{3}{2}} \bigoplus_{L = 0}^{2} I_{J}^{\mathcal{K}_{1}\mathcal{K}_{2}\mathcal{K}_{3}} \otimes \frac{I_{L}^{\mathcal{H}}}{d_{L}} \otimes Y^{\{2\}}_{JL} \\
Y^{\{1\}} &= \bigoplus_{L = \frac{1}{2}}^{\frac{3}{2}} \frac{I_{L}^{\mathcal{H}_{1}\mathcal{H}_{1}}}{d_{L}} \otimes Y^{\{1\}}_{L}.  
\end{align}
The positivity condition (\ref{eq:sdp-31-0}) is equivalent to 
\begin{equation}
\widetilde{\Pi}^{(i)}_{JL} \geq 0,
\end{equation}
for $ i = 1,2 $ and $ J,L = 0,1,2$.  
Equation (\ref{eq:sdp-31-1}) is same as Eq.\,(\ref{eq:sdp-211-1}) in Condiguration $1$.  
For Eq.\,(\ref{eq:sdp-31-2}), the LHS is the same as the on in Eq.\,(\ref{eq:tr-h_4}) and we have
\begin{align}
I_{\mathcal{K}_{1}\mathcal{K}_{2}\mathcal{K}_{3}} \otimes Y^{\{1\}} = \bigoplus_{J = \frac{1}{2}}^{\frac{3}{2}}\bigoplus_{L = \frac{1}{2}}^{\frac{3}{2}} I_{J}^{\mathcal{K}} \otimes \frac{I_{L}^{\mathcal{H}}}{d_{L}} \otimes (I_{\mathcal{V}^{[3]}_{J}} \otimes Y^{\{1\}}_{L}).  
\end{align}
Therefore Eq.\,(\ref{eq:sdp-31-2}) is equivalent to
\begin{align}
I_{\mathcal{V}^{[3]}_{J}}\otimes Y^{\{1\}}_{\frac{1}{2}} &= (I_{\mathcal{V}^{[3]}_{J}} \otimes P^{[4]}_{0,\frac{1}{2}}) Y_{J0}^{\{2\}} (I_{\mathcal{V}^{[3]}_{J}} \otimes P^{[4] \dagger}_{0,\frac{1}{2}}) + 
(I_{\mathcal{V}^{[3]}_{J}} \otimes P^{[4]}_{1,\frac{1}{2}}) Y_{J1}^{\{2\}} (I_{\mathcal{V}^{[3]}_{J}} \otimes P^{[4] \dagger}_{1,\frac{1}{2}}) \\
I_{\mathcal{V}^{[3]}_{J}}\otimes Y^{\{1\}}_{\frac{3}{2}} &= (I_{\mathcal{V}^{[3]}_{J}} \otimes P^{[4]}_{1,\frac{3}{2}}) Y_{J1}^{\{2\}} (I_{\mathcal{V}^{[3]}_{J}} \otimes P^{[4] \dagger}_{1,\frac{3}{2}}) + 
(I_{\mathcal{V}^{[3]}_{J}} \otimes P^{[4]}_{2,\frac{3}{2}}) Y_{J2}^{\{2\}} (I_{\mathcal{V}^{[3]}_{J}} \otimes P^{[4] \dagger}_{2,\frac{3}{2}})
\end{align}
for $J= \frac{1}{2}, \frac{3}{2}$.  Equation (\ref{eq:sdp-31-3}) is equivalent to
\begin{equation}
\sum_{L = \frac{1}{2}}^{\frac{3}{2}} \tr Y^{\{1\}}_{L} = 1.  
\end{equation}

%%%%%%%%%%%%%%%%%%%%%%%%%%%%%%%%%%%%%%%%%%%%%%%%%%%%%%%%%%%%%%%%%%%%%%%%%%%%%%%%%%%%
\subsection{Concurrency Pattern (131)}
The SDP for Concurrency Pattern (131) is 
\begin{align}
\text{maximize}\quad & p_{succ} = \frac{1}{2} \tr \left[  M_{1}^{\langle j\rangle} \widetilde{\Pi}_{1} +  M_{2}^{\langle j\rangle} \widetilde{\Pi}_{2} \right] \notag \\
\text{subject to}\quad & \widetilde{\Pi}_{i} \geq 0, \ i = 1,2 \label{eq:sdp-4-0} \\
&\widetilde{\Pi}_{1} + \widetilde{\Pi}_{2} = I_{\mathcal{K}_{1}\mathcal{K}_{2}\mathcal{K}_{3}\mathcal{K}_{4}} \otimes X \label{eq:sdp-4-1}  \\
&\tr X = 1,  \label{eq:sdp-4-2}
\end{align}
We rewrite Eqs.\,(\ref{eq:sdp-4-1}) - (\ref{eq:sdp-4-2}) in the multiplicity subspaces.  We may assume without loss of generality that
\begin{align}
\widetilde{\Pi}_{i} &= \bigoplus_{J,L = 0}^{2} I_{J}^{\mathcal{K}} \otimes \frac{I_{L}^{\mathcal{H}}}{d_{L}} \otimes \widetilde{\Pi}^{(i)}_{JL} \\
X &= \bigoplus_{L = 0}^{2} \frac{I_{L}^{\mathcal{H}}}{d_{L}} \otimes X_{L}.  
\end{align}
The positivity condition (\ref{eq:sdp-4-0}) is equivalent to 
\begin{equation}
\widetilde{\Pi}^{(i)}_{JL} \geq 0,
\end{equation}
for $ i = 1,2 $ and $ J,L = 0,1,2$.  
Equation (\ref{eq:sdp-4-1}) is equivalent to
\begin{equation}
\widetilde{\Pi}^{(1)}_{JL} + \Pi^{(2)}_{JL} = I_{\mathcal{V}^{[4]}_{J}} \otimes X_{L} 
\end{equation}
for $J,L = 0,1,2$ and Eq.\,(\ref{eq:sdp-4-2}) is to
\begin{align}
\sum_{L = 0}^{2} \tr X_{L} &= 1. 
\end{align}